\documentclass[article]{jfm}

\usepackage{graphicx}
\usepackage{newtxtext}
\usepackage{newtxmath}
\usepackage{natbib}
\usepackage{hyperref}
\usepackage{subcaption}
\usepackage{caption} 
\usepackage{floatrow}
\usepackage{stackengine} 
\usepackage{hyperref}
\usepackage{todonotes}
\usepackage{xcolor}
\usepackage{upgreek}

\definecolor{darkred}{RGB}{139,0,0}
\definecolor{lightsalmon}{RGB}{255,160,122}
\definecolor{tan}{RGB}{210,180,140}
\definecolor{uColor}{RGB}{69, 1, 36}
\definecolor{vColor}{RGB}{200, 15, 90}
\definecolor{wColor}{RGB}{255, 153, 51}
\definecolor{QI}{RGB}{68, 1, 84}
\definecolor{QII}{RGB}{48, 103, 141}
\definecolor{QIII}{RGB}{53, 183, 121}
\definecolor{QIV}{RGB}{253, 231, 37}
\definecolor{richPurple}{rgb}{0.5, 0.1, 0.7}

\let\oldautoref\autoref
\renewcommand{\autoref}[1]{%
  \def\figureautorefname{Figure}
  \oldautoref{#1}%
}

\hypersetup{
    colorlinks = true,
    linkcolor  = blue,
    urlcolor   = blue,
    citecolor  = black,
}

\newcommand{\RomanNumeralCaps}[1]

\title{Secondary flows drive triboelectric powder charging in pneumatic conveying}

\author{
Gizem Ozler\aff{1,2}
\corresp{\email{gizem.oezler@ptb.de}}
\and Holger Grosshans\aff{1, 2}
}

\affiliation{
\aff{1}Physikalisch- Technische Bundesanstalt (PTB), Braunschweig, Germany
\aff{2}Otto von Guericke University of Magdeburg, Institute of Aparatus and Environmental Technology, Magdeburg, Germany
}

\begin{document}
\maketitle

\begin{abstract}
Highly resolved simulations reveal the fundamental influence of the carrier fluid's flow dynamics on triboelectric powder charging.
We found that particles transported through a square-shaped duct charge faster than in a channel flow caused by secondary flows that led to more severe particle-wall collisions.
Specifically, particles with $St$~=~4.69 achieve 85\% of their equilibrium charge approximately 1.5 times faster in duct flow than in channel flow.
Also, charge distribution is more uniform in a duct cross-section compared to a channel cross-section. 
In channel flow, particles are trapped near the walls and collide frequently due to limited movement in the wall-normal direction, causing localized charge buildup. 
In contrast, duct flow promotes better mixing through secondary flows, reducing repeating collisions and providing uniform charge distribution across the cross-section.
Upon charging, electrostatic forces significantly reshape particle behavior and distribution. 
Once the powder achieves half of its equilibrium charge, particles increasingly accumulate at the wall, leading to a reduced concentration in the central region. 
These changes in particle distribution have a noticeable impact on the surrounding fluid phase and alter the overall flow dynamics.
These findings open the possibility for a new measure to control powder charging by imposing a specific pattern.
\end{abstract}

\section{Introduction}
\label{sec:headings}
Triboelectric charging plays a role in a wide range of contexts, from the static electricity generated when removing synthetic garments to handling industrial powders. In nature, particles triboelectrically charge during volcanic eruptions~\citep{Jam00, Mat06}, dust devils, and sandstorms~\citep{Sto69}. Their charging directly influences particle dynamics and leads to measurable environmental consequences. For instance, triboelectric effects enable volcanic ash to ascend into the ionosphere, impacting Earth's climate systems~\citep{Gen18}.

Amid industrial operations, particles electrify particularly during pneumatic conveying~\citep{Kli18}.
To date the control of powder electrification is carried out empirically~\citep{Nom03} and even basic concepts are still being debated~\citep{Wei15}.
As a result, hazardous spark discharges occur that are a source of fatal dust explosions, loss of lives, and economic damage~\citep{Glor03,Osh11}.

Past efforts to control powder charging primarily focused on the material properties of the particles and the piping system.
However, the scientific advances in recent years thanks to the development of new highly-sensitive measurement apparatus and detailed theoretical models (an extensive review is provided, e.g., by~\citet{Lacks19}) indicate that a conclusive understanding of triboelectricity cannot be achieved in the near future.
\citet{Mat03} demonstrated that even under identical impact conditions, the charge transfer between a particle and a target during contact is not reproducible.
This observation is not surprising considering that the contact potential of a surface is far from being constant but instead a temporally varying mosaic of oppositely charged regions of nanoscopic dimensions~\citep{Bay11}.
Similar astonishing is the occurrence of bipolar charge in powders~\citep{Zhao03,Bil14}.
Also, it was recently demonstrated that triboelectric charging is time-dependent, irrespective of the precise charge carrier or mechanism.
Moreover, shaker experiments revealed that competing processes act over multiple timescales in non-equilibrium~\citep{Shi18,Yin18a,Yin18b}.
This and the above-discussed findings imply that the outcome of a contact charging event may be inherently unstable, and non-reproducible.
Moreover, the complexity of the underlying physics prohibited the formulation of a generally valid theoretical model.
For this reason, the most common measure aiming to control particle charging, namely the manipulation of their surfaces, is so far of limited success.

However, there is evidence that processes acting on the macroscopic powder level may be decisive for the charging rate of particle-laden flows.
Empirical analysis of powder charging during pneumatic conveying using Faraday cages revealed global trends, e.g., the relation between the flow velocity, solid mass loading, or particle material with the total powder charge~\citep{Wata06,Nda11,Pel18,Schw17}.
The disadvantage of this method is that it is not capable of providing a detailed view of the electrification process and the mechanisms that cause these trends cannot be expected.

For this purpose, the development of predictive numerical tools seems more promising.
Large-Eddy Simulations (LES), where the large turbulent scales are resolved, confirmed the strong dependence of powder charging on the flow Reynolds number, the mass flow rate of the powder, and the pipe diameter~\citep{Lim12,Kor14,Gro16}.
However, LES is not free of potential errors, especially regarding the dynamics in the near-wall regions where charge separation takes place.
Direct Numerical Simulations (DNS) where all turbulence scales are resolved on the grid gave evidence that the charging rate of powder may be determined by the occurrence of small-scale flow mechanisms~\citep{Gro17a}.
More specifically, in a turbulent channel flow of $Re_\uptau=180$ at moderate Stokes numbers and low particle volume fractions, the electric charge builds up but cannot escape the viscous sublayer due to the turbophoretic drift.
In this case, the particles close to the wall reach at some point equilibrium charge, but most particles in the bulk remain uncharged.
The complete flow can only charge if either the Stokes number or the particle number density is increased, thus, giving rise to particle-bound charge transport or inter-particle charge diffusion, respectively.

In addition, flow patterns influence powder charging. 
For example, wall-bounded turbulence was found to suppress bipolar particle charging ~\citep{Simon}.
Moreover, turbulence causes mid-sized particles to accumulate the most negative charge and not the smallest ones, as previously assumed.

\begin{figure}[t]
    \centering
    \captionsetup[subfigure]{labelformat=empty}
    \includegraphics[width=0.7\textwidth]{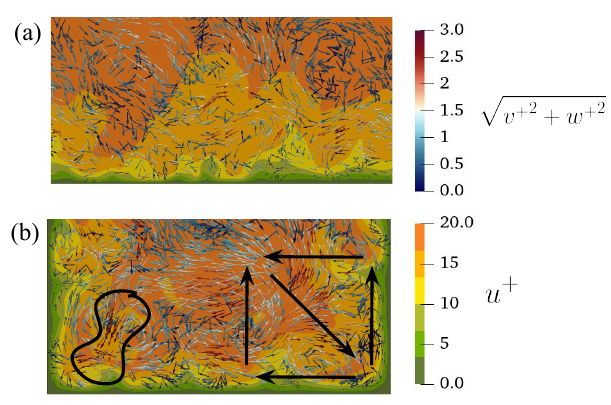}
    \caption{Normalized stream-wise velocity contours and cross velocity vectors of fluid in channel (a) and duct (b) flows' cross-section. The lower left half of the channel and lower half of the duct are shown. Arrows represent the direction of the secondary flows. The confined area shows fluid acceleration to the corners due to coinciding vortices.}
    \label{fig:cross}
\end{figure}

Channel flows, with their predominantly one-dimensional nature, are mainly affected by proximity to the walls. 
Conversely, duct flows lack a uniform direction, resulting in cross-sectional secondary flow structures, as illustrated in Figure~\ref{fig:cross}\textcolor{blue}{(b)}.
Although weaker, these secondary flows significantly impact particle behavior by altering fluid velocity gradients. 
The interaction of cross vortices accelerates particles toward the walls, increasing their collision frequency. 
This secondary flow drives particles from the corners along the walls towards the center, enhancing uniform mixing within the duct.
In contrast, in channels, particles at the wall are driven towards the bulk of the flow primarily through a random sequence of small-scale turbulent motions.
Thus, a greater number of particles are concentrated near the wall which reduces particle-bound charge transport.

\captionsetup{justification=justified, width=1\linewidth}
\vspace{0.5\baselineskip}

Upon charging, particles experience substantial changes in their behavior.
Measurements of the mean electric field generated by charged dust particles indicate strengths that could be comparable to gravitational forces, potentially leading to the elevation of dust in the presence of electric fields~\citep{Kok2008}.
\citet{Gro21} employed DNS to explore how electrostatic forces affect wall turbulence in a square duct with monodisperse charged particles.
Their findings revealed an active suppression of the vortical motion of particles within secondary flow structures by electrostatic forces. 
Consequently, there was a significant increase in particle number density at the bisectors of the walls.
Another study, utilizing direct numerical simulations of mono and bidisperse particles at a friction Reynolds number $Re_\tau=550$ in turbulent channel flow, reveals significant alterations in particle behavior when the electrostatic Stokes number is on the order of $O(10^{-1})$ ~\citep{Zhang23}.
These changes in the particles' behaviors due to the charging of powder result in alterations in the fluid field.
A more recent DNS study of turbulent channel flow at $Re_\tau=540$ involving bidisperse particles shows that inter-particle electrostatic forces significantly impact turbulent characteristics. Specifically, the presence of electrostatic forces alters both the intensity and structure of turbulence. They found that the inner-scaled mean stream-wise fluid velocity decreases in charged particle flows, which suggests an increase in fluid friction velocity ~\citep{Zhang24}.

Based on these recent findings, we elaborate in this study on a new way to solve the problem, namely the possible control of the triboelectric charging of powder flows by imposing certain flow dynamics.

To this end, direct numerical simulations using the tool~\citet{pafiX} are performed.
Therein, the flow dynamics is solved in an Eulerian framework.
Each particle is tracked individually as a Lagrangian point by solving Newton’s equations of motion where fluid drag, electrostatic, and collisional forces are accounted for.
A full description of the solver and the numerical scheme and their validation are given in Refs.~\citep{Gro17a,Gro21}.
 
The computation of electrostatic forces between charged particles is challenging due to the limitations of existing computational methods. Coulomb’s law calculates the forces between each particle directly but becomes computationally expensive for dense particulate flows as the cost scales quadratically with the number of particles, $O(N^{2})$.
Gauss's law uses charge density instead of calculating individual particle interactions, making it efficient for dense flows. However, its accuracy relies on grid resolution, which must be at least comparable to the particle size to accurately capture electrostatic interactions between nearby particles.

To overcome the difficulties mentioned above, several computational techniques have been developed. One example is the Fast Multipole Method (FMM), which reduces computational costs in pairwise calculations to $O(N\textup{log}N)$ by grouping distant particles~\citep{Rok90}. The combined effects of the grouped particles are then approximated by using fast multipole expansions. This approach is often applied to molecular dynamic simulations~\citep{Boa92,Din92}. Recently, \citet{Rua24} used this approach in particle-laden turbulent flow simulations to study the influence of charge segregation on the dynamics of bidisperse inertial particles. Although commonly used, this approach has challenges, such as numerical instability and a high error constant~\citep{Dar00}. Addressing these challenges requires additional techniques that add complexity to the implementation of the approach to numerical simulations.

Another approach, the Particle-Mesh (PM) method projects the particle charges onto a grid and solves the Poisson equation to determine the electric potential. This method assumes that the electrostatic forces between nearby particles are negligible compared to the collective effect of distant particles. While this makes it computationally efficient, it causes significant errors, especially for dense systems or systems where particles cluster~\citep{Yao18}.

The Particle-Particle Particle-Mesh (P3M) method \citep{Hoc88} builds on the Particle-Mesh (PM) method. It uses a computational grid to approximate long-range forces by solving the Poisson equation in Fourier space while short-range forces are calculated using Coulomb’s law. To avoid double-counting from both ranges, the long-range potential is modified with a correction term, and the computational cost scales with $O(N\textup{log}N)$. The Particle-Particle Particle-Mesh (P3M) method was initially designed to predict interactions between ions in electrolytic solutions. However, it has also been applied to simulations of charged inertial particles in a Taylor-Green vortex and an isotropic turbulent flow~\citep{Yao18}. 
While the method is useful for mono-dispersed, uniform particulate flows, and homogeneous charge distributions as in ionic solutions, it faces challenges when particles are clustered or when high precision is needed, which is the case in particle-laden flows.
Moreover, this method's use of Fourier transforms limits its application to the periodic boundary conditions~\citep{Gro17c}.

The Pseudo Particle Method (PPM), developed by \citet{Bou23}, estimates the short-range interactions via the sum of particle-particle interactions using Coulomb's law, similar to P3M. However, it differs from the P3M approach, since long-range interactions are also approximated using Coulomb’s law, but in an approximate form through pseudo-particles. In this approach, the distant particles are grouped into clusters (pseudo-particles), and their collective influence is calculated based on Coulomb's law.
The computational efficiency of the PPM scales as $O(N^{1.5})$.

In this work, we used the hybrid approach proposed by ~\citep{Gro17c}. In this approach, interactions between neighboring particles (particles in the same computational cells) are computed directly using Coulomb’s law, and superposed on the particle in question; while the collective effects of particles in other cells are approximated using Gauss’s law. This approach brings similar accuracy to direct calculation using Gauss's law and reduces computational costs by a factor of eight. Also, it is more suitable for wall-bounded flows as compared to the P3M method.

As discussed above, there is no consensus regards the physics of the charging of individual particles.
Only a few Computational Fluid Dynamics (CFD) models are available, namely the condenser model for conductive~\citep{Soo71,Mas76,John80}, and insulating~\citep{Gro17b} surfaces, and models relying on the surface state theory~\citep{Low86a,Low86b,Duff08,Lacks07,Kon17}.
All these models handle only very specific situations, require tuning, and are, thus, not able to generally predict particle charging.

In the current simulations, we employ a simple approach that funds upon two basic relations which are commonly observed under nearly all circumstances. 
First, particles gain charge upon contact with another object in their contact area.
Second, with repeating contact, the charge reaches asymptotically a certain saturation value or equilibrium charge, $q_\mathrm{eq}$. 
This simple model is sufficient for the purpose of this study since the conclusions drawn herein are largely independent of the precise amount of exchanged charge.

This paper is organized as follows. 
Section 2 details the mathematical model and computational methods employed. 
Section 3 describes the numerical setup of simulations. 
Section 4 presents and analyzes the numerical results. 
Finally, Section 5 offers concluding remarks.

\section{Mathematical model and numerical methods} 

We used the tool~\citet{pafiX} for conducting particle-laden flow simulations. 
The subsequent section provides an overview of the mathematical model and computational methods employed in pafiX.

The system's governing equations consist of three interconnected components: (i) the Navier-Stokes equations describe the flow of the carrier gas, (ii) Gauss's law governs the electrostatic field, and (iii) Newton's law of motion controls the movement of particles. 
While components (i) and (ii) are established within the Eulerian framework, the equations governing particle motion are addressed using the Lagrangian approach. 
The model accounts for four-way coupling between the fluid and particle phases, which means momentum exchange in both directions and interactions among individual particles are considered.

For the constant-density carrier gas flow, the Navier-Stokes equations are expressed as follows:
\begin{equation}
\mathbf{\nabla} \cdot\textbf{\textit{u}}=0 ,
\label{eq:Nav-1}
\end{equation}
\begin{equation}
\frac{\partial {\textbf{\textit{u}}}}{\partial t}+
(\textbf{\textit{u}} \cdot \mathbf{\nabla})\textbf{\textit{u}}=-\frac{1}{\rho_\text f}{\nabla}P+\nu\nabla ^{2}{\textbf{\textit{u}}}+\textbf{\textit{F}}_\text s + \textbf{\textit{F}}_\text f.
\label{eq:Nav-2}
\end{equation}

In these equations, $\textbf{\textit{u}}$ represents the velocity vector of the fluid, $\rho_\text f$ denotes the fluid density, $P$ indicates the dynamic pressure, and $\nu$ represents the kinematic viscosity. 
Equations \ref{eq:Nav-1} and \ref{eq:Nav-2} use central difference schemes of second-order accuracy to compute spatial derivatives.
The temporal derivative in equation \ref{eq:Nav-2} is integrated via an implicit second-order scheme using a variable time step.

The source term, $\textbf{\textit{F}}_\text s$, accounts for momentum transfer from the particles to the fluid. 
It is defined as the negative sum of the fluid forces acting on the particles within the control volume. This negative sign indicates that $\textbf{\textit{F}}_\text s$ represents the feedback force from the particles, which acts in the direction opposite to the fluid forces.
The source term can be expressed as

\begin{equation}
\textbf{\textit{F}}_\text s = -\frac{\rho_\text p}{\rho_\text f}\omega \sum_{i=1}^{N} \textbf{\textit{f}}_{\text{fl},i}\,
\end{equation}
where $N$ is the number of particles in the same computational cell, $\omega$ is the local particle volume fraction, $\rho_\text p$ is the particle density, and $\textbf{\textit{f}}_{\text{fl},i}$ is the acceleration due to fluid forces acting on individual particles. The fluid forces include drag and lift components and are detailed later in this section.
$\textbf{\textit{F}}_\text f$ denotes an external force applied to counterbalance the momentum dissipation of the fluid caused by wall friction.

The presence of charged particles generates an electric field, represented by $\textbf{\textit{E}}$, which is related to the electric potential $\varphi_\text{el}$ through the gradient as
\begin{equation}
\textbf{\textit{E}} = -\nabla \varphi_\text{el}.
\label{eq:ElectricField}
\end{equation}
This electric potential $\varphi_\text{el}$ is then determined using Gauss' law, which is
\begin{equation}
\nabla^2 \varphi_\text{el} = -\frac{\rho_{\text{el}}}{\varepsilon}.
\label{eq:GaussLaw}
\end{equation}

In this equation, $\rho_{\text{el}}$ represents the electric charge density and $\varepsilon$ is the permittivity of the fluid. 
If no external electric field is present, the electric charge density results directly from the positions of the individual particles and their charge.
The electric permittivity ($\varepsilon$) of the solid-gas mixture is approximated using the value of free space permittivity ($8.85 \times 10^{-12}$~F/m) due to the negligible solid volume fraction, as suggested by previous studies ~\citep{Rokkam10,Rivas07}.
Second-order central difference discretization is applied to the left-hand side of equation \ref{eq:GaussLaw}.

Our simulations assume rigid, spherical particles with uniform density. 
These particles are smaller than the computational grid cells, allowing us to track them individually using a point-mass approach within the Lagrangian framework.
Newton's second law of motion then governs the acceleration of each particle,

\begin{equation}
\frac{\text{d}  {\textbf{\textit{u}}}_\text p}{\text{d} t} = \textbf{\textit{f}}_{\text{d}} + \textbf{\textit{f}}_{\text{l}} + \textbf{\textit{f}}_{\text{el}} + \textbf{\textit{f}}_{\text{coll}}\,.
\label{eq:Newton}
\end{equation}

Here $\textbf{\textit{u}}_\text p$ is the particle velocity, $\,\textbf{\textit{f}}_{\text {d}}$ is the aerodynamic drag, $\,\textbf{\textit{f}}_{\text {l}}$ is the acceleration due to lift force, $\,\textbf{\textit{f}}_{\text {el}}$ the particle acceleration due to the electric field, and $\,\textbf{\textit{f}}_{\text {coll}}$ is the collisional acceleration.
The particle trajectories are solved by a second-order Crank-Nicolson scheme.

The aerodynamic acceleration is given as
\begin{equation}
\textbf{\textit{f}}_{\text {d}}=-\frac{3\rho_\text f}{8\rho_\text p r_\text p}C_\text d\left| \textbf{\textit{u}}_\text {rel} \right|\textbf{\textit{u}}_\text {rel}\,,
\label{eq:Drag}
\end{equation}
where $\textbf{\textit{u}}_{\text{rel}}$ is the relative velocity between the particle and the fluid, and $r_{\text{p}}$ is the radius of particles.
Therein the drag coefficient is based on Putnam's correlation
\begin{equation}
C_{\text {D}} = \left\{ \begin{array}{cl}
\frac{24}{Re_\text p} \left( 1 + \frac{1}{6} {Re_\text p}^{2/3} \right) &  \ {Re_\text p} \leq 1000 \\
0.424 &  \ {Re_\text p} > 1000 \, ,
\end{array} \right. 
\label{eq:Putnam}
\end{equation}
where $Re_\text p$ is the particle's Reynolds number, $\left| \textbf{\textit{u}}_\text {rel} \right| d_\text{p}/\nu$.

The acceleration due to lift force according to \citet{Saf65} adjusted by \citet{Mei92} reads

\begin{equation}
\textbf{\textit{f}}_{\text {l}} = \frac{1.54\, \rho_\text f \,\sqrt{\nu}}{ \rho_\text p \, r_\text p} (u-u_\text {p})\,\sqrt{|\nabla\textit{u}|}\,\text{sign}(\nabla\textit{u})\,C_{\text {L}},
\end{equation}
where
\begin{equation}
C_{\text {L}} =
\begin{cases}
(1 - 0.3314 \, \sqrt{\alpha}\,) e^{-0.1 Re_\text p} + 0.3314 \sqrt{\alpha} & {Re_\text p} \leq 40 \\
0.0524 \sqrt{\alpha {Re_\text p}} & {Re_\text p}> 40.
\end{cases}
\end{equation}
Here, $\alpha$ is the dimensionless shear rate, defined as the ratio $Re_\text{s} / Re_\text{p}$, and $e$ is Euler's number. The shear Reynolds number, $Re_\text{s}$, is given by $4r_\text{p}^2\dot{\gamma}/ \nu$, where $\dot{\gamma}$ is the shear rate.

The acceleration due to electrostatic force on a particle with charge $Q$ is calculated using the hybrid method proposed by~\citet{Gro17c}. The method combines Coulombic interactions with $N$ neighboring particles, \,\,\(\textbf{\textit{f}}_{\text{el,C}}\)\,, and long-range forces determined by Gauss's law, \,\(\textbf{\textit{f}}_{\text{el,G}}\). Here, neighboring particles, $N$, correspond to particles in the same computational cell. The acceleration due to total electrostatic force is expressed as
\begin{equation}
\textbf{\textit{f}}_{\text{el}} = \textbf{\textit{f}}_{\text{el,C}} + \textbf{\textit{f}}_{\text{el,G}} \,,
\end{equation}
and the acceleration due to Coulombic force between the particle and each of its $N$ neighboring particles is calculated using Coulomb’s law,
\begin{equation}
\textbf{\textit{f}}_{\text{el,C}} = \sum_{n=1}^{N} \frac{Q \,Q_{\text{n}}\, \textbf{\textit{z}}_\text{n}}{4 \pi \varepsilon |\textbf{\textit{z}}_{\text{n}}|^3\textit{m}_{\text{p}}}.
\end{equation}
In this expression, $Q_{\text{n}}$ represents the charge of the neighboring particles, the vector \(\textbf{\textit{z}}_{\text{n}}\) points from the center of each neighboring particle to the particle in question, $\textit{m}_{\text{p}}$ denotes the mass of the particle.
On the other hand, acceleration due to far-field forces arise from the collective electric field of distant charges and are computed using Gauss's law as
\begin{equation}
\textbf{\textit{f}}_{\text{el,G}} = \frac{Q \textbf{\textit{E}}}{\textit{m}_{\text{p}}} =-\frac{Q\nabla \varphi_\text{el}}{\textit{m}_{\text{p}}}.
\end{equation}

The transfer of charge from a wall to a particle during impact, $\Delta q_\mathrm{pw}$\,, is given by
\begin{equation}
\Delta q_\mathrm{pw} = C_\mathrm{1} \left( q_\mathrm{eq} - q_\mathrm{n} \right) , 
\end{equation}
and during contact with another particle by

\begin{equation}
\Delta q_\mathrm{n} = -\Delta q_\mathrm{m} = C_\mathrm{2} \left( q_\mathrm{m} - q_\mathrm{n} \right) .
\end{equation}%
In these equations, $q_\mathrm{n}$ and $q_\mathrm{m}$ denote the charge of the particles before the impact.

The charging model is designed to be straightforward by focusing solely on the particle’s current charge. This approach leads to fast charging rates since limiting parameters in more sophisticated models, such as contact area, and probability of collision with an uncharged patch on another particle’s surface~\citep{Kor14}, are not involved. To reduce the charging rate and ensure a more realistic simulation, we use proportionality factors, $C_\mathrm{1}$ and $C_\mathrm{2}$. The value of 0.2 was determined through simulations to achieve an optimal charge transfer rate, ensuring that the flow can adapt to new charge conditions and maintaining the accuracy of the simulation.

Particle collisions are addressed using a variant of the hard-sphere approach called the ray-casting method ~\citep{Roth82}. 
This method dictates a simple collision rule: upon contact with a wall, the particle's normal velocity component reverses direction, while the tangential component remains unchanged.

Adhesion force is an important factor to consider in particle collisions with their magnitude comparable to that of electrostatic forces. However, adhesive forces act over much shorter ranges compared to electrostatic forces. In other words, the duration that the adhesive forces act on the particle is much shorter, compared to electrostatic forces.  Consequently, the adhesion force is neglected in our simulations. Nevertheless, energy dissipation from van der Waals forces may cause particles to stick to the wall at low impact velocities. This effect is not considered in this study.

In fact, particle collisions are affected by many factors such as surface roughness, particle and wall elasticity, electrostatic forces, and hydrodynamic effects. These factors are often represented by the restitution coefficient, which measures the ratio of the normal components of particle velocity before and after a collision~\citep{Sardina2012}. In this study, we use a restitution coefficient of 1, indicating perfectly elastic collisions. While this is an idealization, it is suitable for many applications for materials with high restitution coefficients. Research, including studies by~\citet{Li2001}, and our previous DNS results~\citep{Ozler2023} shows that using a coefficient of 0.9 for inelastic collisions leads to only minor differences in particle dispersion compared to a coefficient of 1. Many other studies in the literature adopted elastic collisions due to their negligible impact on particle behavior \citep{Zhang23, Zhang24, Mot22, Joh20}

The solution algorithm implemented in the pafiX software uses a parallel Fortran 90 code. 
This code is parallelized with the Message Passing Interface (MPI) and scales up to 256 processors.
Refer to ~\citep{Gro17a} for code validation, scalability studies, and a more detailed explanation of the mathematical model and numerical methods.

\section {Numerical setup}
\begin{figure}[t]
    \centering
    \captionsetup[subfigure]{labelformat=empty}
    \includegraphics[width=0.7\textwidth]{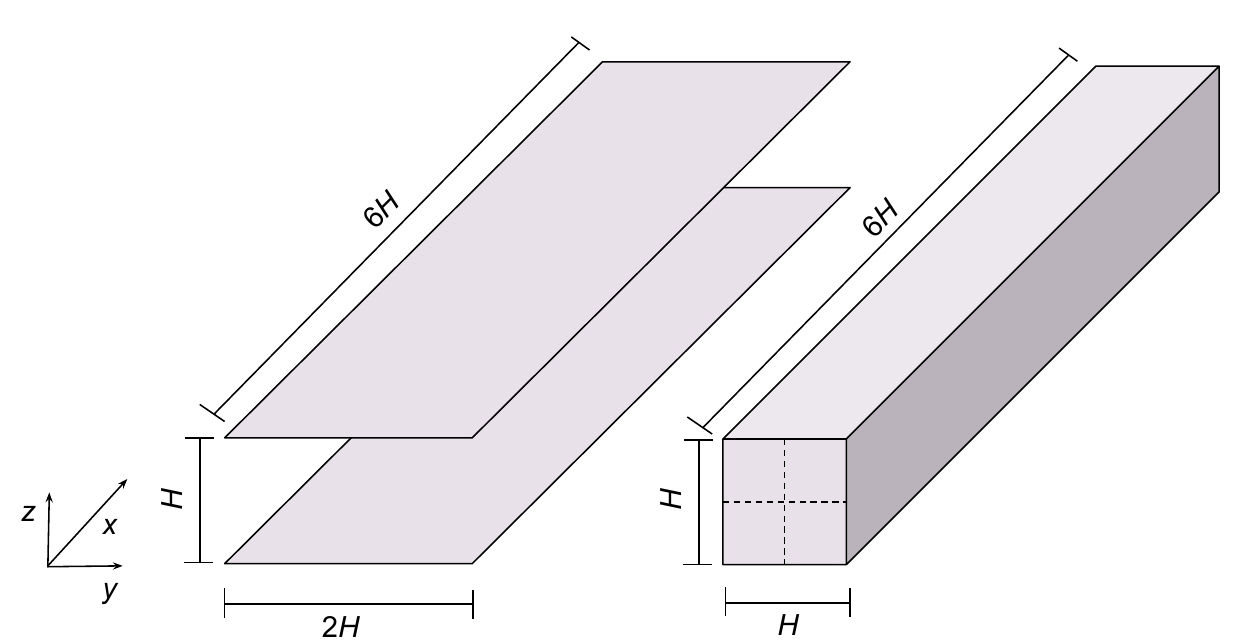}
    \caption{Dimensions of channel and duct flow containers. Dashed lines on the duct container show the bisectors.}
    \label{fig:dimensions}
\end{figure}

We investigated the impact of flow patterns on charging by simulating turbulent particle-laden flows with identical parameters in both a square-shaped duct and a channel.
The computational domains are illustrated in Figure~\ref{fig:dimensions}.
The flow direction is oriented along the x-axis. 
Gravity is omitted to isolate the effects of cross-sectional flows in duct flows.

A constant pressure gradient enforces a flow of $Re_{\mathrm{\tau}}\,$=\,180 with a particle number density, $n_\mathrm{p}$, of $1.0 \times 10^{8}$ $\mathrm{m}^{-3}$.
The friction Reynolds number is defined as $Re_{\mathrm{\tau}}=\frac{u_{\tau} H/2}{\nu}$ with $u_{\tau}$ being the friction velocity, $H$ is the width, and $\nu$ is the kinematic viscosity.
Periodic boundaries are assumed for the duct in the stream-wise direction and the channel in the stream-wise and homogeneous span-wise direction.
At the non-periodic walls, no-slip boundary conditions for the fluid are imposed.
It is assumed that the duct is grounded and fully conductive i.e. zero electrical potential at the walls imposed.

\captionsetup{justification=justified, width=1\linewidth}

The duct has a spanwise dimension of $H$, while the channel’s spanwise dimension is set to 2$H$. This choice ensures that the computational domain is sufficiently large to capture the full statistical properties of turbulent flow in the spanwise direction.~\citet{Kim1987} showed that, for a friction Reynolds number of $Re_{\mathrm{\tau}}\,$=\,180, spanwise velocity correlations become negligible at separations of 3 times the channel half-width. Setting the channel’s spanwise dimension to 2$H$ ensures the domain size is adequate for accurately representing fully developed turbulence, thus improving the precision of turbulence modeling without constraining the flow.

The selection of meshes for both geometries follows a grid resolution study conducted by ~\citet{Gro23}, aiming to optimize the comparability of simulations.

In homogeneous directions (i.e., $x$ and $y$ in the channel, and $x$ in the duct), cells are uniformly distributed. In inhomogeneous directions, a non-uniform mesh is used, with cell sizes decreasing toward the walls to resolve relevant flow structures.  To refine the cells near the wall, we applied the cosine stretching function~\citep{Kim1987} to define $N$ grid points, calculating their positions using

\begin{equation} z_j = \frac{H}{2} \cos\left(\frac{\pi (j - 1)}{N - 1}\right), \quad j = 1, \ldots, N. \end{equation}

This function is also applied in the y-direction for duct flow. The minimum grid spacing at the wall-adjacent cells is $\Delta z^+ = 0.0434$, while the maximum spacing at the center plane of the channel is $\Delta z^+ = 3.9544$. All cases are computed using 256 $\times$ 144 $\times$ 144 grid points.

Particles of uniform size are introduced into the system only after the fluid phase has reached a fully developed state.
They are randomly distributed, with velocities matching the fluid velocity at their respective locations. 

We assume an equilibrium charge of \( q_\mathrm{eq} = 126 \, \text{fC} \), which is about 10\% of the maximum charge a 50 $\upmu$m particle can hold, according to~\citep{Mat18}. We chose this lower charge level to prevent particles from sticking to the walls, which occurred at higher charges and was not suitable for simulation. This charge keeps the particles airborne, allowing us to investigate the effect of flow dynamics on charging.
The particle charging model is activated once the particle phase converges completely with the fluid flow.

We investigated eight distinct cases, including four scenarios of duct flow and four of channel flow.
Table 1 summarizes the properties of the investigated cases.
Stokes number, $St$, is defined as $St~=~{\tau_{\mathrm{p}} u_{\tau}^{2}}/{\nu}$, where $\tau_{\mathrm{p}}~=~{\rho_{\mathrm{p}}d_{\mathrm{p}}^{2}}/{18\rho_{\mathrm{f}}\nu}$ is the particle response time. 

\begin{table}[t]
\centering
\caption{Summary of simulation conditions.}
\noindent\rule[0.5ex]{0.70\linewidth}{1pt} 
\setlength{\tabcolsep}{1pt}
\begin{tabular}{@{\extracolsep{8pt}}ccrrr}
\multicolumn{1}{c}{\textup{Case}} & \multicolumn{1}{c}{Flow} & \multicolumn{1}{c}{$St$} & \multicolumn{1}{c}{$\,\rho_{\mathrm{p}}$ (kg/m$^3$)} \vspace{0.15cm}\\
1 & channel & 4.69 & 500 \\ 
2 & channel & 9.38 & 1000 \\
3 & channel & 18.75 & 2000 \\
4 & channel & 37.50 & 4000 \\ 
5 & duct & 4.69 & 500  \\
6 & duct & 9.38 & 1000 \\
7 & duct & 18.75 & 2000 \\
8 & duct & 37.50 & 4000 
\end{tabular}
\noindent\rule[0.5ex]{0.70\linewidth}{1pt} 
\end{table}

\section {Results and discussion}
This section presents our findings in two subsections. 
The first subsection analyzes how different flow patterns impact powder charging rates and distributions. 
The second subsection evaluates the effects of electrostatic charges on flow dynamics.

\subsection{Effect of flow patterns on powder charging}
Simulations reveal that flow patterns markedly impact powder charging, influencing both the rate and distribution of charge. 
Flow patterns alter particle trajectories, affecting collision rates with the walls and the migration of particles from the wall to the bulk fluid. 
Specifically, wall collisions affect the charging rate, while particle migration influences charge distribution.
Since we simulated a dilute flow with a particle volume fraction of $6.54\times10^{-6}$, particle-particle collisions are found to be negligible compared to particle-wall collisions.
The total average number of particle-wall collisions until the powder reaches equilibrium charge is on the order of $10^4$, while a particle experiences only a few particle-particle collisions. For this reason, the discussion below will focus on particle-wall collisions.

\captionsetup{justification=justified, width=1\linewidth}
\begin{figure}[t]
    \centering
    \captionsetup[subfigure]{labelformat=empty}
    \begin{subfigure}[b]{0.48\textwidth}
        \centering
        \raisebox{-0.5\height}{\stackinset{l}{0.5ex}{t}{0.5ex}{\text{(a)}}{\includegraphics[width=\textwidth]{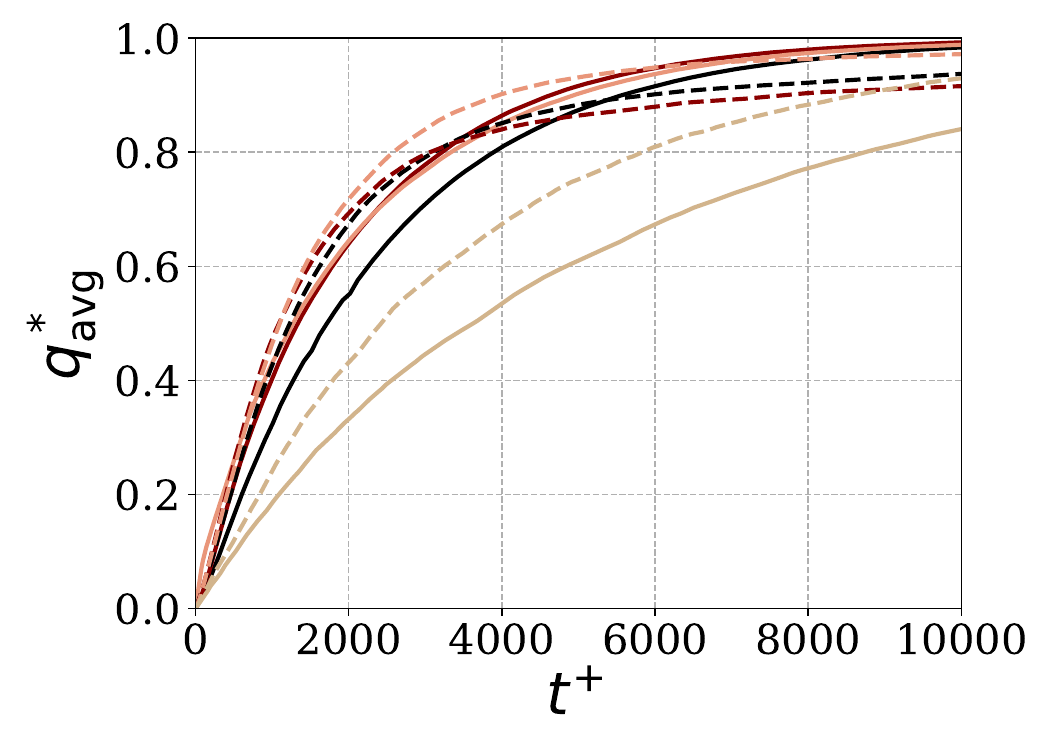}}}
        \caption{}
        \label{fig:q_t}
    \end{subfigure}
    \hfill
    \begin{subfigure}[b]{0.48\textwidth}
        \centering
        \raisebox{-0.5\height}{\stackinset{l}{0.5ex}{t}{0.5ex}{\text{(b)}}{\includegraphics[width=\textwidth]{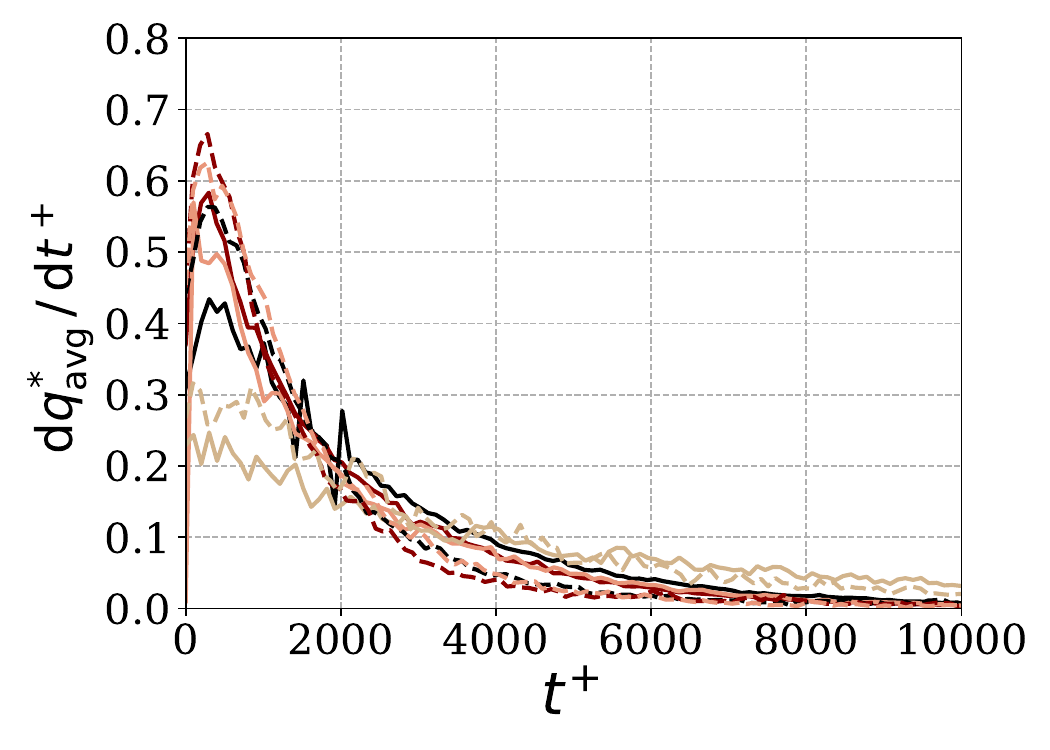}}}
        \caption{}
        \label{fig:dqdt_t}
    \end{subfigure}
    \vspace{-1.5\baselineskip} 
    \caption{Temporal evolution of the average powder charge normalized by the equilibrium charge (a). The charging rate of powder (b). Lines represent channel flow, and dashed lines represent duct flow. Colors indicate Stokes number; 
        (\,\textcolor{black}{\rule[0.2ex]{0.5cm}{1pt}}\,) $St\,=\,37.50$,\,
        (\,\textcolor{darkred}{\rule[0.2ex]{0.5cm}{1pt}}\,) $St\,=\,18.75$,\,
        (\,\textcolor{lightsalmon}{\rule[0.2ex]{0.5cm}{1pt}}\,)$St\,=\,9.38$,\,
        (\,\textcolor{tan}{\rule[0.2ex]{0.5cm}{1pt}}\,)$St\,=\,4.69$.}
    \label{fig:charge}
\end{figure}

The average powder charge, normalized by the equilibrium charge, is depicted in figure~\ref{fig:q_t}.
The average powder charge is the sum of all particles' charges divided by the total number of particles. 
The equilibrium charge is $q_{\mathrm{eq}}=126$~fC.
The temporal axis is normalized as $t^{+}=t\,u_{\tau }^{2}\,/\,\nu$, where $u_{\tau}$ is the wall friction velocity and $\nu$ is the kinematic viscosity.

Figure~\ref{fig:q_t} shows that duct flow generates a higher average powder charge than channel flow at the same Stokes number.
However, a switch occurs as approaching the equilibrium charge, where the average charge in channel flow becomes higher.
Specifically, for $St$~=~9.38, 18.75, and 37.50, the average charge switches when it reaches 0.95, 0.82, and 0.88, respectively. 
For $St$~=~4.69, the charges switch at $q_\mathrm{avg}^*$~=~0.99 (not shown in the figure). 
This switch is attributed to the accumulation of uncharged particles at the duct corners, which will be discussed in detail in subsequent sections. 
It is important to note that this switch has minimal impact on overall powder charging, as most of the powder is already charged by this stage and the charging rate is significantly reduced close to the equilibrium charge.

The difference in powder charging between channel and duct flow is most pronounced for $St$~=~4.69. 
In duct flow, particles with this Stokes number reach half their equilibrium charge at $t^\mathrm{+}$~=~2427, compared to $t^\mathrm{+}$~=~3619 in channel flow. 
For $St$~=~37.50, the difference is smaller: duct flow reaches half of the equilibrium charge at $t^\mathrm{+}$~=~1239, while channel flow does so at $t^\mathrm{+}$~=~1720. 
This disparity occurs because low Stokes number particles, having lower inertia, closely follow flow patterns and are more influenced by the flow conditions.

In both channel and duct flow, the average powder charge is lowest for $St$~=~4.69.
In channel flow $St$~=~9.38 and $St$~=~18.75 show a similar trend. 
In duct flow, the same trend is observed until $q_\mathrm{avg}^*$ reaches 0.6, after which the average charge for $St$~=~9.38 becomes higher than for $St$~=~18.75.
For $St$~=~37.50, the powder charge is lower than that for $St$~=~9.38 and $St$~=~18.75 in both duct and channel flow. 
Therefore, the relationship between the Stokes number and the average powder charge is not straightforward.

Figure~\ref{fig:dqdt_t} illustrates the temporal evolution of powder charging rates.

At later stages, in all cases, the charging rate consistently decreases over time; however, distinct patterns in the charging rate emerged initially.
For $St$~=~4.69, both channel and duct flow steadily decrease from the start. 
In contrast, for $St$~=~9.38, $St$~=~18.75, and $St$~=~37.50, the charging rate initially increases and reaches a peak before decreasing.

It is important to note that differences in initial patterns may be influenced by the time step size used for post-processing simulation results. 
While the same time step was employed across all simulations, it might be too coarse in certain cases to accurately capture the initial charging patterns.
Typically, the initial rise in the charging rate is due to electrostatic forces. 
In the beginning, when the powder is uncharged, the charging events, i.e., particle-wall contacts, are only driven by the aerodynamic acceleration of the particles toward the walls.
When the first particles receive charges, image charges at the conductive walls appear. 
Thus, the particle gets attracted to its charge, accelerates back to the wall, and undergoes another charge event.
In other words, the frequency of particle-wall collision increases.

In all cases, the rate of charge transfer reduces over time.
This reduction is due to the charge the particles already have accumulated. 
Since each particle can hold only a certain amount of charge, the precharge reduces the net charge exchange during a particle-wall contact. 
Thus, the increase in the charge event frequency is counter-balanced by the decrease in the charge transfer during each event, as elaborated in the following.

\captionsetup{justification=justified, width=1\linewidth}
\begin{figure}[t]
    \centering
    \captionsetup[subfigure]{labelformat=empty} 
    \begin{subfigure}[b]{0.48\textwidth}
        \centering
        \raisebox{-0.5\height}{\stackinset{l}{0.5ex}{t}{0.5ex}{\text{(a)}}{\includegraphics[width=\textwidth]{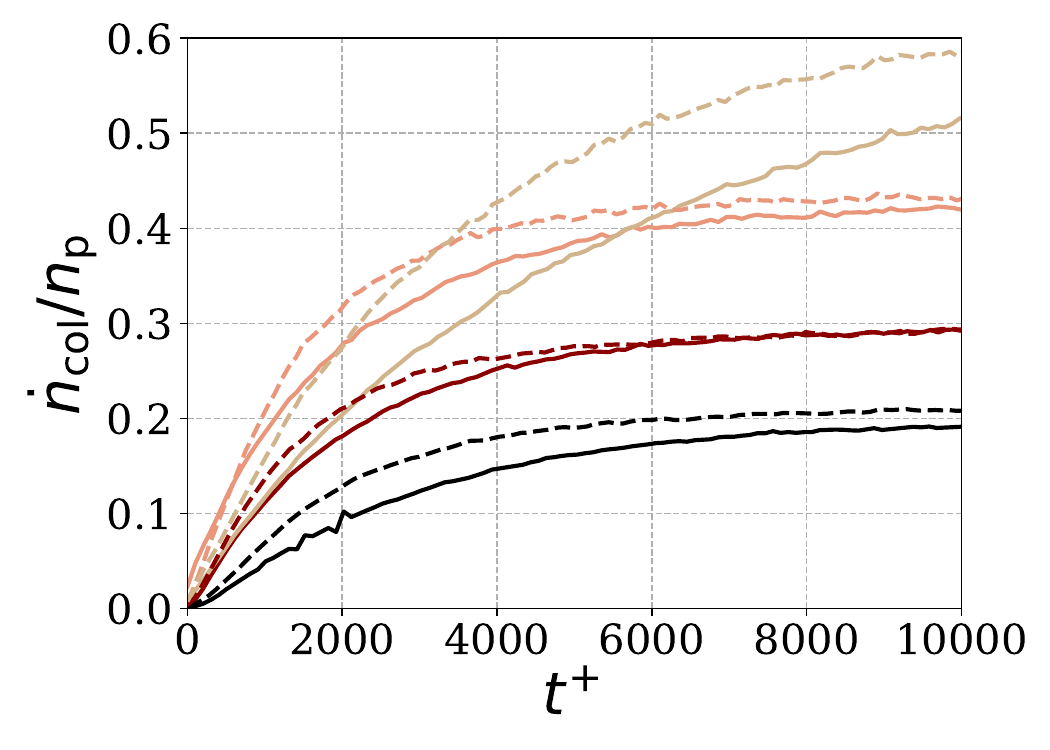}}}
        \caption{}
        \label{fig:col}
    \end{subfigure}
    \hfill
    \begin{subfigure}[b]{0.48\textwidth}
        \centering
        \raisebox{-0.5\height}{\stackinset{l}{0.5ex}{t}{0.5ex}{\text{(b)}}{\includegraphics[width=\textwidth]{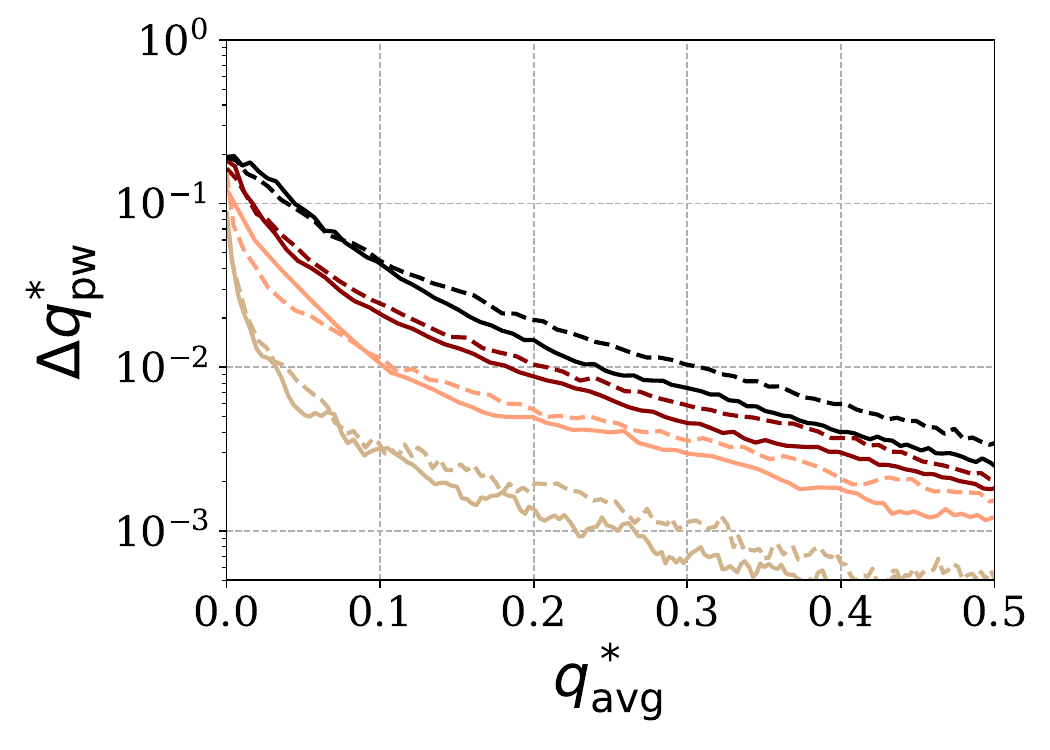}}}
        \caption{}
        \label{fig:impact}
    \end{subfigure}
    \vspace{-1.5\baselineskip}
    \caption{Rate of average particle-wall collisions (a), the charge transferred during one particle-wall collision, impact charge, over time (b). Lines represent channel, dashed lines represent duct flow. Colors indicate Stokes number; 
        (\,\textcolor{black}{\rule[0.2ex]{0.5cm}{1pt}}\,) $St\,=\,37.50$,\,
        (\,\textcolor{darkred}{\rule[0.2ex]{0.5cm}{1pt}}\,) $St\,=\,18.75$,\,
        (\,\textcolor{lightsalmon}{\rule[0.2ex]{0.5cm}{1pt}}\,)$St\,=\,9.38$,\,
        (\,\textcolor{tan}{\rule[0.2ex]{0.5cm}{1pt}}\,)$St\,=\,4.69$.}
    \label{fig:collision}
\end{figure}

The charging rate of powder depends on (i) the frequency of particle charging events at the walls, and (ii) the charge transferred during each collision event. 
To understand these dynamics, we investigate the rate of wall collisions over time and the impact charge, depicted in figure~\ref{fig:collision}.

Figure~\ref{fig:col} illustrates the average wall collision rate of particles in duct and channel flow across different Stokes numbers. 
Particles collide with walls more frequently in duct flow compared to channel flow at the same Stokes number, contributing to the higher charging rates in duct flows. 
This indicates that particles in duct flow have increased velocities in the wall-normal and span-wise directions, leading to more frequent wall collisions. 
This behavior is linked to cross-sectional secondary flows in duct flow, which will be discussed later in the paper.

As expected, low-inertia particles ($St$~=~4.69) are most impacted by the flow, leading to higher collision rates compared to particles with higher inertia. 
This observation explains the difference in charging rates depicted in Figure~\ref{fig:q_t}.

In total, for $St$~=~4.69, most collisions occur. 
However, as the charge transferred in each collision event diminishes over time, wall collisions in the initial time steps are relevant for powder charging. 
For $t^\mathrm{+}< 2000$, particles of $St$~=~9.38 collide at the highest rate with walls.
Collision events are less frequent for $St$~=~18.75 and 37.50, despite their powder charging rates being comparable to those of $St$~=~9.38.

\begin{figure}[t]
    \centering
    \captionsetup[subfigure]{labelformat=empty}
    \begin{subfigure}[b]{0.49\textwidth}
        \centering
        \raisebox{-0.5\height}{\stackinset{l}{0.1ex}{t}{0.1ex}{\text{(a)}}{\includegraphics[width=\textwidth]{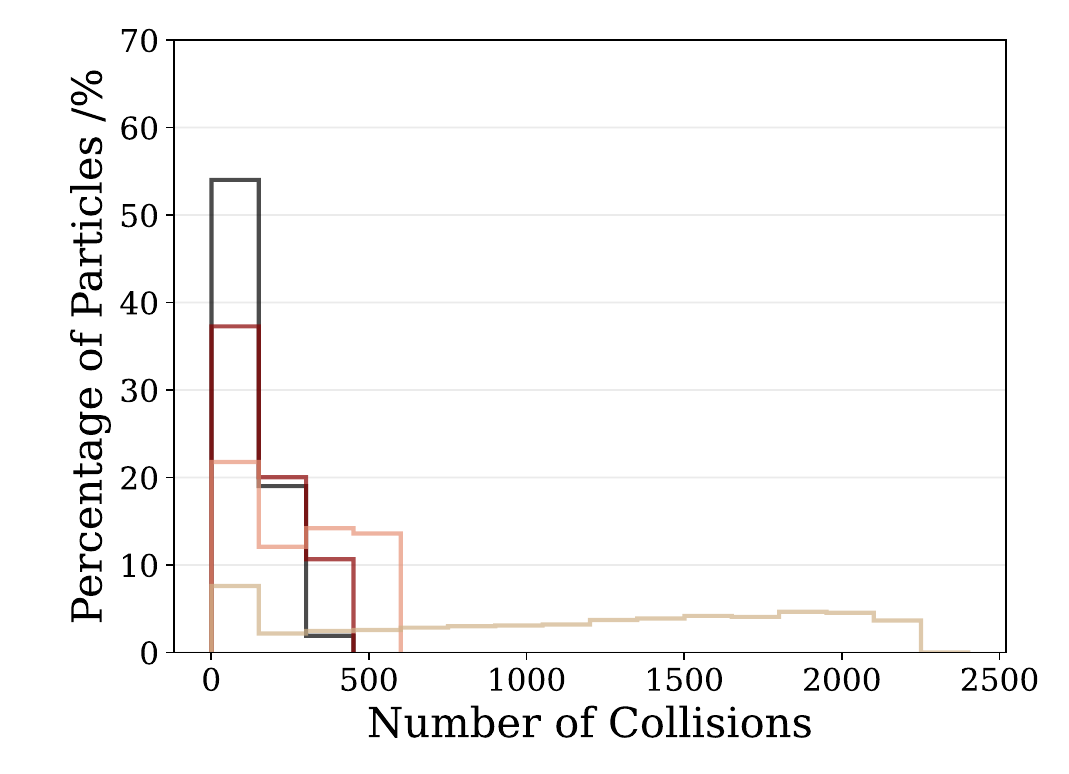}}}
        \caption{}
        \label{fig:col_freq1}
    \end{subfigure}
    \hfill
    \begin{subfigure}[b]{0.49\textwidth}
        \centering
        \raisebox{-0.5\height}{\stackinset{l}{0.1ex}{t}{0.1ex}{\text{(b)}}{\includegraphics[width=\textwidth]{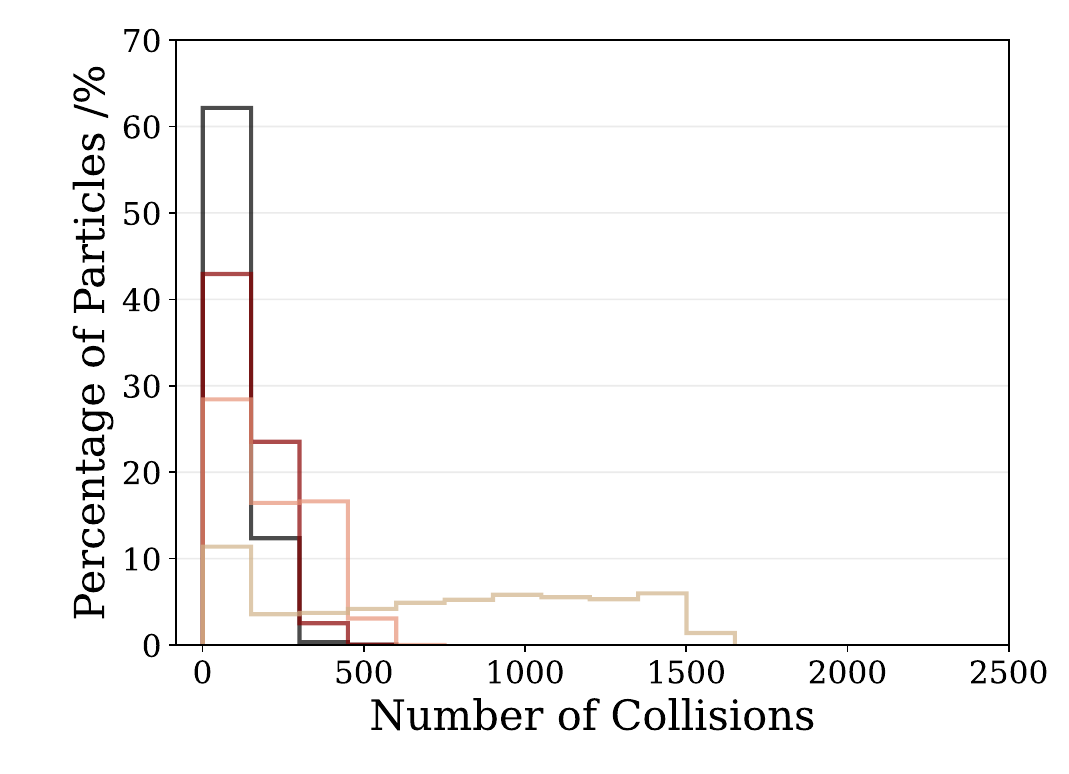}}}
        \caption{}
        \label{fig:col_freq2}
    \end{subfigure}
    \vspace{-1.5\baselineskip}
    \caption{Frequency of wall collision for channel (a) and duct (b) flows. It shows the percentage of particles and the corresponding number of collisions with the wall until the average powder charge reaches half of the equilibrium charge, $q^*_{\textup{avg}}=0.5$.
The histograms exclude particles that did not collide with a wall yet. Colors indicate Stokes number; 
        (\,\textcolor{black}{\rule[0.2ex]{0.5cm}{1pt}}\,) $St\,=\,37.50$,\,
        (\,\textcolor{darkred}{\rule[0.2ex]{0.5cm}{1pt}}\,) $St\,=\,18.75$,\,
        (\,\textcolor{lightsalmon}{\rule[0.2ex]{0.5cm}{1pt}}\,)$St\,=\,9.38$,\,
        (\,\textcolor{tan}{\rule[0.2ex]{0.5cm}{1pt}}\,)$St\,=\,4.69$. }

    \label{fig:col_freq}
\end{figure}

The rate of particle-wall collisions does not fully assess their impact on the charging rate. 
Some particles may accumulate near-wall and collide with the wall consistently without exchanging any charge, while others may experience fewer collisions as they travel along the domain but transfer a significant amount of charge per one collision.
In such cases, the charge transferred during each collision (figure~\ref{fig:impact}) and the collision frequency of individual particles (figure~\ref{fig:col_freq}) indicate whether most particles are interacting with the wall.

Figure~\ref{fig:impact} illustrates the charge transferred as a result of a single particle-wall collision, averaged over particles vs. the corresponding average powder charge. 
In each case, the impact charge decreases over time as particles accumulate charge.

In duct flow, impact charges are significantly higher than in channel flow.
Therefore, in duct flow, both the higher frequency of collisions and the increased impact charge contribute to the higher charging rates. In both channel and duct flow, the charge transferred during each collision is the highest for $St$~=~37.50, followed by $St$~=~18.75.
The lowest impact charges are seen for  $St$~=~9.38 and  $St$~=~4.69.

The collision frequency of particles is shown in figure~\ref{fig:col_freq}.
Compared to duct flow, in channel flow repetitive collisions occur more frequently. 
In channel flow, individual particles can undergo up to 2250 collisions, whereas in duct flow, the number is restricted to a maximum of 1,500 collisions per particle.
Specifically, in channel flow, approximately 50\%, 35\%, 20\%, and 8\% of particles collide with the wall 125 times or less for $St$~=~37.50, 18.75, 9.38, and 4.69, respectively. Conversely, in duct flow, these percentages are 60\%, 40\%, 26\%, and 11\% for the same respective Stokes numbers. 
This suggests that in duct flow, more particles experience a particle-wall collision, leaving fewer particles collision-free.

Both channel and duct flow exhibit the effects of particle inertia on collision frequency.
For $St$~=~4.69, a minority of particles undergo a significant number of wall collisions.
Specifically, approximately 2.5\% of particles experience 2250 collisions in channel flow, while 2\% of particles undergo 1500 collisions in duct flow.
Conversely, with an increase in Stokes number, such as for $St$~=~37.50, a larger proportion of particles collide with the wall fewer times.

These observations suggest that particles exhibit a more uniform motion in duct flow compared to channel flow. 
In duct flow, this uniform motion is primarily driven by secondary flow, which both brings particles closer to the wall and sweeps them along.

The differences related to the Stokes number are related to the interplay between turbophoresis and particle inertia.
In turbulent flows, particles experience turbophoresis, causing them to migrate towards regions of decreasing turbulence, typically near the walls. 
Low-inertia particles tend to be trapped and accumulate in the near-wall region, resulting in repeated wall collisions. 
Conversely, particles with higher Stokes numbers can escape these regions due to their high inertia, moving freely in the wall-normal direction and limiting repeated collisions.

\begin{figure}[t]
    \centering
    \captionsetup[subfigure]{labelformat=empty}
    \includegraphics[width=0.95\textwidth]{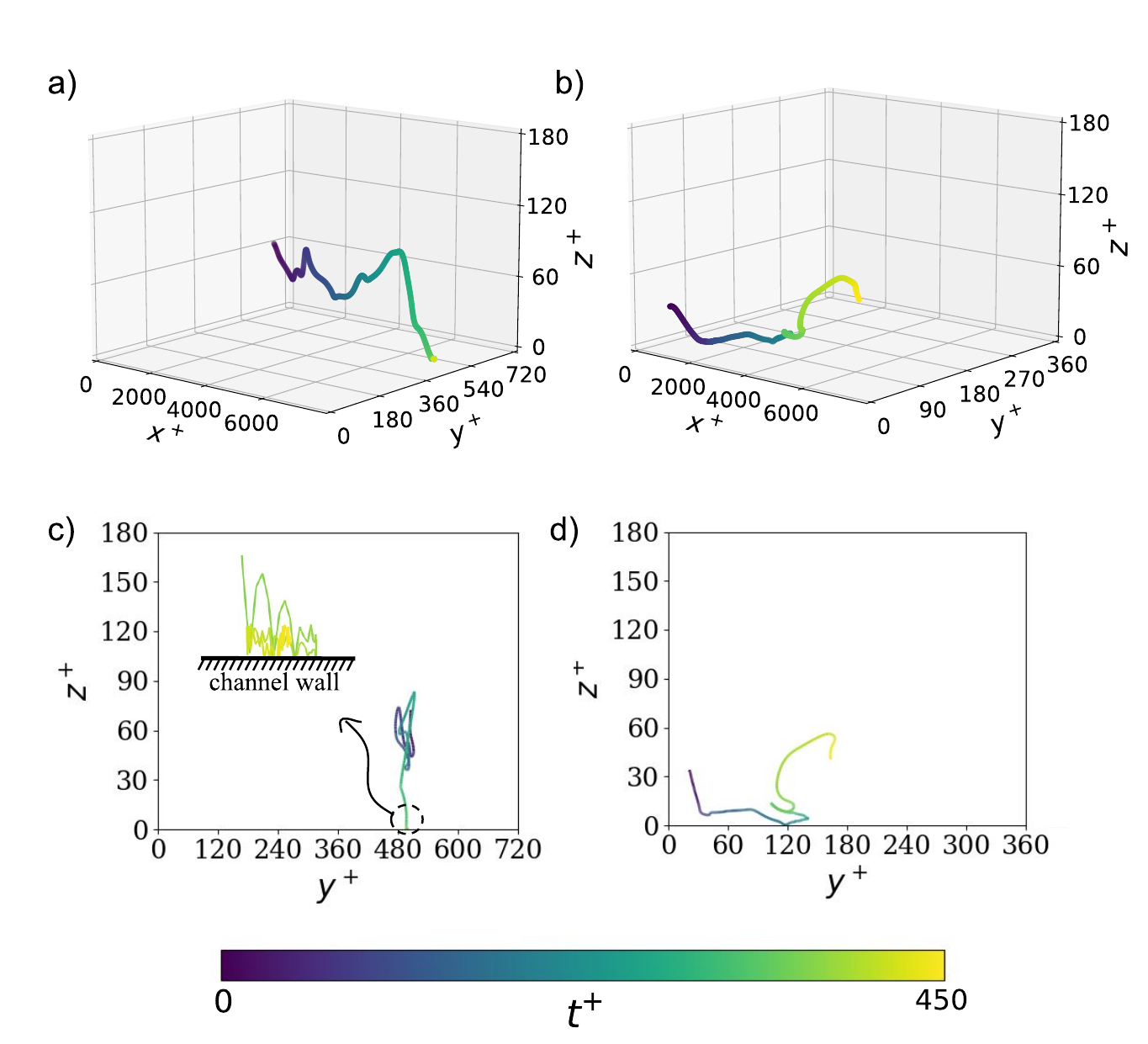}
    \caption{Typical trajectories of particles with $St$~=~4.69 in a channel (a),(c) and duct flow (b),(d). Figures (a, b) show 3D trajectories, Figures (c, d) show the trajectories in the cross-section. Color bar shows normalized time.}
    \label{fig:trajectories}
\end{figure}

Figure~\ref{fig:trajectories} illustrates typical trajectories of particles with $St$~=~4.69 in a channel and a duct flow. 
In channel flow, a particle is trapped in the near-wall region, shown in Figure~\ref{fig:trajectories}\textcolor{blue}{(c)}, due to the influence of near-wall structures.
This leads to a series of repetitive collisions with the wall. 
After several collisions, the particle reaches the equilibrium charge, where it no longer accumulates additional charge. 

\begin{figure}[t]
    \centering
    \captionsetup[subfigure]{labelformat=empty}
    \captionsetup{justification=justified, width=1\linewidth}
    \begin{subfigure}[b]{0.48\textwidth}
        \centering
        \raisebox{-0.5\height}{\stackinset{l}{0.5ex}{t}{0.5ex}{\text{(a)}}{\includegraphics[width=\textwidth]{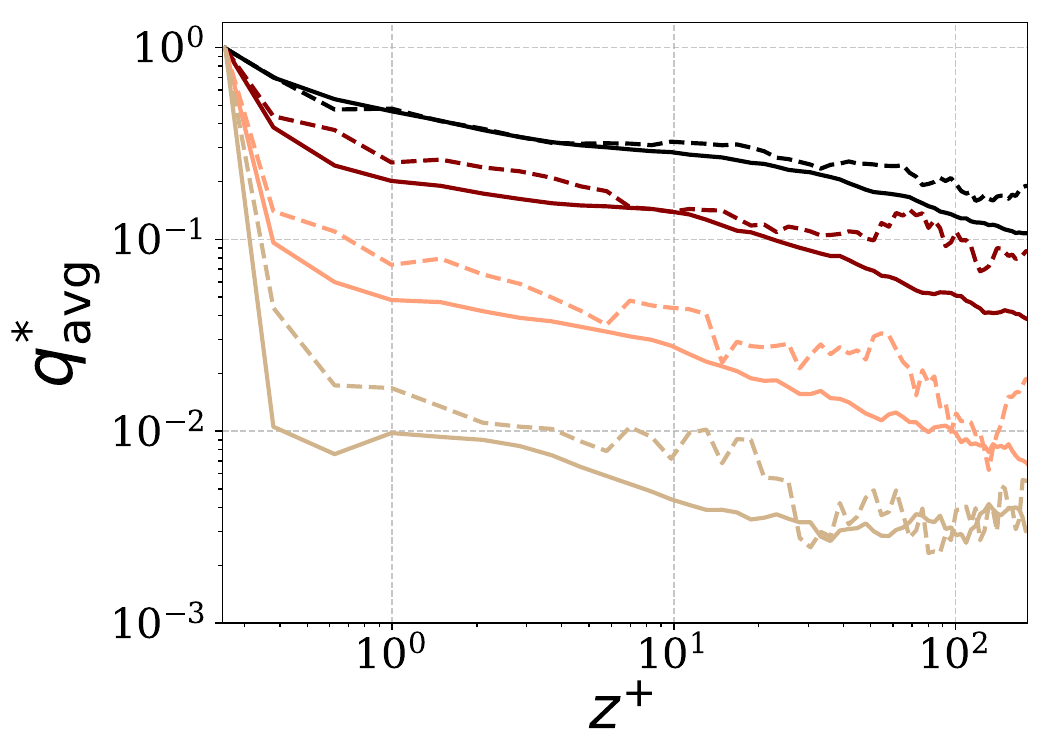}}}
        \caption{}
        \label{fig:q_z}
    \end{subfigure}
    \begin{subfigure}[b]{0.48\textwidth}
        \centering
        \raisebox{-0.5\height}{\stackinset{l}{0.1ex}{t}{0.1ex}{\text{(b)}}{\includegraphics[width=\textwidth]{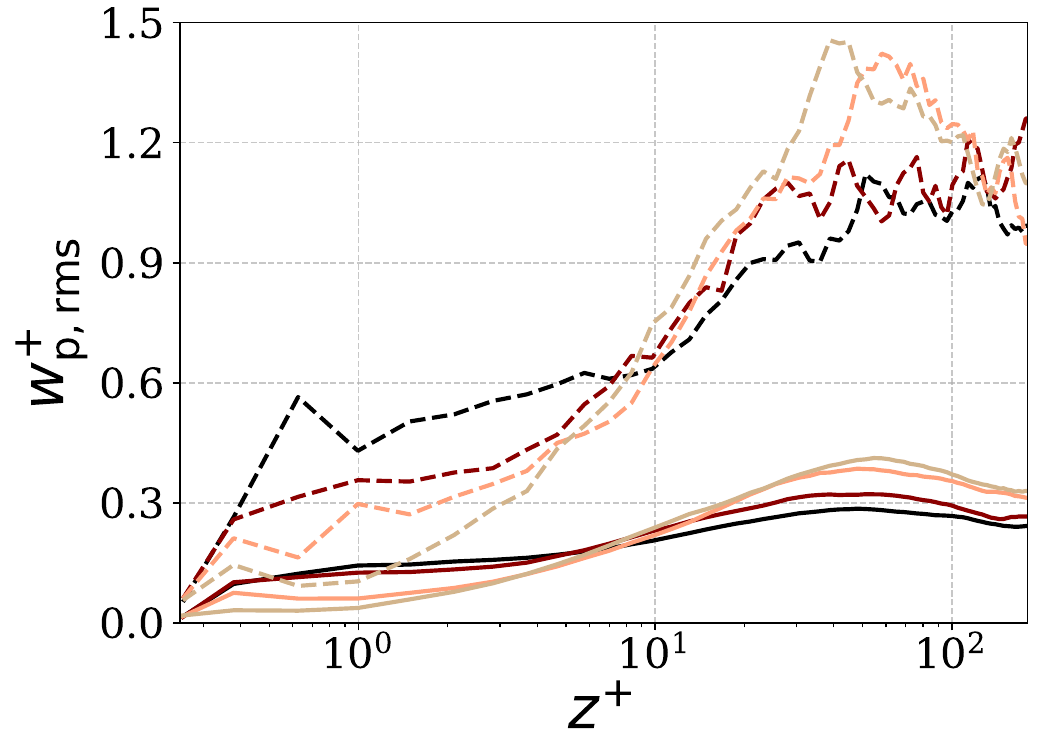}}}
        \caption{}
        \label{fig:wprms_z}
    \end{subfigure}
    \vspace{-1.5\baselineskip}
\caption{
        Powder charge distribution along the wall-normal direction in channel and duct flow (a), rms wall-normal particle velocities (b). Particle charge is normalized by the equilibrium charge. Figures show the data when the average powder charge is half of the equilibrium charge, $q^*_{\textup{avg}}=0.5$. The wall-normal profiles of the along the bisector,  $y^{+\textup{}}= 180$, are depicted for duct flow.
        Lines represent channel, dashed lines represent duct flow. Colors indicate Stokes number; 
        (\,\textcolor{black}{\rule[0.2ex]{0.5cm}{1pt}}\,) $St\,=\,37.50$,\,
        (\,\textcolor{darkred}{\rule[0.2ex]{0.5cm}{1pt}}\,) $St\,=\,18.75$,\,
        (\,\textcolor{lightsalmon}{\rule[0.2ex]{0.5cm}{1pt}}\,)$St\,=\,9.38$,\,
        (\,\textcolor{tan}{\rule[0.2ex]{0.5cm}{1pt}}\,)$St\,=\,4.69$.}
        
    \label{fig:qvw}
\end{figure}

In contrast, in duct flow, the same particle demonstrates significantly greater mobility, shown in Figure~\ref{fig:trajectories}\textcolor{blue}{(d)}. 
Following a collision, it can escape the near-wall regions and navigate within the cross-section. 
The particle is capable of moving along the wall and migrating back and forth toward the wall.
An increase in mobility is important since it promotes uniform charge distribution across the domain shown in Figure~\ref{fig:q_z}.

Figure~\ref{fig:q_z} depicts the distribution of powder charge along the wall-normal axis in duct and channel flows.
In each case, the powder charge is highest at the wall and gradually decreases towards the center. 
For example, the charge at the wall is two orders of magnitude higher than the charge at the center for $St$~=~4.69.

This pattern arises because charging begins at the wall, where particles first contact the surface. 
Charge is then carried towards the center by particles moving from the wall into the bulk fluid.
Since duct flow promotes uniform mixing of particles, it results in a more even charging pattern compared to channel flow with the same Stokes number.

Stokes number notably influences charge profiles, with the most uniform distribution observed for $St$~=~37.50. 
As Stokes number decreases to $St$~=~4.69, the powder charge reduces in the center noticeably. 
Specifically, particles with $St$~=~37.50 carry a charge approximately two orders of magnitude higher than those with $St$~=~4.69, observed in both duct and channel flow.

Figure~\ref{fig:wprms_z} illustrates the root mean square (rms) of wall-normal particle velocities. 
Particle velocity is normalized with the wall friction velocity, $u_{\tau}$.

In duct flow, particle velocity fluctuations are markedly higher than in channel flow due to cross-sectional vortices. These vortices, which scale with duct width, induce significant acceleration of particles towards the duct walls. Consequently, particles in ducts are subjected to more frequent and intense accelerations toward the walls, resulting in higher rms velocity fluctuations.

The differences between duct and channel flows become more pronounced as moving toward the center of the flow, although substantial differences persist even near the wall. 
For example, at $z^+=10$, the fluctuations in duct flow are 3 times greater than in channel flow. 
This indicates that particles near the wall in ducts are more frequently swept from the near-wall region than in channel flow, where particle mobility near the wall is significantly constrained.

Particles with higher inertia, such as $St$~=~37.50, exhibit significant velocity fluctuations near the wall ($z^+ < 5$). 
In contrast, particles with lower inertia, such as $St$~=~4.69, demonstrate the lowest fluctuation velocities, indicating their tendency to remain close to the wall for extended periods and experience repetitive collisions.
These observations support the anticipated charge profiles, suggesting a more uniform charge distribution in duct flows and higher Stokes numbers.

\captionsetup[subfigure]{labelformat=empty, skip=0pt} 
\captionsetup{justification=justified, width=\linewidth} 
\begin{figure}[t]
    \begin{subfigure}{0.33\textwidth}
        \centering
        \raisebox{0.005\height}{\stackinset{l}{0.05\textwidth}{t}{0.1ex}{\text{(a)}}{\includegraphics[width=\textwidth]{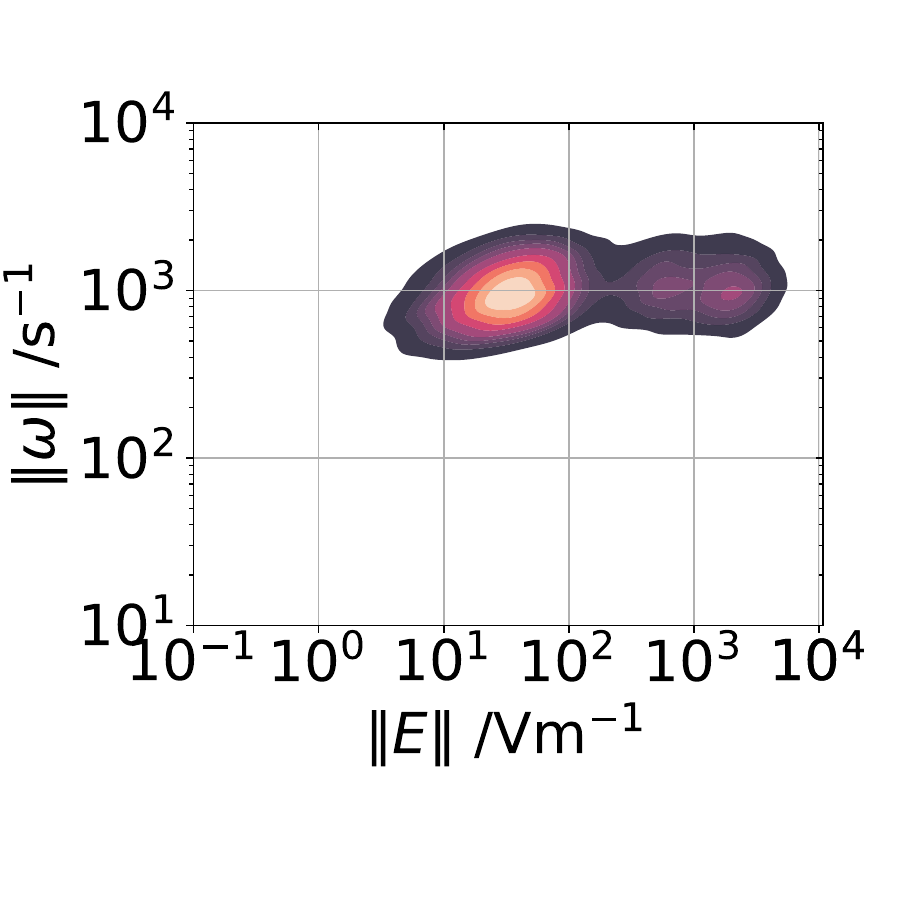}}}\vspace{-0.4cm}
        \caption{channel}
        \vspace{0.1cm}
        \caption{$z^{+}\leq 5$}
        \label{fig:channel_wall}
    \end{subfigure}%
    \begin{subfigure}{0.33\textwidth}
        \centering
        \raisebox{0.005\height}{\stackinset{l}{0.05\textwidth}{t}{0.1ex}{\text{(b)}}{\includegraphics[width=\textwidth]{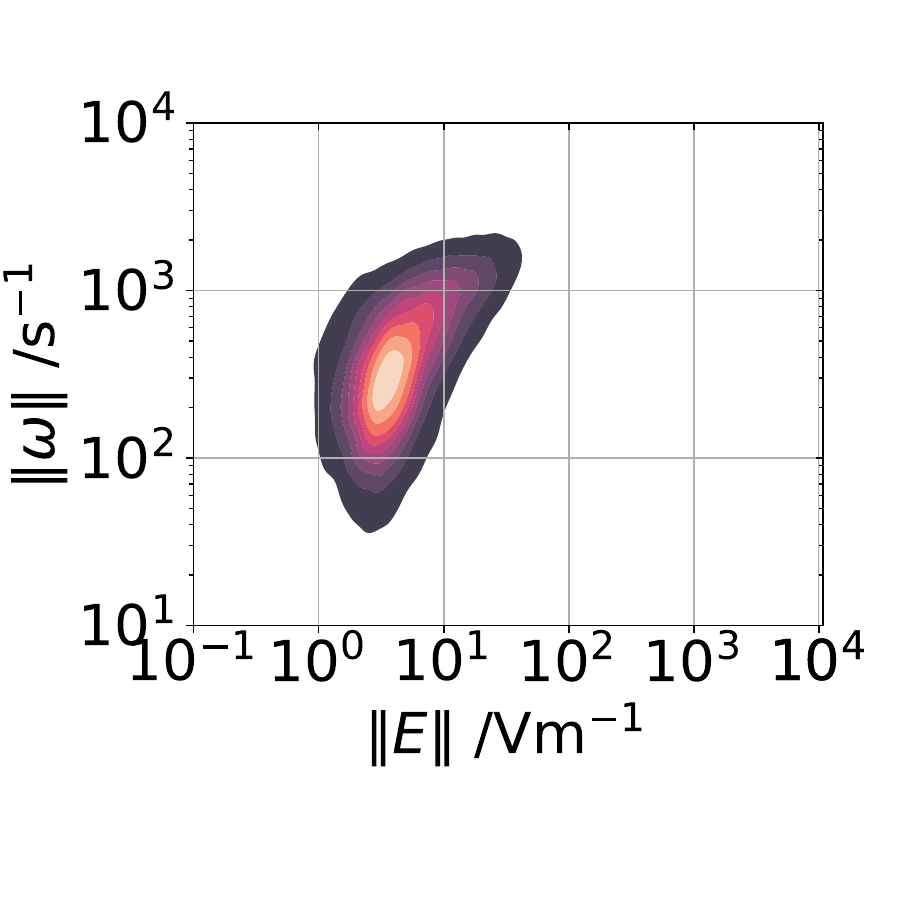}}}\vspace{-0.4cm}
        \caption{channel}
        \vspace{0.1cm}
        \caption{$5 \leq z^{+}\leq 30$}
        \label{fig:channel_buffer}
    \end{subfigure}%
    \begin{subfigure}{0.33\textwidth}
        \centering
        \raisebox{0.005\height}{\stackinset{l}{0.05\textwidth}{t}{0.1ex}{\text{(c)}}{\includegraphics[width=\textwidth]{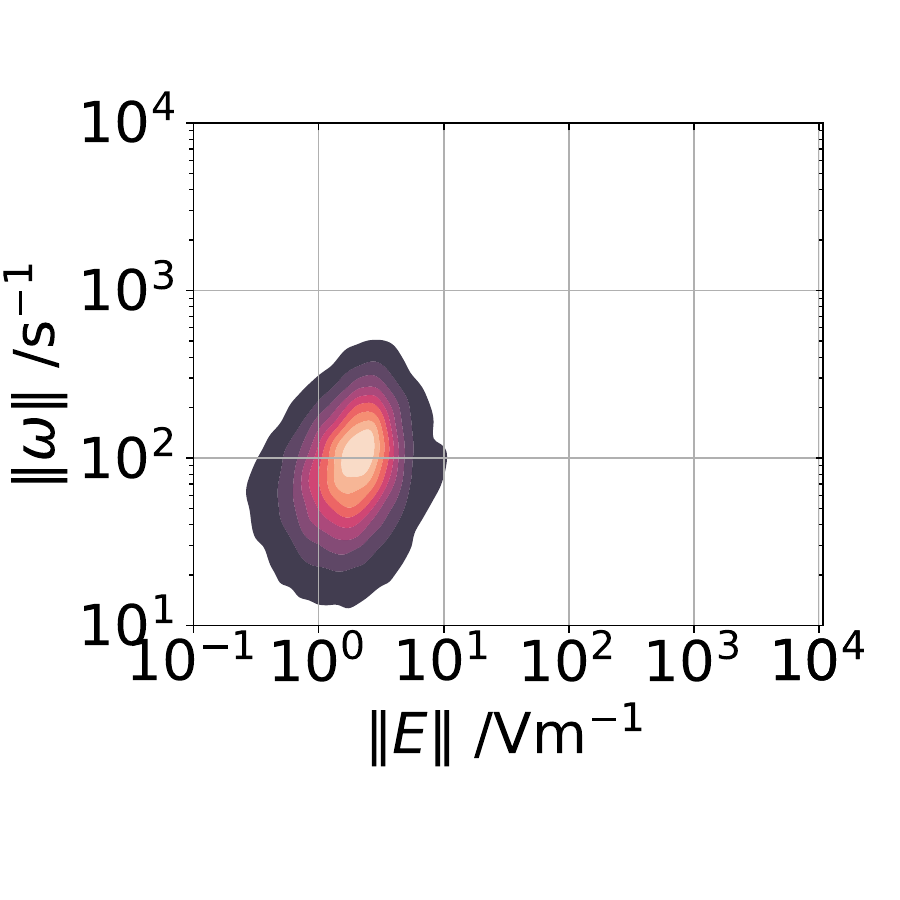}}}\vspace{-0.4cm}
        \caption{channel}
        \vspace{0.1cm}
        \caption{$z^{+} \geq 30 $}
        \label{fig:channel_bulk}
    \end{subfigure}

    \begin{subfigure}{0.33\textwidth}
        \centering
        \raisebox{0.005\height}{\stackinset{l}{0.05\textwidth}{t}{0.1ex}{\text{(d)}}{\includegraphics[width=\textwidth]{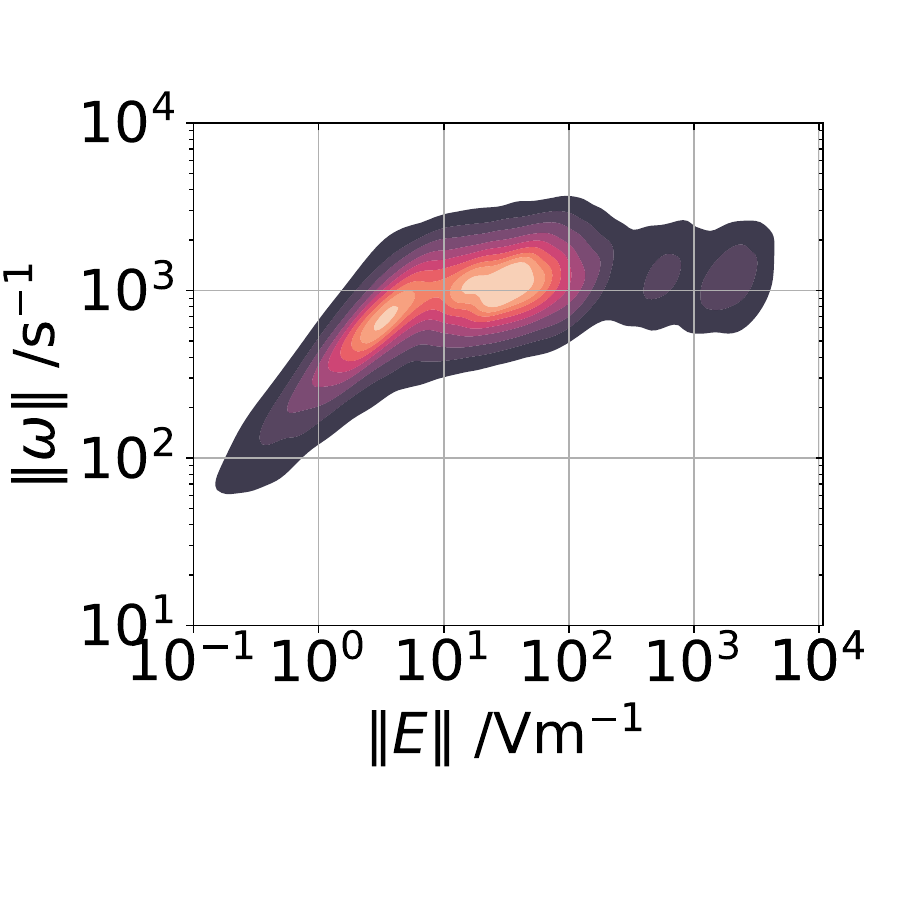}}}\vspace{-0.4cm}
        \caption{duct}
        \vspace{0.1cm}
        \caption{$z^{+}\leq 5$}
        \label{fig:duct_wall}
    \end{subfigure}%
    \hfill
    \begin{subfigure}{0.33\textwidth}
        \centering
        \raisebox{0.005\height}{\stackinset{l}{0.05\textwidth}{t}{0.1ex}{\text{(e)}}{\includegraphics[width=\textwidth]{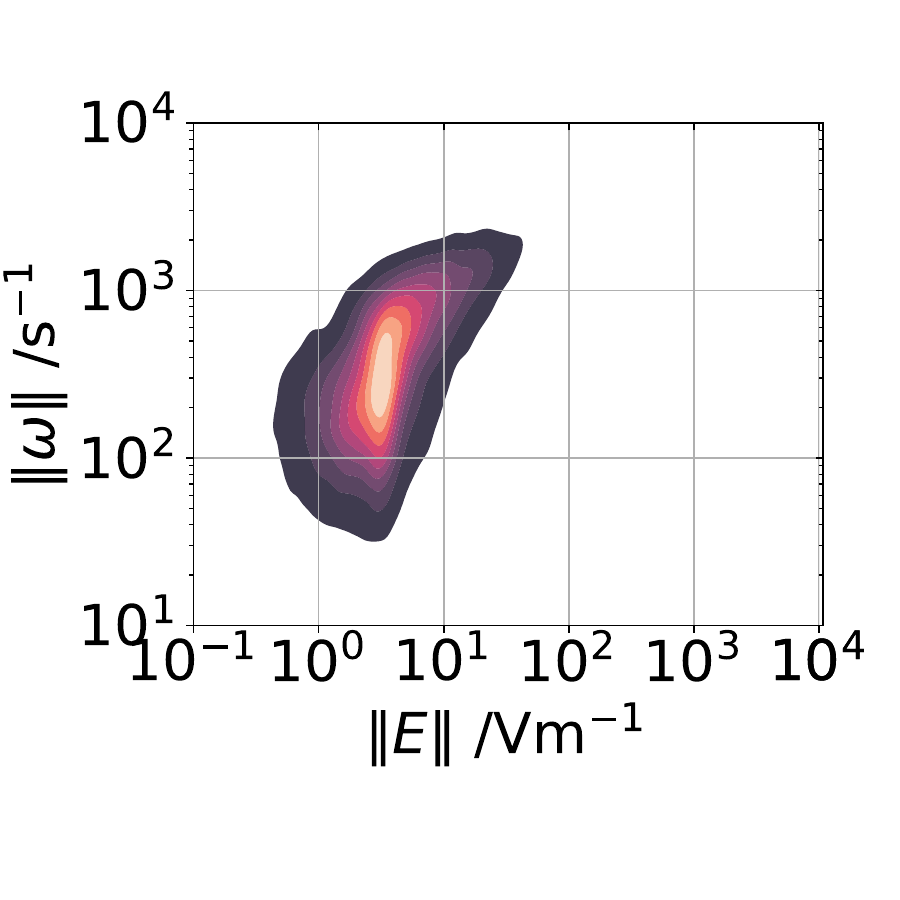}}}\vspace{-0.4cm}
        \caption{duct}
        \vspace{0.1cm}
        \caption{$5 \leq z^{+}\leq 30$}
        \label{fig:duct_buffer}
    \end{subfigure}%
    \hfill
    \begin{subfigure}{0.33\textwidth}
        \centering
        \raisebox{0.005\height}{\stackinset{l}{0.05\textwidth}{t}{0.1ex}{\text{(f)}}{\includegraphics[width=\textwidth]{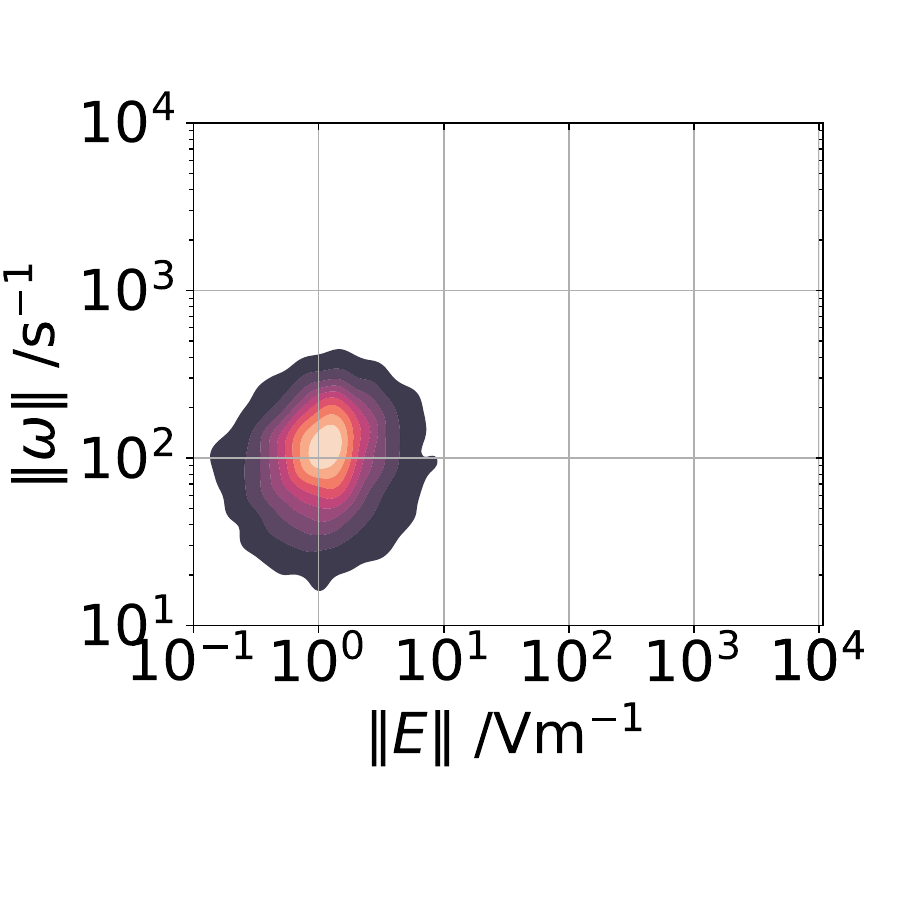}}}\vspace{-0.4cm}
        \caption{duct}
        \vspace{0.1cm}
        \caption{$z^{+} \geq 30 $}
        \label{fig:duct_bulk}
    \end{subfigure}
    \caption{Magnitude of electric field strength in low vorticity ($||\omega||$~=~$10^{\,2}$~$\mathrm{s}^{-1}$) and high vorticity ($||\omega||$~=~$10^{\,3}$~$\mathrm{s}^{-1}$) regions for channel and duct flow for particles of $St =4.69$. Figures show the data when the average powder charge is half of the equilibrium charge, $q^*_{\textup{avg}}=0.5$.}
    \label{fig:vor_1}
\end{figure}
\captionsetup[figure]{labelfont=default}

Figure~\ref{fig:vor_1} illustrates the probability density distribution of vorticity and electric field for $St = 4.69$ in the wall-resolved region, buffer layer, and bulk flow. 
For both channel and duct flow, the electric field strength peaks at the walls and shifts to lower values toward the bulk region. This aligns with the charge profiles shown in Figure~\ref{fig:q_z}, where the average charge of the powder is highest at the walls.

In the wall-resolved region, channel flow consists solely of high-vorticity zones, while duct flow contains both high- and low-vorticity regions, with low vorticity resulting from the duct corners. The vorticity profiles for channel and duct flows show similar patterns in the buffer layer and bulk flow.

In the channel's wall-resolved region, electric field strength ranges from 3.49 to \( 5.54 \times 10^3 \, \text{V/m} \). The maximum occurrence is at a vorticity of \( 1 \times 10^3 \, \text{s}^{-1} \) and an electric field strength of \( 3.22 \times 10^1 \, \text{V/m} \), followed by a secondary peak at around \( 1 \times 10^3 \, \text{V/m} \).
In duct flow, electric field strength varies from \( 1.34 \times 10^{-1} \) to \( 4.26 \times 10^3 \, \text{V/m} \). Duct flow features regions of low electric field strength in the wall-resolved region, unlike in channel flow. High-vorticity regions in the duct have stronger electric fields, while low-vorticity areas have much weaker electric fields. The highest occurrence in duct flow appears at vorticity of \( 1 \times 10^3 \, \text{s}^{-1} \) and an electric field strength of \( 9.53 \times 10^1 \, \text{V/m} \), with another peak at around \( 6.57 \times 10^2 \, \text{s}^{-1} \) and \( 3.38 \, \text{V/m} \). Buffer layer and bulk region show similar trends for channel and duct flow.

\captionsetup[subfigure]{labelformat=empty, skip=0pt}
\captionsetup{justification=justified, width=\linewidth}

\begin{figure}[t]
    \begin{subfigure}{0.33\textwidth}
        \centering
        \raisebox{0.005\height}{\stackinset{l}{0.05\textwidth}{t}{0.1ex}{\text{(a)}}{\includegraphics[width=\textwidth]{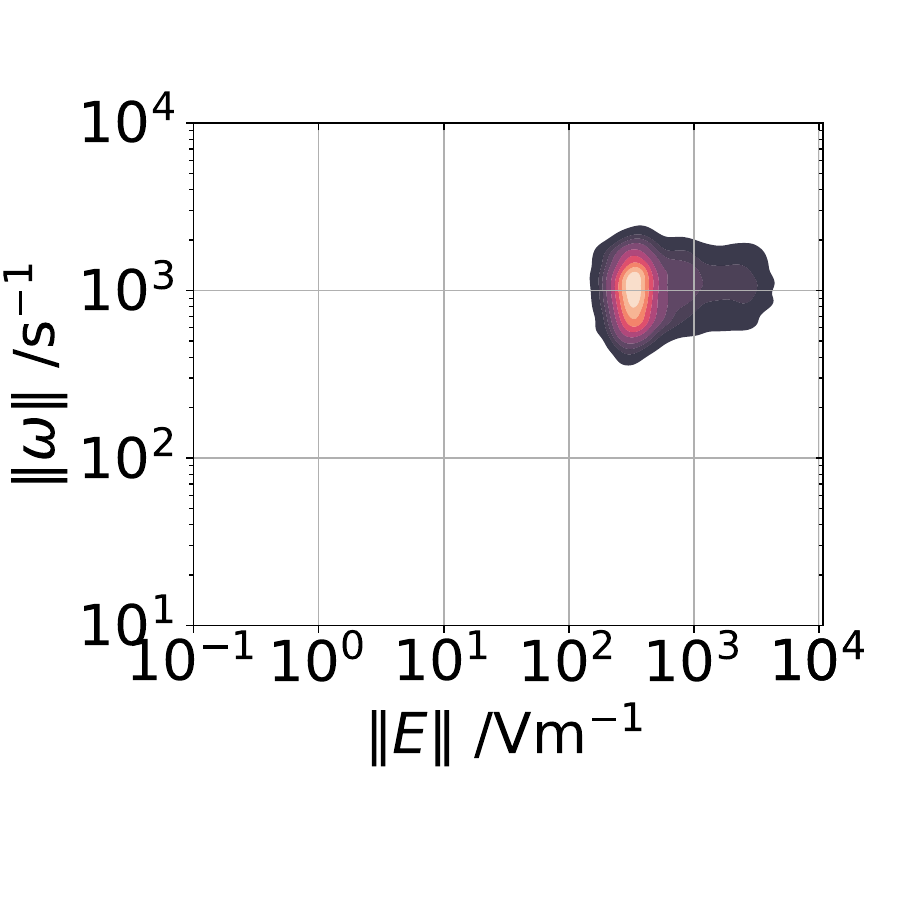}}}\vspace{-0.4cm}
        \caption{channel}
        \vspace{0.1cm}
        \caption{$z^{+}\leq 5$}
        \label{fig:channel_wall}
    \end{subfigure}%
    \begin{subfigure}{0.33\textwidth}
        \centering
        \raisebox{0.005\height}{\stackinset{l}{0.05\textwidth}{t}{0.1ex}{\text{(b)}}{\includegraphics[width=\textwidth]{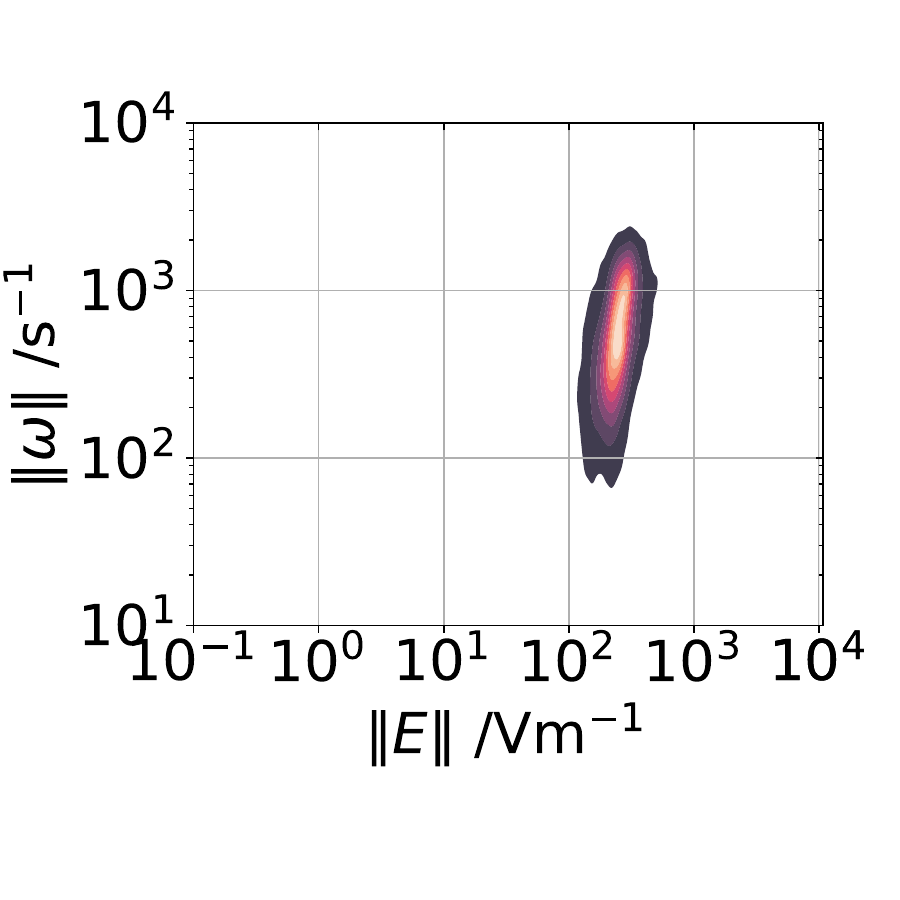}}}\vspace{-0.4cm}
        \caption{channel}
        \vspace{0.1cm}
        \caption{$5 \leq z^{+}\leq 30$}
        \label{fig:channel_buffer}
    \end{subfigure}%
    \begin{subfigure}{0.33\textwidth}
        \centering
        \raisebox{0.005\height}{\stackinset{l}{0.05\textwidth}{t}{0.1ex}{\text{(c)}}{\includegraphics[width=\textwidth]{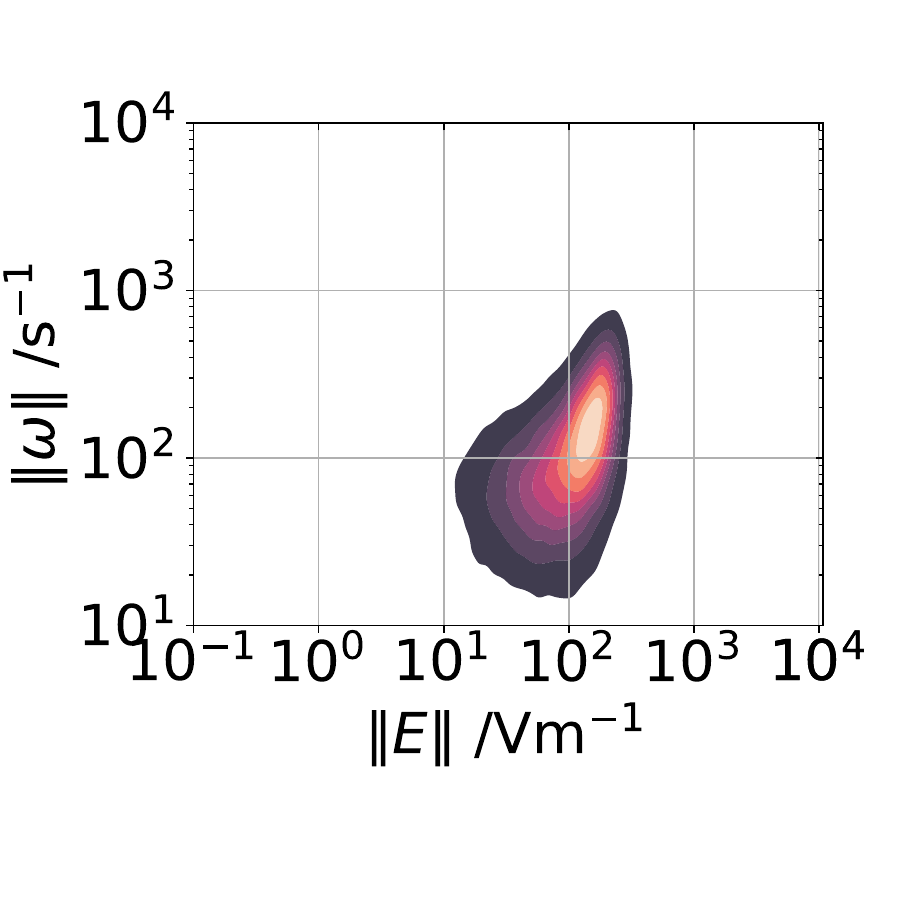}}}\vspace{-0.4cm}
        \caption{channel}
        \vspace{0.1cm}
        \caption{$z^{+} \geq 30 $}
        \label{fig:channel_bulk}
    \end{subfigure}

    \begin{subfigure}{0.33\textwidth}
        \centering
        \raisebox{0.005\height}{\stackinset{l}{0.05\textwidth}{t}{0.1ex}{\text{(d)}}{\includegraphics[width=\textwidth]{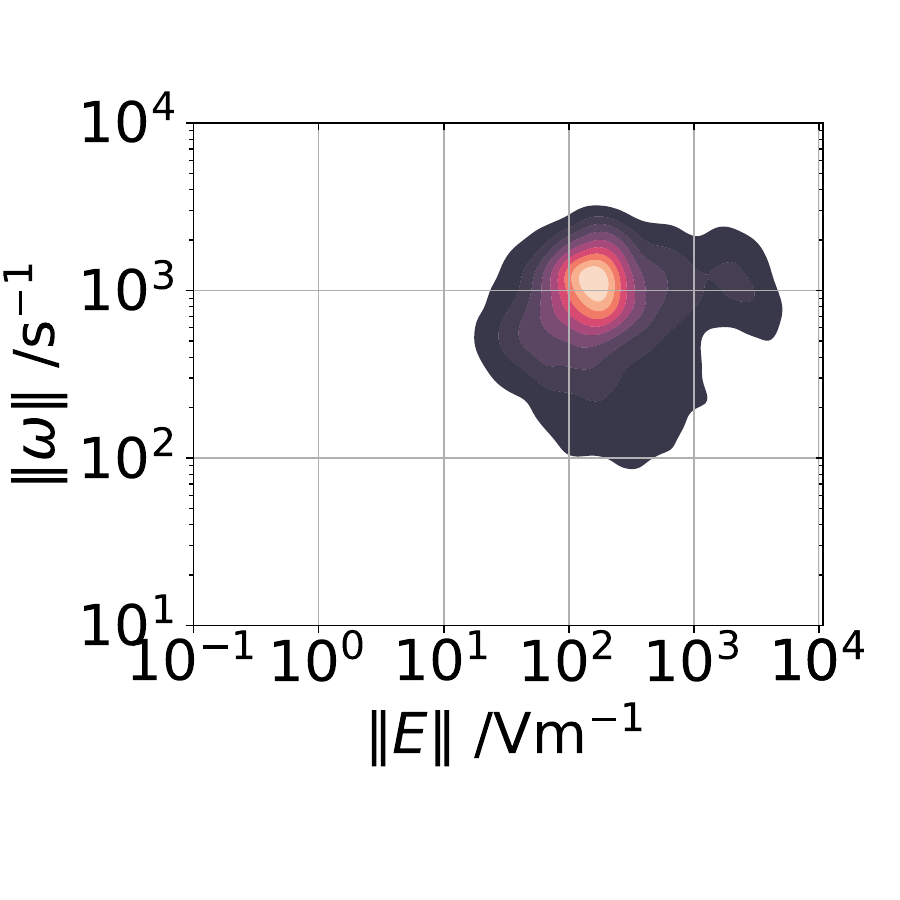}}}\vspace{-0.4cm}
        \caption{duct}
        \vspace{0.1cm}
        \caption{$z^{+}\leq 5$}
        \label{fig:duct_wall}
    \end{subfigure}%
    \hfill
    \begin{subfigure}{0.33\textwidth}
        \centering
        \raisebox{0.005\height}{\stackinset{l}{0.05\textwidth}{t}{0.1ex}{\text{(e)}}{\includegraphics[width=\textwidth]{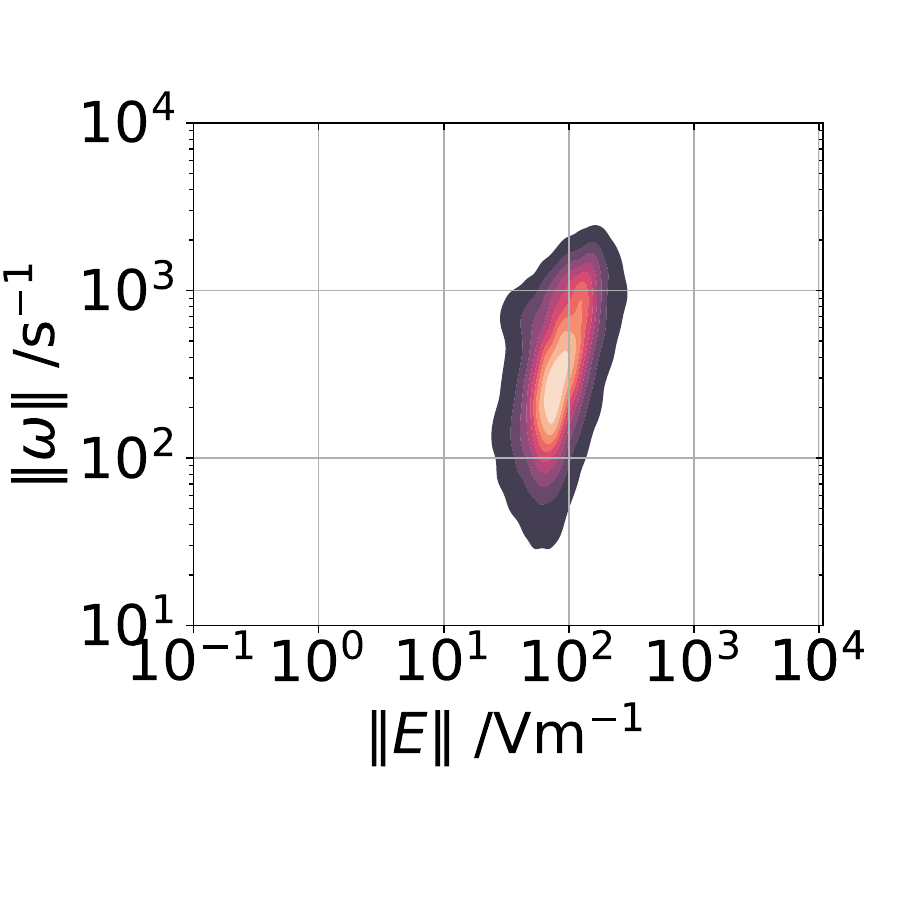}}}\vspace{-0.4cm}
        \caption{duct}
        \vspace{0.1cm}
        \caption{$5 \leq z^{+}\leq 30$}
        \label{fig:duct_buffer}
    \end{subfigure}%
    \hfill
    \begin{subfigure}{0.33\textwidth}
        \centering
        \raisebox{0.005\height}{\stackinset{l}{0.05\textwidth}{t}{0.1ex}{\text{(f)}}{\includegraphics[width=\textwidth]{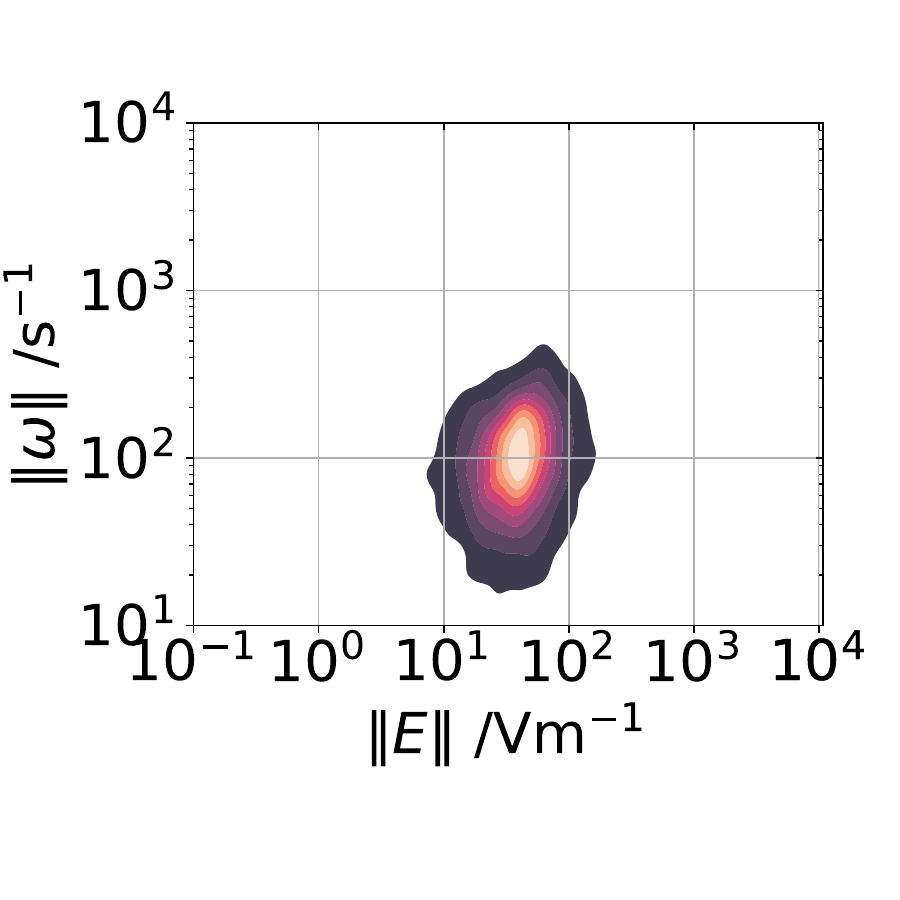}}}\vspace{-0.4cm}
        \caption{duct}
        \vspace{0.1cm}
        \caption{$z^{+} \geq 30 $}
        \label{fig:duct_bulk}
    \end{subfigure}
    \caption{Magnitude of electric field strength in low vorticity ($||\omega||$~=~$10^{\,2}$~$\mathrm{s}^{-1}$) and high vorticity ($||\omega||$~=~$10^{\,3}$~$\mathrm{s}^{-1}$) regions for channel and duct flow for particles of $St =37.50$. Figures show the data when the average powder charge is half of the equilibrium charge, $q^*_{\textup{avg}}=0.5$.}
    \label{fig:vor_2}
\end{figure}
\captionsetup[figure]{labelfont=default}

Figure~\ref{fig:vor_2} presents the probability density distribution for $St=37.50$, for the wall-resolved, buffer, and bulk regions. For both flow types, electric field strength reaches its peak in the wall-resolved region and diminishes in the bulk, similar to the behavior seen at $St=4.69$. However, compared to $St=4.69$, the lower threshold of electric field strength is higher, which aligns with Figure~\ref{fig:q_z}, indicating higher average powder charges for $St=37.50$.

In the wall-resolved region of channel flow, the electric field strength ranges from \( 1.56 \times 10^2 \) to \( 4.48 \times 10^3 \, \text{V/m} \) in high-vorticity areas. In duct flow, both high- and low-vorticity regions are present near the wall, with electric field strengths ranging from \( 1.56 \times 10^1 \) to \( 6.85 \times 10^3 \, \text{V/m} \), indicating a broader range compared to the channel.
Compared to $St = 4.69$, the electric field strength is not reduced in low-vorticity regions and exhibits a more uniform distribution across varying vorticity levels. This suggests that the distribution of charged particles along the wall significantly differs with different Stokes numbers.

In the buffer layer, both flow types exhibit regions of high and low vorticity. In the high vorticity region, electric field strength is higher, with channel flow demonstrating higher values than duct flow.

In the bulk region, channel flow maintains only stronger electric fields in high-vorticity areas, while duct flow shows a uniform electric field distribution across vorticity levels.

\subsection{Effects of electrostatic charge on fluid and particle dynamics}
In this section, we investigate the impact of triboelectric charging on powder flow dynamics. 
Electrostatic forces significantly influence particle behavior and distribution within the domain. 
This interaction, in turn, affects the fluid phase and alters the overall flow characteristics.

\captionsetup[subfigure]{labelformat=empty}
\captionsetup{justification=justified, width=1\linewidth}
\begin{figure}[htbp]
    \centering
    \begin{subfigure}{0.45\textwidth}
        \centering
        \raisebox{-0.5\height}{\stackinset{l}{0.1ex}{t}{0.1ex}{\text{(a)}}{\includegraphics[width=0.95\textwidth]{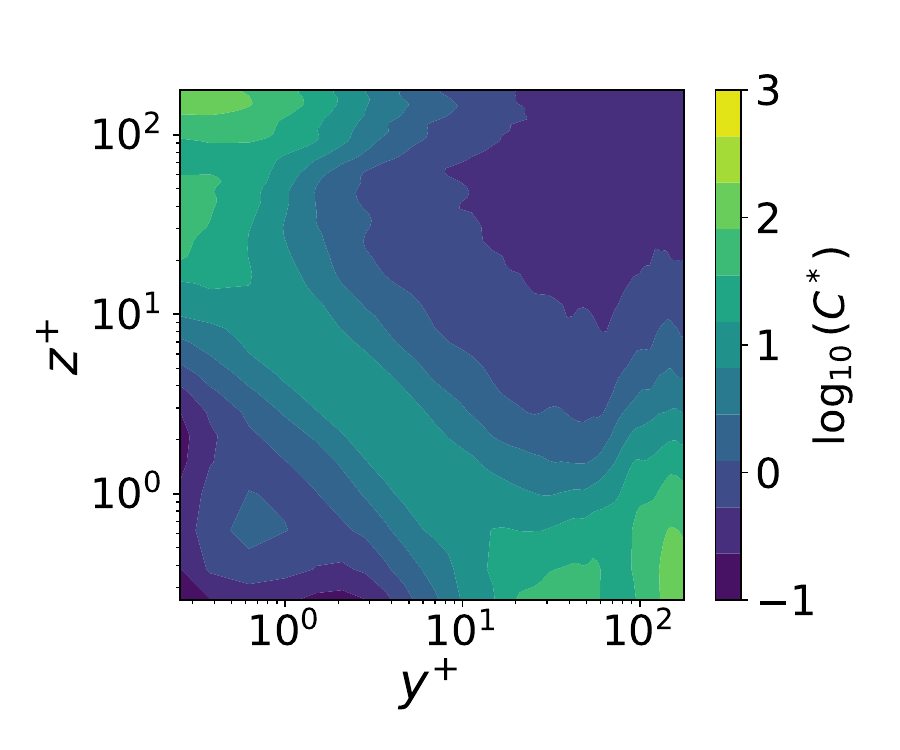}}}
        \vspace{-0.25cm}
        \captionsetup{justification=justified, width=1\linewidth}
        \caption{uncharged, $St$~=~4.69}
        \vspace{0.30cm}
        \label{fig:con1}
    \end{subfigure}
    \hfill
    \begin{subfigure}{0.45\textwidth}
        \centering
        \raisebox{-0.5\height}{\stackinset{l}{0.1ex}{t}{0.1ex}{\text{(b)}}{\includegraphics[width=0.95\textwidth]{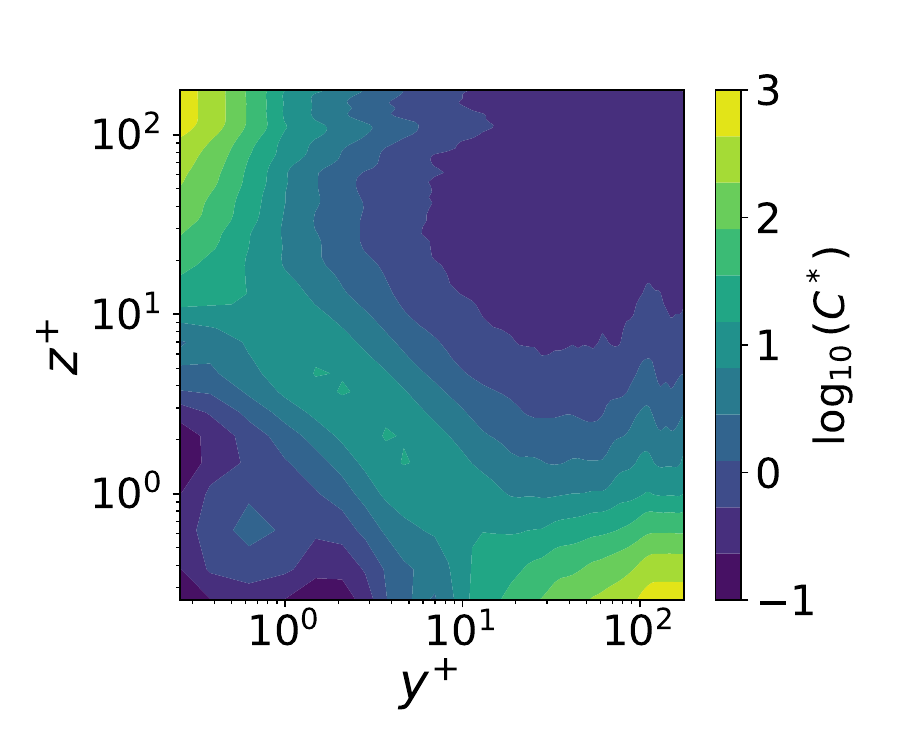}}}
        \vspace{-0.25cm}
        \captionsetup{justification=justified, width=1\linewidth}
        \caption{charged, $St$~=~4.69}
        \vspace{0.30cm}
        \label{fig:con2}
    \end{subfigure}
    \\
    \vspace{-0.38cm}
    \begin{subfigure}{0.45\textwidth}
        \centering
        \raisebox{-0.5\height}{\stackinset{l}{0.1ex}{t}{0.1ex}{\text{(c)}}{\includegraphics[width=0.95\textwidth]{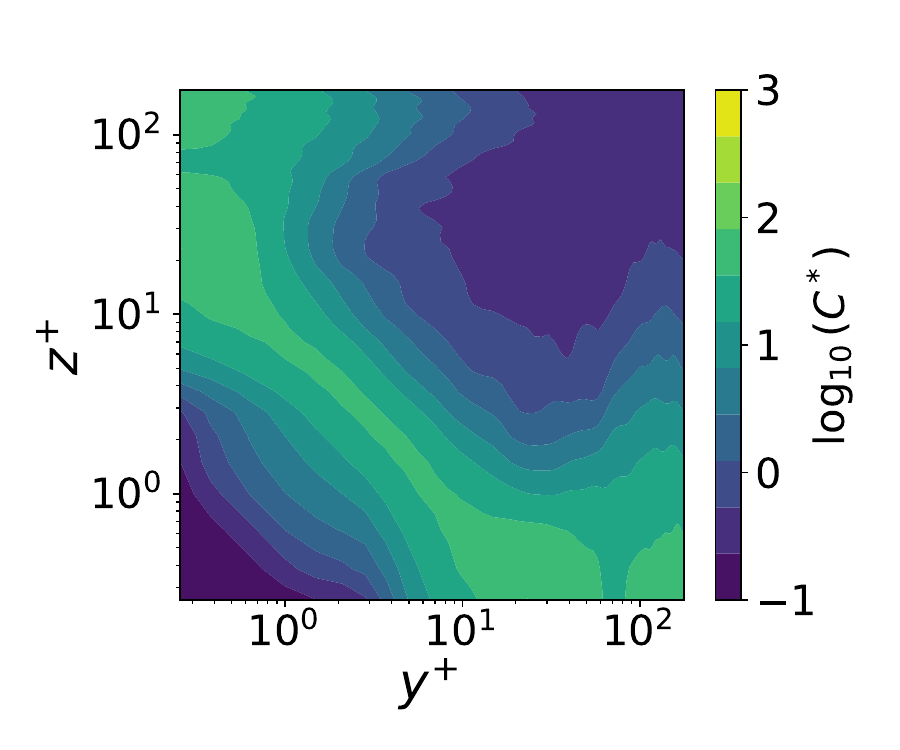}}}
        \vspace{-0.25cm}
        \captionsetup{justification=justified, width=1\linewidth}
        \caption{uncharged, $St$~=~9.38}
        \vspace{0.30cm}
        \label{fig:con3}
    \end{subfigure}
    \hfill
    \begin{subfigure}{0.45\textwidth}
        \centering
        \raisebox{-0.5\height}{\stackinset{l}{0.1ex}{t}{0.1ex}{\text{(d)}}{\includegraphics[width=0.95\textwidth]{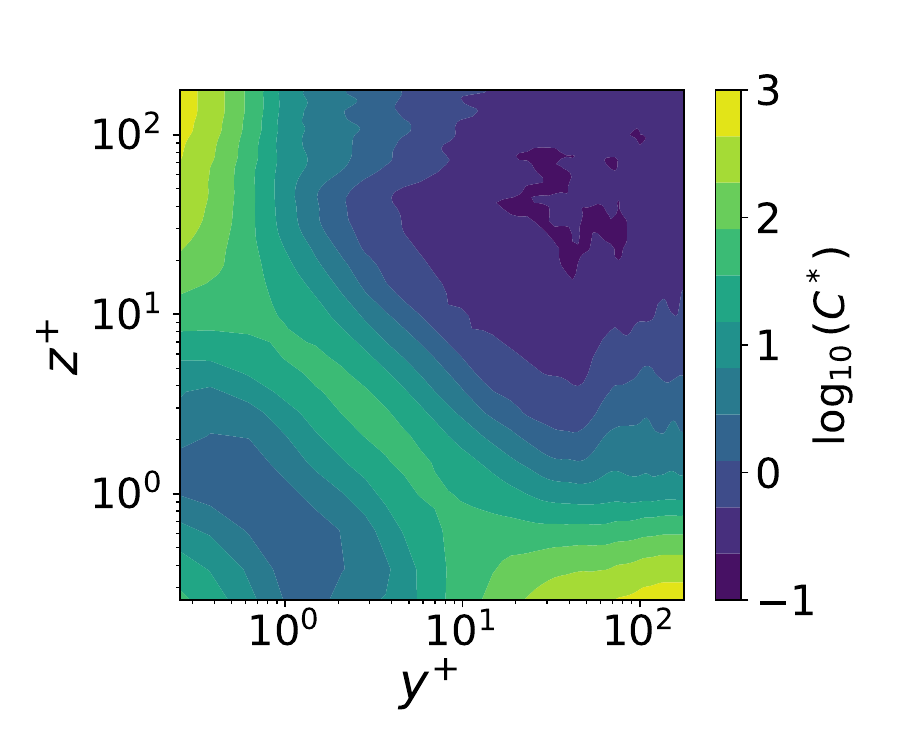}}}
        \vspace{-0.25cm}
        \captionsetup{justification=justified, width=1\linewidth}
        \caption{charged, $St$~=~9.38}
        \vspace{0.30cm}
        \label{fig:con4}
    \end{subfigure}
    \\
    \vspace{-0.38cm}
    \begin{subfigure}{0.45\textwidth}
        \centering
        \raisebox{-0.5\height}{\stackinset{l}{0.1ex}{t}{0.1ex}{\text{(e)}}{\includegraphics[width=0.95\textwidth]{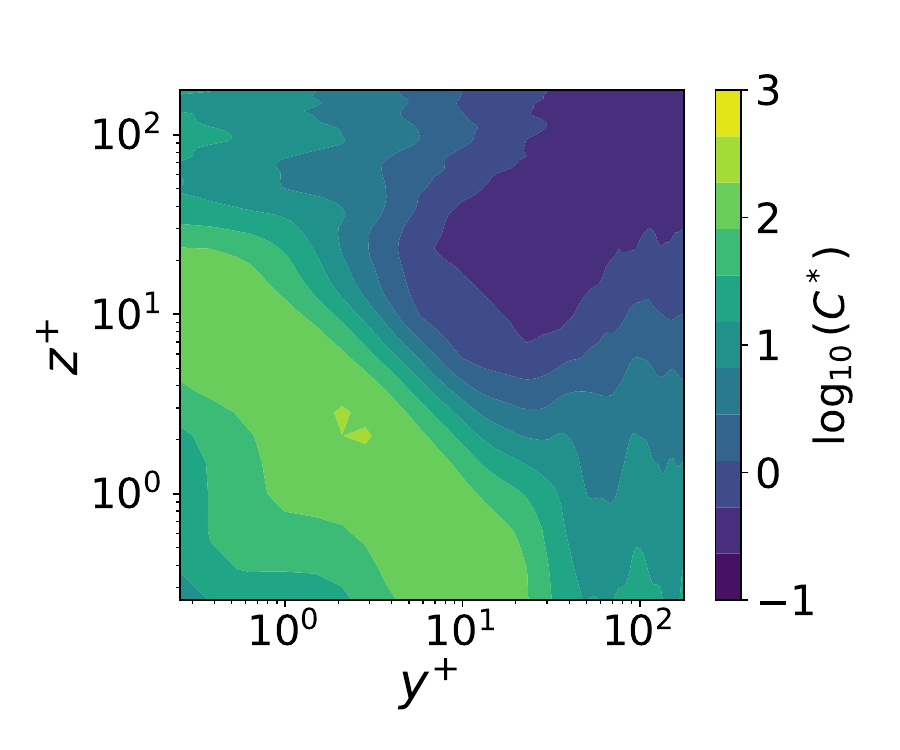}}}
        \vspace{-0.25cm}
        \captionsetup{justification=justified, width=1\linewidth}
        \caption{uncharged, $St$~=~18.75}
        \vspace{0.30cm}
        \label{fig:con5}
    \end{subfigure}
    \hfill
    \begin{subfigure}{0.45\textwidth}
        \centering
        \raisebox{-0.5\height}{\stackinset{l}{0.1ex}{t}{0.1ex}{\text{(f)}}{\includegraphics[width=0.95\textwidth]{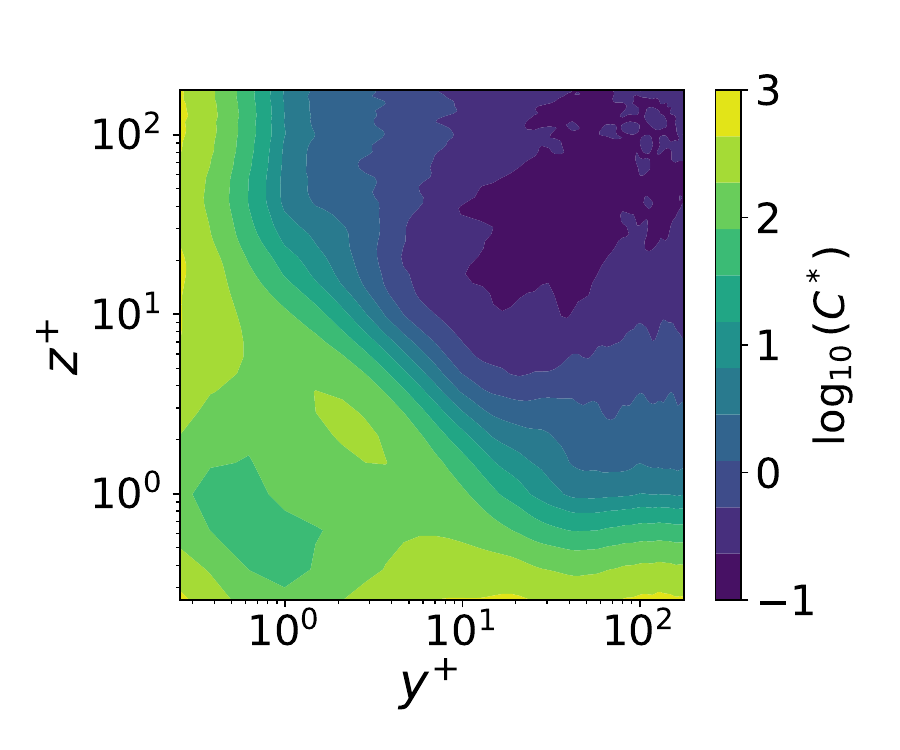}}}
        \vspace{-0.25cm}
        \captionsetup{justification=justified, width=1\linewidth}
        \caption{charged, $St$~=~18.75}
        \vspace{0.30cm}
        \label{fig:con6}
    \end{subfigure}
    \\
    \vspace{-0.38cm}
    \begin{subfigure}{0.45\textwidth}
        \centering
        \raisebox{-0.45\height}{\stackinset{l}{0.1ex}{t}{0.1ex}{\text{(g)}}{\includegraphics[width=0.95\textwidth]{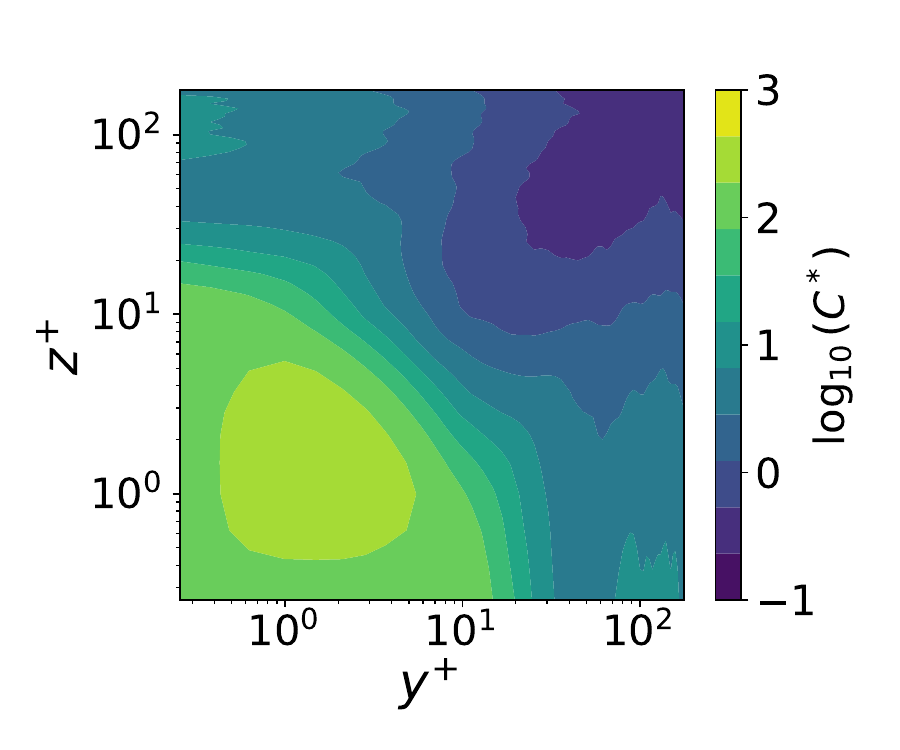}}}
        \vspace{-0.25cm}
        \captionsetup{justification=justified, width=1\linewidth}
        \caption{uncharged, $St$~=~37.50}
        \vspace{0.30cm}
        \label{fig:con7}
    \end{subfigure}
    \hfill
    \begin{subfigure}{0.45\textwidth}
        \centering
        \raisebox{-0.45\height}{\stackinset{l}{0.1ex}{t}{0.1ex}{\text{(h)}}{\includegraphics[width=0.95\textwidth,]{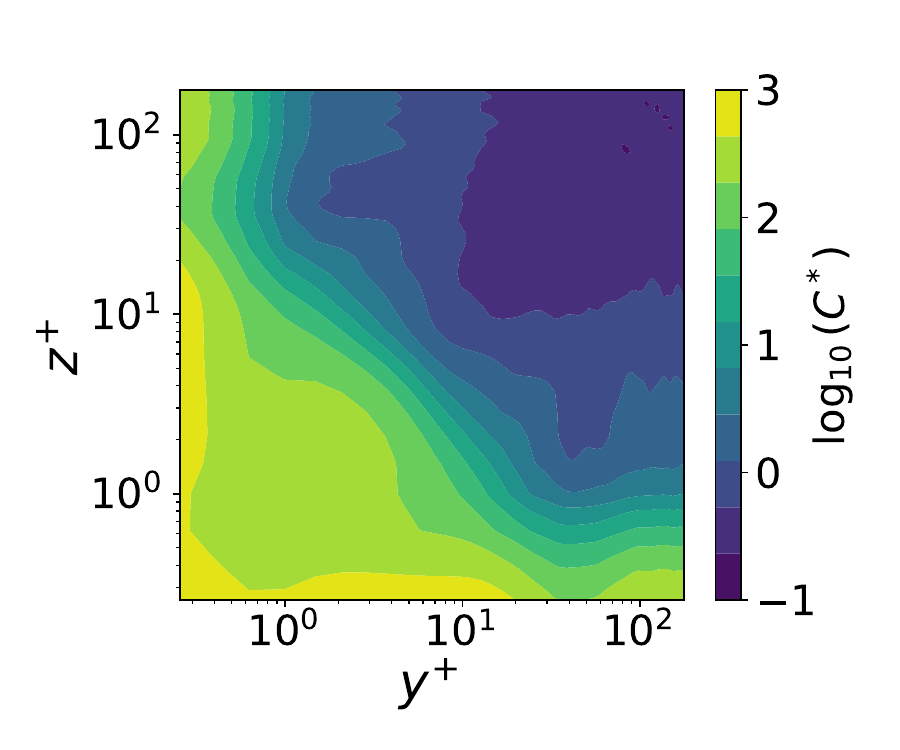}}}
        \vspace{-0.25cm}
        \captionsetup{justification=justified, width=1\linewidth}
        \caption{charged, $St$~=~37.50}
        \vspace{0.3cm}
        \label{fig:con8}
    \end{subfigure}
    \vspace{-0.4cm}
    \caption{Normalized particle concentration profiles of uncharged and charged duct flow. For the charged case, the average powder charge in the duct is half of the equilibrium charge, $q^*_{\textup{avg}}=0.5$.\,\,Due to symmetry, only one quadrant of the square-duct is shown.}
    \label{fig:concentration}
\end{figure}
\captionsetup[figure]{labelfont=default}
Figure~\ref{fig:concentration} illustrates concentration profiles within duct flow, comparing scenarios with uncharged particles to those where the powder has reached half of its equilibrium charge ($q^*_{\textup{avg}}=0.5$). 
In all scenarios, particles accumulate at the duct walls, with fewer particles found at the center of the duct. 
The Stokes number of the particles affects the specific locations of this accumulation along the wall.

For uncharged particles, those with $St = 4.69$ accumulate at the wall bisectors and within the $z^+ = 20$ range to $z^+ = 60$. Particles with $St = 9.38$ collect at the wall bisectors and between $z^+ = 10$ and $z^+ = 60$. In contrast, particles with $St = 18.75$ and $St = 37.50$ concentrate at the corners of the duct.

The preferred accumulation of particles is directly related to the interplay between particles' inertia and the secondary flow structures.
Particles with higher inertia, $St$~=~18.75, 37.50 accumulate mainly in the stagnation region (corners). 
This occurs because, beyond a specific point near the corner where the secondary flow changes direction, the secondary flow velocities aren't strong enough to maintain the motion of high-inertia particles parallel to the walls, leading them to accumulate more within the stagnation region ~\citep{Wan20}.
In contrast, low-inertia particles, $St$~=~4.69, closely follow the secondary flow, avoiding collection at the corner.

When particles acquire charge, they undergo significant changes due to the electrostatic forces exerted by mirror charges on the walls. 
Charged particles also generate electric fields in their vicinity, capable of repelling other charged particles. 
These interactions influence their trajectories and alter the resulting particle concentration profiles.

Results show that, when particles are charged, their accumulation at the wall increases across all scenarios. 
However, the distribution along the wall varies. 
In the uncharged case, with $St$~=~4.69, particles accumulate primarily at the wall bisectors and within the $z^+ = 20$ range to $z^+ = 60$. 
Particles with $St = 9.38$ collect at the wall bisectors and between $z^+ = 10$ and $z^+ = 60$.
Upon charging, accumulation becomes solely concentrated at the wall bisectors.
Compared to $St$~=~4.69, the profile along the wall is more uniform when $St$~=~9.38.
In cases where $St$~=~18.75 and 37.50, charged particles accumulate significantly at the corners of the duct. This increased concentration is due to electrostatic forces from mirror charges on both walls, which attract particles toward these corners.

\begin{figure}[t]
    \centering
    \captionsetup[subfigure]{labelformat=empty} 
    \begin{subfigure}[b]{0.48\textwidth}
        \centering
        \raisebox{-0.5\height}{\stackinset{l}{0.5ex}{t}{0.5ex}{\text{(a)}}{\includegraphics[width=\textwidth]{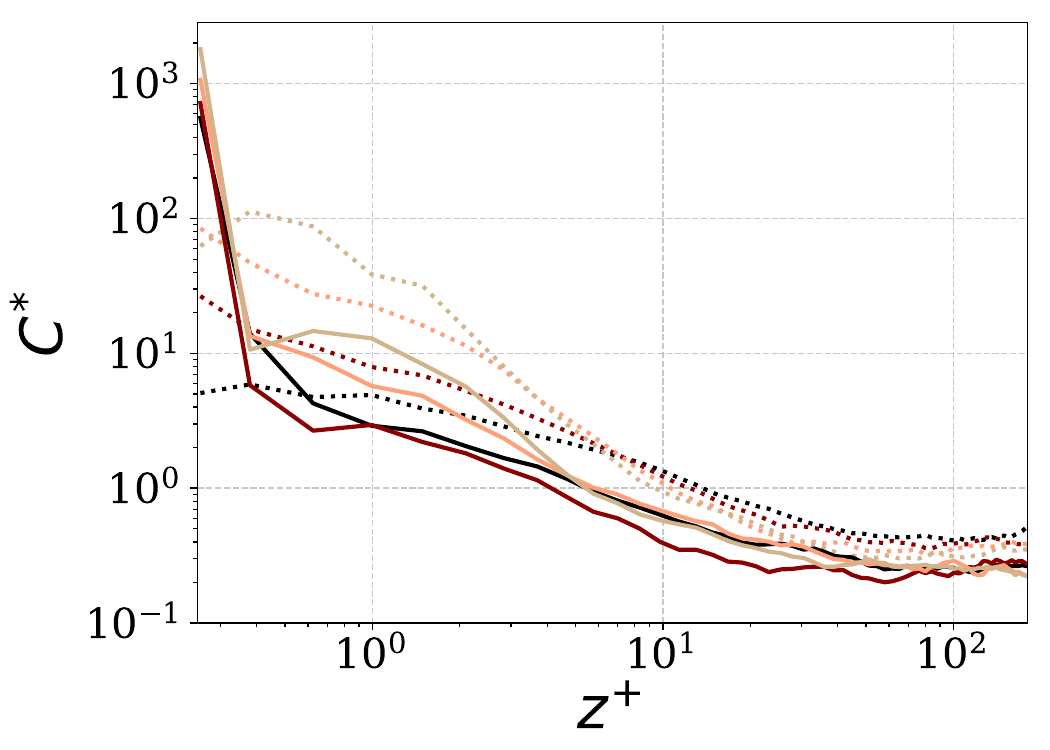}}}
        \caption{}
        \label{fig:C_duct}
    \end{subfigure}
    \hfill
    \begin{subfigure}[b]{0.48\textwidth}
        \centering
        \raisebox{-0.5\height}{\stackinset{l}{0.5ex}{t}{0.5ex}{\text{(b)}}{\includegraphics[width=\textwidth]{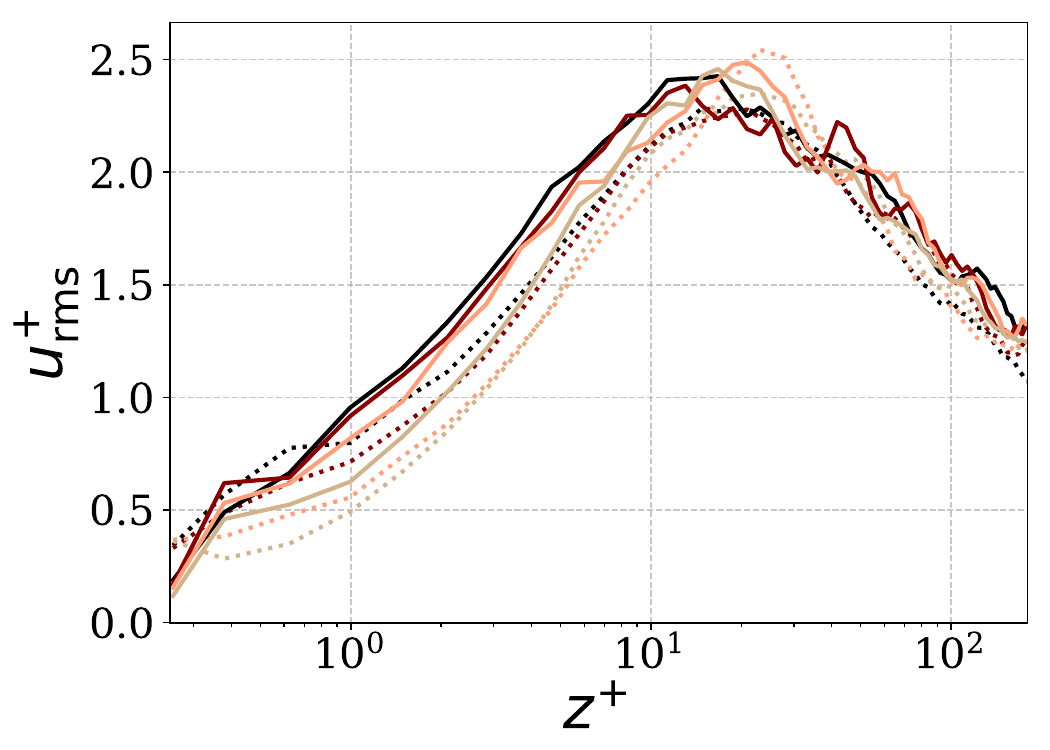}}}
        \caption{}
        \label{fig:urms_duct}
    \end{subfigure}
    \vspace{-1.5\baselineskip}
    \caption{Normalized particle concentration profiles (a), rms stream-wise fluid velocity (b), for uncharged and charged duct flow. The wall-normal profiles along the bisector,  $y^{+\textup{}}= 180$, are depicted for duct flow. The uncharged case is shown by line points while the charged case is shown by lines. For the charged case, the average powder charge in the domain is half of the equilibrium charge, $q^*_{\textup{avg}}=0.5$.  for duct flow. Colors indicate Stokes number; 
        (\,\textcolor{black}{\rule[0.2ex]{0.5cm}{1pt}}\,) $St\,=\,37.50$,\,
        (\,\textcolor{darkred}{\rule[0.2ex]{0.5cm}{1pt}}\,) $St\,=\,18.75$,\,
        (\,\textcolor{lightsalmon}{\rule[0.2ex]{0.5cm}{1pt}}\,)$St\,=\,9.38$,\,
        (\,\textcolor{tan}{\rule[0.2ex]{0.5cm}{1pt}}\,)$St\,=\,4.69$.}
    \label{fig:cu_duct}
\end{figure}

Figure~\ref{fig:C_duct} shows the concentration profiles along the bisector, $y^+=~180$, of the duct for charged and uncharged particles.
In all scenarios, particles accumulate the most at the duct walls, with the most pronounced accumulation observed for $St$~=~4.69. 
Notably, in the uncharged case, the accumulation for $St$~=~4.69 occurs not directly at the wall but at $z^+~=~0.38$.
However, when particles are charged, they accumulate directly at the wall. 
This shift demonstrates that electrostatic forces have a stronger influence than aerodynamic forces in determining particle location. 
In other words, the attractive electrostatic forces from the mirror charges at the wall bring particles from the near-wall region directly to the wall.

As Stokes number increases, the concentration at the duct wall decreases. 
This trend is similar for charged and uncharged particles, although less pronounced for charged particles. 
For uncharged particles, the wall concentration is around 20 times higher for $St$~=~4.69 compared to $St$~=~37.50. 
When particles are charged, this concentration ratio reduces to 4.
Therefore, electrostatic forces dominate aerodynamic forces in the near-wall region.

When particles gain charge, there is a significant increase in their accumulation directly at the wall. 
For instance, the particle concentration at the wall increased approximately 30 times for $St$~=~4.69 compared to the uncharged condition. 
Similarly, for $St$~=~9.38, 18.75, and 37.50, the concentration increased by 10, 20, and 100 times, respectively. 
Consequently, this wall accumulation reduces particle concentration at the center of the duct.

Figure~\ref{fig:urms_duct} displays the rms fluid velocity in the stream-wise direction for both the uncharged and charged cases.

When charged, particles tend to move away from the center of the duct. 
This migration reduces the number of particles in the center, which, in turn, allows for higher turbulence and fluctuations in the fluid velocity. 
As a result, the root mean square (rms) fluid velocity, which measures the intensity of these fluctuations, increases at the center of the duct.

In contrast, at the duct walls, the increase in particle concentration due to charging has the opposite effect. 
The high concentration of particles at the wall can dampen or suppress the turbulence in this region. 
This results in a decrease in the rms fluid velocity near the wall, as the increased particle presence reduces the intensity of turbulent fluctuations in that area.
These findings align with~\citet{Zhang24}'s research on DNS of turbulent channel flow, which showed that the rms of stream-wise fluctuating velocity decreases with particle presence and is further suppressed for charged particles, indicating that inter-particle electrostatic forces inhibit turbulence.

\captionsetup[subfigure]{labelformat=empty}
\captionsetup{justification=justified, width=1\linewidth}
\begin{figure}[ht]
    \centering
    \begin{subfigure}{0.49\textwidth}
        \centering
        \raisebox{-0.5\height}{\stackinset{l}{0.1ex}{t}{0.1ex}{\text{(a)}}{\includegraphics[width=\textwidth]{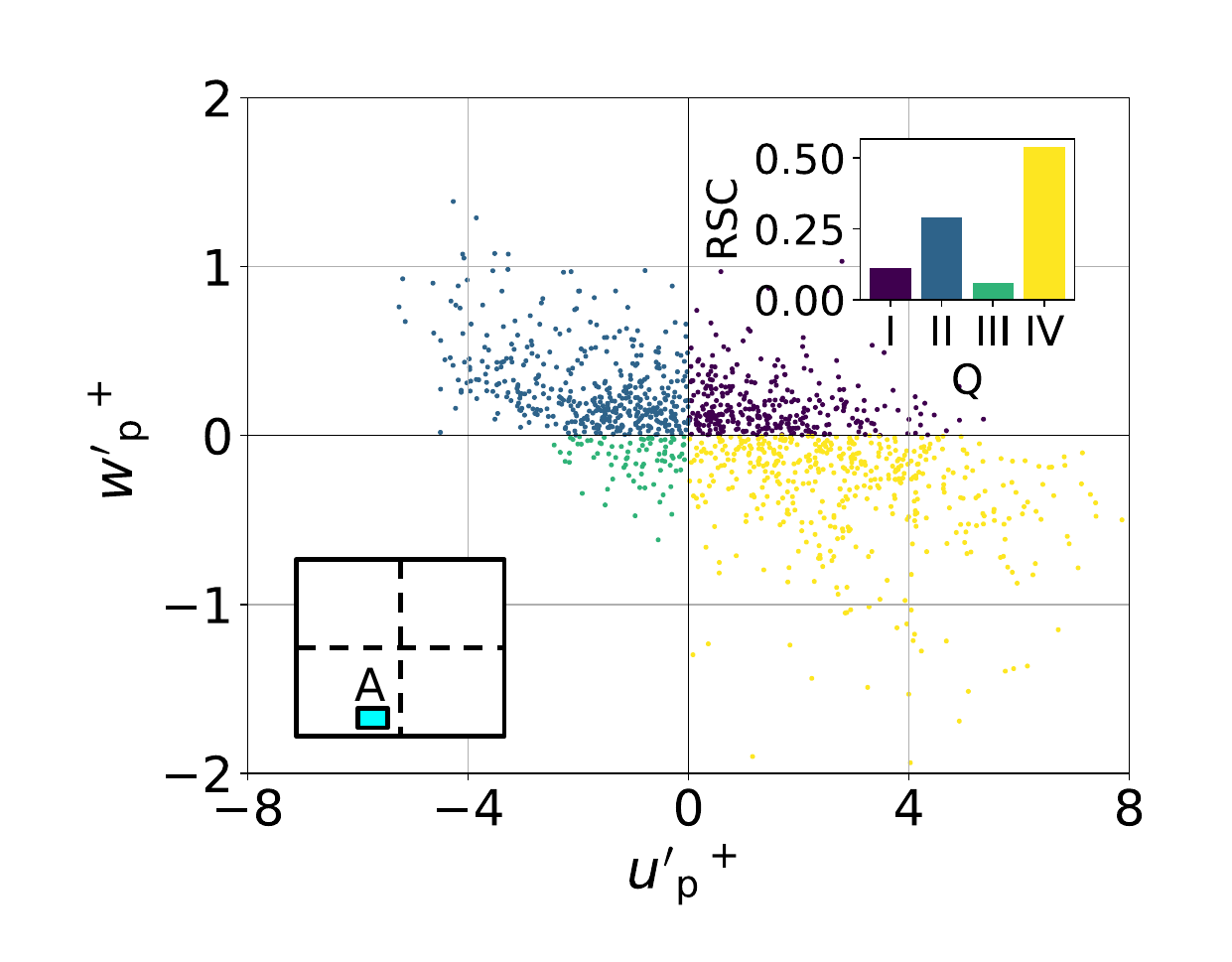}}}
        \caption{\,\,\,\,\,\,\,\,\,\,uncharged, $155\leq y^{+\textup{}}\leq 180$}
        \label{fig:QA_duct_center_uncharged}
    \end{subfigure}
    \hfill
    \begin{subfigure}{0.49\textwidth}
        \centering
        \raisebox{-0.5\height}{\stackinset{l}{0.1ex}{t}{0.1ex}{\text{(b)}}{\includegraphics[width=\textwidth]{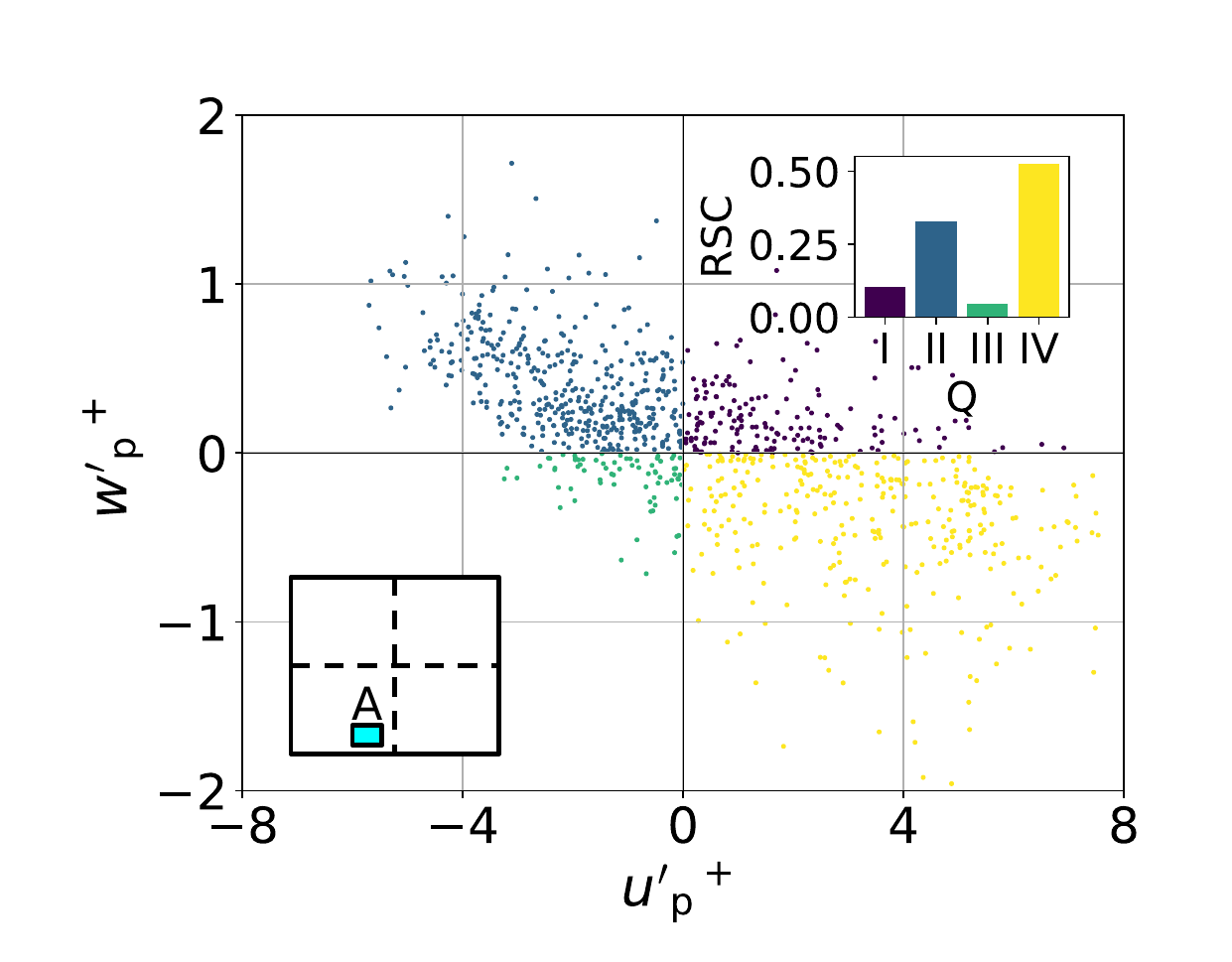}}}
        \caption{\,\,\,\,\,\,\,\,\,\,charged, $155\leq y^{+\textup{}}\leq 180$}
        \label{fig:QA_duct_center_charged}
    \end{subfigure}
    \\
    \begin{subfigure}{0.49\textwidth}
        \centering
        \raisebox{-0.5\height}{\stackinset{l}{0.1ex}{t}{0.1ex}{\text{(c)}}{\includegraphics[width=\textwidth]{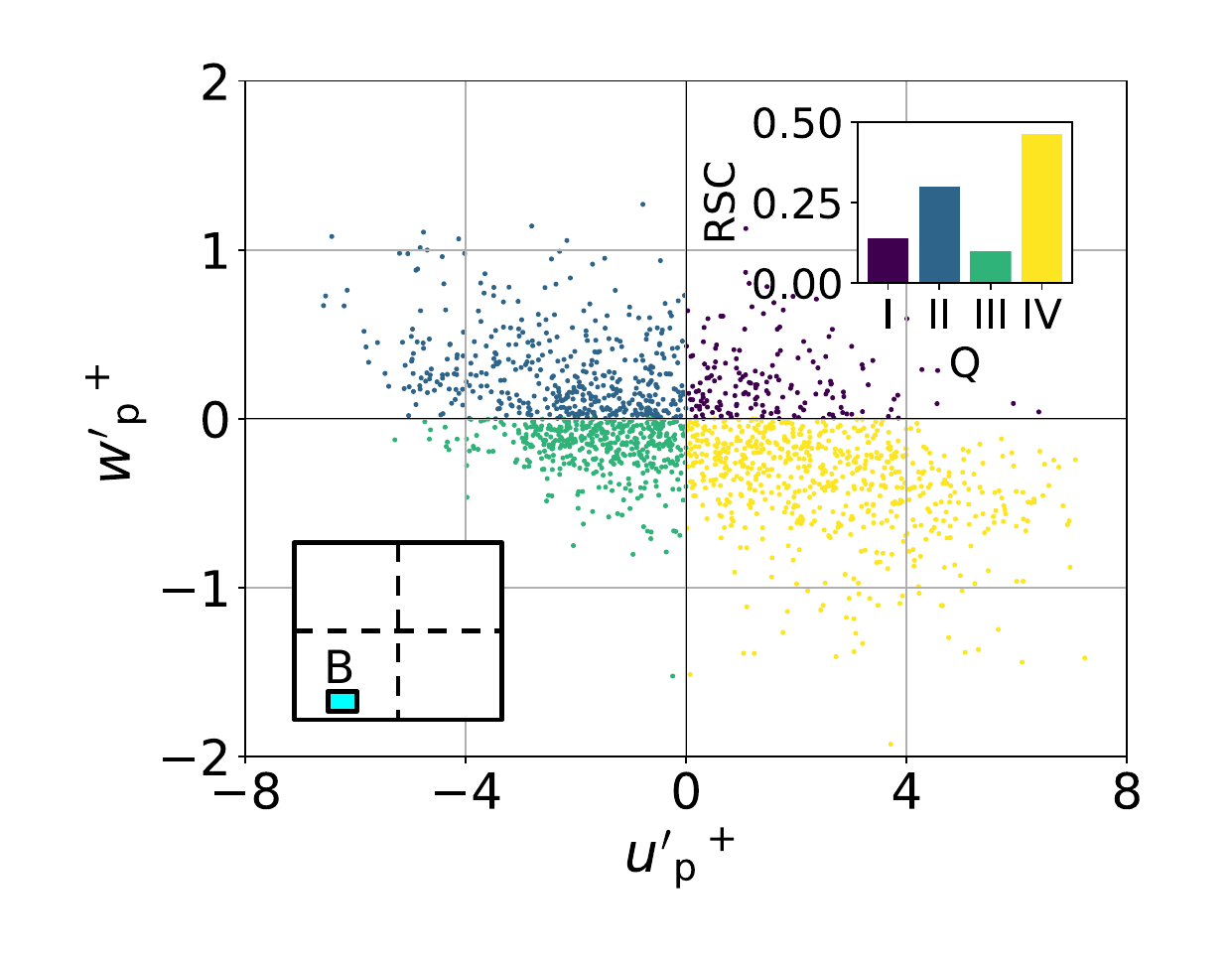}}}
        \caption{\,\,\,\,\,\,\,\,\,\,\,uncharged, $77.5\leq y^{+\textup{}}\leq 102.5$}
        \label{fig:QA_duct_bulk_uncharged}
    \end{subfigure}
    \hfill
    \begin{subfigure}{0.49\textwidth}
        \centering
        \raisebox{-0.5\height}{\stackinset{l}{0.1ex}{t}{0.1ex}{\text{(d)}}{\includegraphics[width=\textwidth]{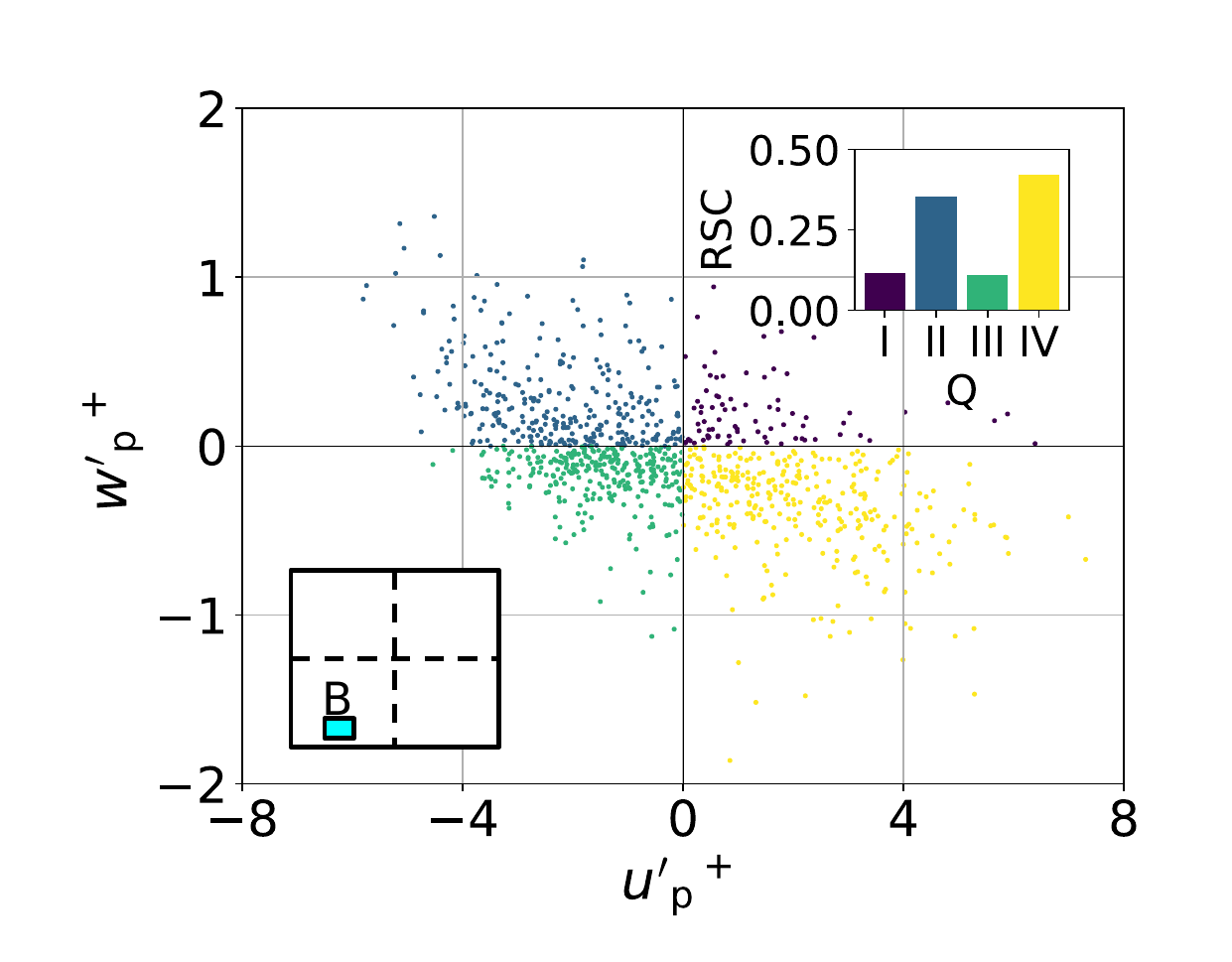}}}
        \caption{\,\,\,\,\,\,\,\,\,\,\,charged, $77.5\leq y^{+\textup{}}\leq 102.5$}
        \label{fig:QA_duct_bulk_charged}
    \end{subfigure}
    \\
    \begin{subfigure}{0.49\textwidth}
        \centering
        \raisebox{-0.5\height}{\stackinset{l}{0.1ex}{t}{0.1ex}{\text{(e)}}{\includegraphics[width=\textwidth]{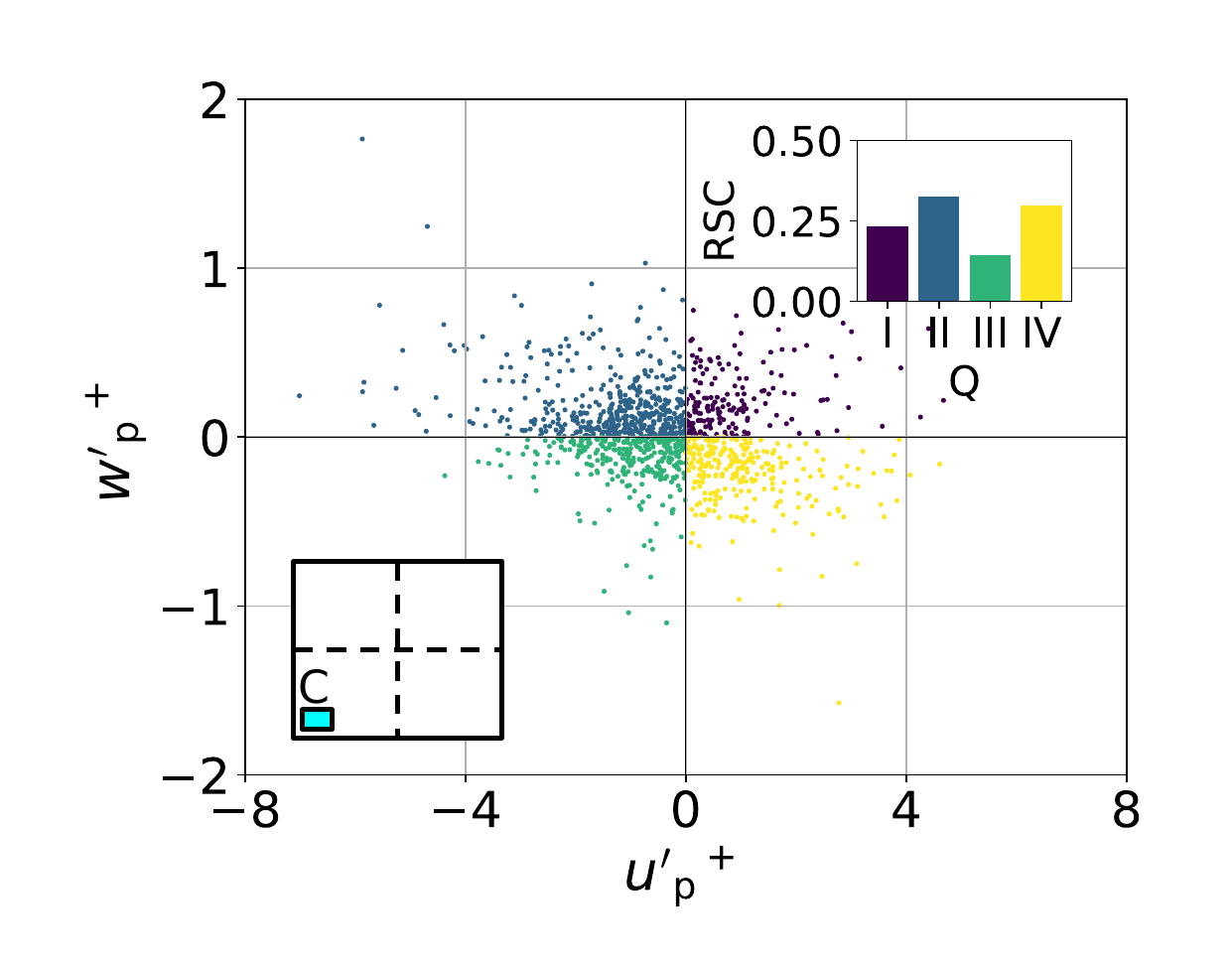}}}
        \caption{\,\,\,\,\,\,\,\,\,\,\,uncharged, $5\leq y^{+\textup{}}\leq 30$}
        \label{fig:QA_duct_corner_uncharged}
    \end{subfigure}
    \hfill
    \begin{subfigure}{0.49\textwidth}
        \centering
        \raisebox{-0.5\height}{\stackinset{l}{0.1ex}{t}{0.1ex}{\text{(f)}}{\includegraphics[width=\textwidth]{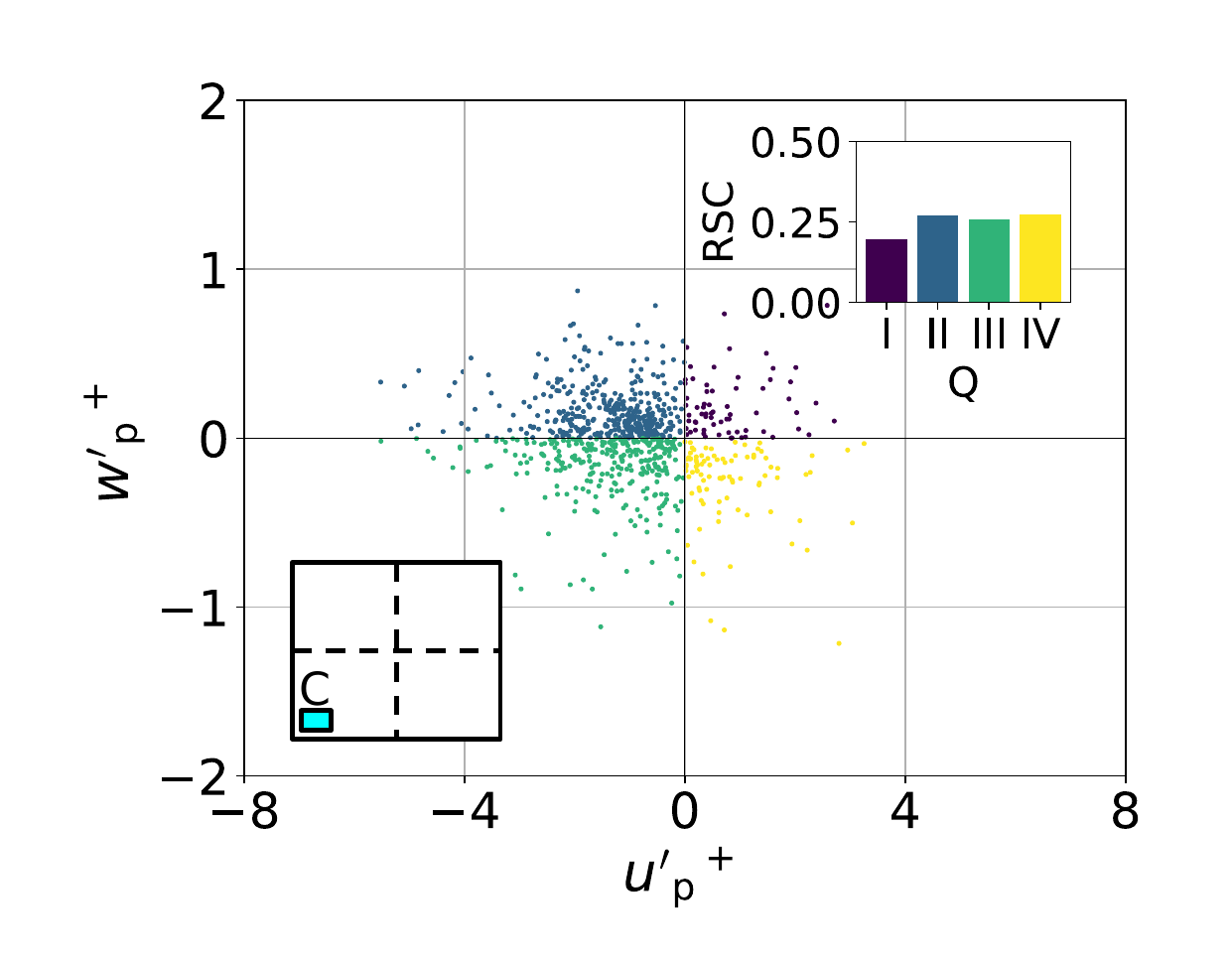}}}
        \caption{\,\,\,\,\,\,\,\,\,\,\,charged, $5\leq y^{+\textup{}}\leq 30$}
        \label{fig:QA_duct_corner_charged}
    \end{subfigure}
    \caption{Quadrant analysis and Reynolds stress contribution (RSC) (upper-right inlet) of uncharged and charged duct flow. The Stokes number is $St$~=~4.69. Figures show the lower-left quarter of the duct in the near-wall region ($5\leq z^{+\textup{}}\leq 30$). For the charged case, the average powder charge is half of the equilibrium charge, $q^*_{\textup{avg}}=0.5$.}
    \label{fig:QA_duct}
\end{figure}

Figure~\ref{fig:QA_duct} presents the quadrant analysis of duct flow, and their contribution to Reynolds stress $\langle \mathit{u}'_{\mathrm{p}}\mathit{w}'_{\mathrm{p}} \rangle$ in the lower-left quarter close to the wall region ($5\leq z^{+\textup{}}\leq 30$). The analysis covers three distinct areas: $155 \leq y^+ \leq 180$ (A),  $77.5 \leq y^+ \leq 102.5$ (B), and  $5 \leq y^+ \leq 30$ (C). The contribution to Reynolds stress for each quadrant is calculated as
\begin{equation}
\textup{Q}_\textup{i}=\frac{1}{\textup{N}_\textup{i}}\sum \left ( u'_\textup{p}w'_\textup{p} \right )_\textup{i}/u_\textup{p,rms}w_\textup{p,rms},
\end{equation}
where $\textup{Q}_\textup{i}$ is the contribution from the $i$th quadrant, $\textup{N}_\textup{i}$ is the number of $i$th quadrant events. Reynolds stress contributions are given as a percentage of the total contributions.

Q2 (ejections) and Q4 (sweeps) events arise from interactions of counter-rotating vortices in wall-bounded turbulent flows. 
In contrast, Q1 and Q3 events do not directly correspond to specific turbulent structures in flows with a single inhomogeneous direction. 
Nevertheless, Q1 and Q3 events significantly influence turbulence dynamics in turbulent duct flows~\citep{Hus93, For18}. 
Thus, while turbulence patterns in a square duct are generally similar to those in a wall-bounded turbulent channel flow, the presence of secondary flows introduces notable differences.

Figure~\ref{fig:QA_duct} shows that overall, Q2 (ejections) and Q4 (sweeps) events contribute significantly to Reynolds stress. 
In region A, Q4 events are responsible for half of the contribution to Reynolds stress alone. 
However, notable increases in the contributions from Q1 and Q3 events appear as we move towards region C. 
Specifically, Q3 events contribute 5.8\% in region A, 9.8\% in region B, and 14.2\% in region C.
Q3 events in region C are driven by interactions between Q2 events from horizontal and vertical walls, which redirect the ejection flow toward the perpendicular wall~\citep{Hus93}. 
Similarly, Q1 event contributions increase from 11.1\%  in region A to 14.0\% in region B and to 23.4\% in region C.

In contrast, Q4 events are most significant in region A and diminish towards the corners, explaining particle accumulation in regions A and B.
Q4 contributions to Reynolds stress are 54.1\% in region A, 46.3\% in region B, and 30.0\% in region C. 
Q2 events are not significantly influenced by proximity to the wall but are slightly reduced in region A.

Upon charging, some particles deposit directly at the walls, which reduces the number of data points.
Charging significantly alters the quadrant analysis, particularly in region C, while its impact in regions A and B is relatively minor.
Upon charging, contributions become almost homogeneously distributed across the quadrants in region C. 
This may stem from the concentrated effect of electrostatic forces at the corners, where attractive forces from image charges on both vertical and horizontal walls amplify the overall electrostatic interaction with particles. 
Specifically, the contribution from Q3 events in region C nearly doubles from 14.2\% to 25.9\%. 
Interestingly, charging also suppresses Q1, Q2, and Q4 events. 
Q4 contributions decrease significantly from 33.0\% to 26.0\%, while Q2 contributions decrease from 32.5\% to 27.0\%, and Q1 drops from 23.4\% to 19.8\%.

\begin{figure}[t]
    \centering
    \captionsetup[subfigure]{labelformat=empty} 
    \begin{subfigure}[b]{0.48\textwidth}
        \centering
        \raisebox{-0.5\height}{\stackinset{l}{0.5ex}{t}{0.5ex}{\text{(a)}}{\includegraphics[width=\textwidth]{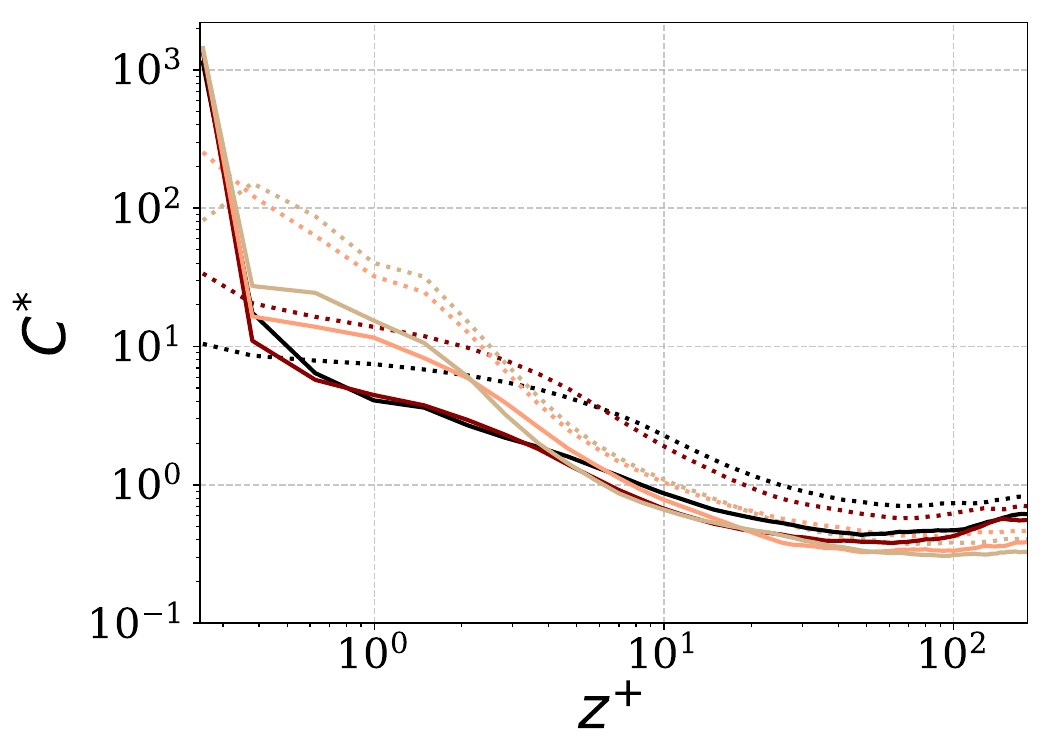}}}
        \caption{}
        \label{fig:C_chan}
    \end{subfigure}
    \hfill
    \begin{subfigure}[b]{0.48\textwidth}
        \centering
        \raisebox{-0.5\height}{\stackinset{l}{0.5ex}{t}{0.5ex}{\text{(b)}}{\includegraphics[width=\textwidth]{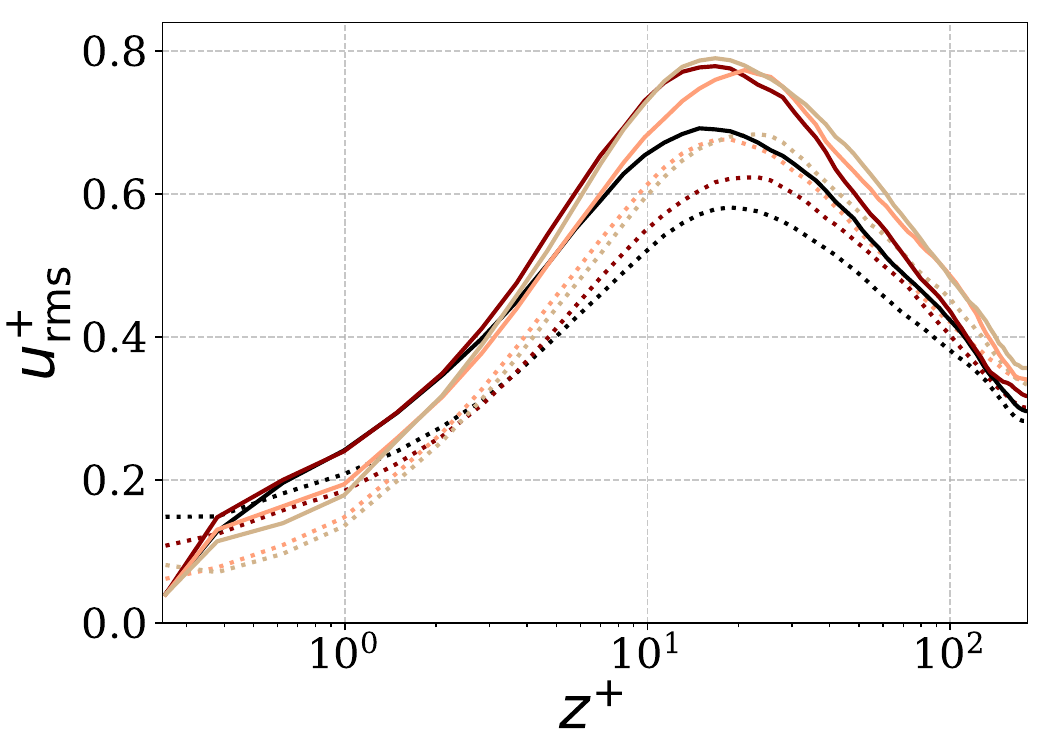}}}
        \caption{}
        \label{fig:urms_chan}
    \end{subfigure}
    \vspace{-1.5\baselineskip}
    \caption{Normalized particle concentration profiles (a), rms stream-wise fluid velocity (b), for uncharged and charged channel flow. The uncharged case is shown by line points while the charged case is shown by lines. For the charged case, the average powder charge in the domain is half of the equilibrium charge, $q^*_{\textup{avg}}=0.5$. Colors indicate Stokes number; 
        (\,\textcolor{black}{\rule[0.2ex]{0.5cm}{1pt}}\,) $St\,=\,37.50$,\,
        (\,\textcolor{darkred}{\rule[0.2ex]{0.5cm}{1pt}}\,) $St\,=\,18.75$,\,
        (\,\textcolor{lightsalmon}{\rule[0.2ex]{0.5cm}{1pt}}\,)$St\,=\,9.38$,\,
        (\,\textcolor{tan}{\rule[0.2ex]{0.5cm}{1pt}}\,)$St\,=\,4.69$.}
    \label{fig:cu_chan}
\end{figure}

Figure~\ref{fig:C_chan} shows normalized particle concentration in the channel flow for uncharged and charged powder.

The concentration profiles in channel flow show a similar pattern to those in duct flows.
Particles are accumulated at the channel walls in uncharged and charged cases.
As particles become charged, there is a significant rise in concentration at the wall, similar to duct flow.
In the charged case, particle concentration directly at the channel walls is less affected by Stokes number than in duct flow.
In both duct and channel flow, particle concentration at the center decreases due to the attraction of charged particles to the walls.

Figure~\ref{fig:urms_chan} shows the rms of fluid velocity in the stream-wise direction for both the uncharged and charged cases.

Overall, fluid velocity fluctuations in the stream-wise direction in channel flow are less pronounced than in duct flow. 
Similar to duct flow, when particles are charged, fluctuations at the wall decrease due to particle accumulation, while fluctuations at the center increase as particles move away from this region.

Thus, while the overall intensity of fluctuations in channel flow is lower than in duct flow, the effect of charging particles on the distribution of these fluctuations remains consistent between the two flow types.

\begin{figure}[t]
    \centering
    \begin{subfigure}{0.49\textwidth}
        \centering
        \raisebox{-0.5\height}{\stackinset{l}{0.1ex}{t}{0.1ex}{\text{(a)}}{\includegraphics[width=\textwidth , height= 4.7cm]{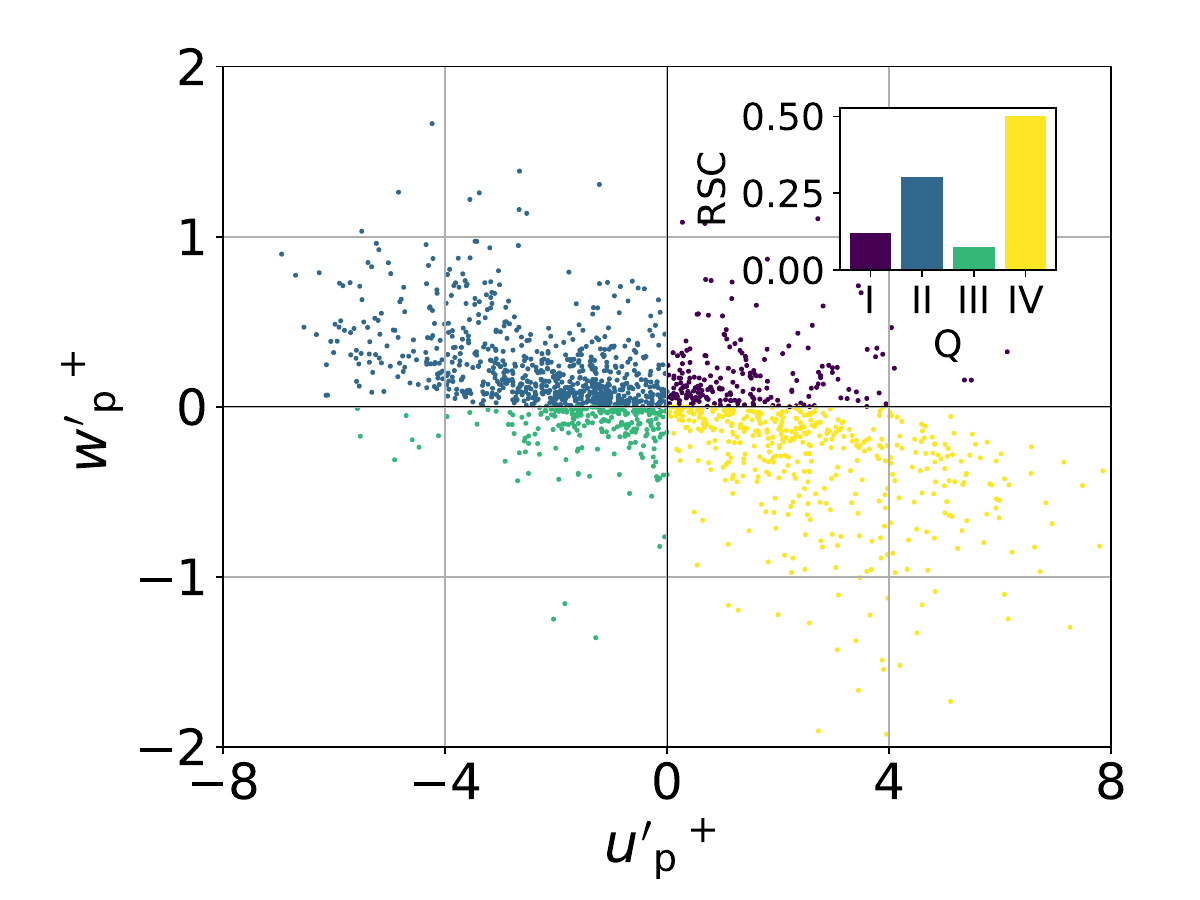}}}
        \caption{}
        \label{fig:QA_chan_uncharged}
    \end{subfigure}
    \hfill
    \begin{subfigure}{0.49\textwidth}
        \centering
        \raisebox{-0.5\height}{\stackinset{l}{0.1ex}{t}{0.1ex}{\text{(b)}}{\includegraphics[width=\textwidth , height= 4.7cm]{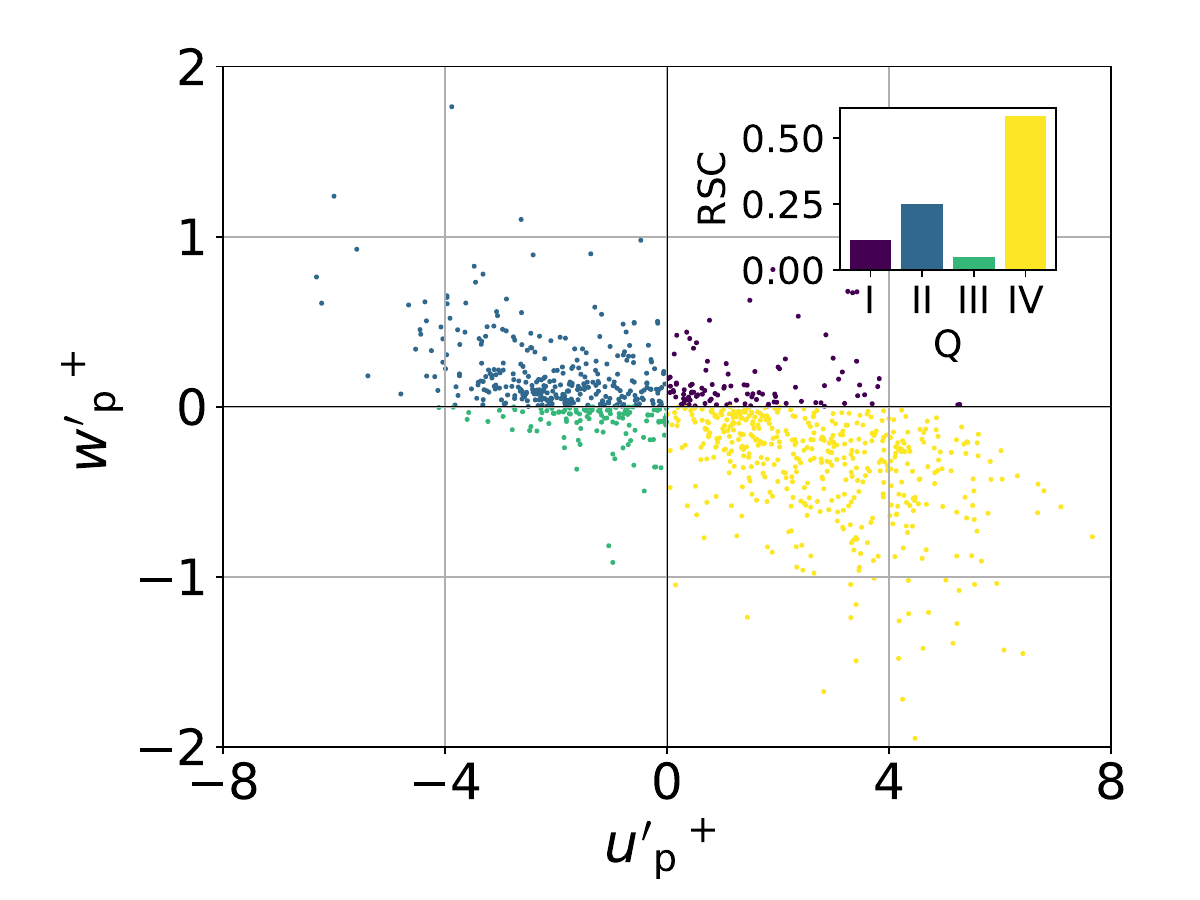}}}
        \caption{}
        \label{fig:QA_chan_charged}
    \end{subfigure}   
    \vspace{-0.5cm}
    \caption{Quadrant analysis and contribution to Reynolds stress (upper-right) of uncharged (a), and charged (b) channel flow. The figure shows the near-wall region ($5\leq z^{+\textup{}}\leq 30$). Stokes number is $St$~=~4.69. For the charged case, the average powder charge is half of the equilibrium charge, $q^*_{\textup{avg}}=0.5$.}
    \label{fig:QA_chan}
\end{figure}

Figure~\ref{fig:QA_chan} shows the quadrant analysis of channel flow at the channel wall for charged and uncharged cases. 
The frequency of Q2 and Q4 events is substantially higher compared to Q1 and Q3 events.
80\% of the Reynolds stress contribution comes from Q2 (ejections) and Q4 (sweeps) events. 
Specifically, Q4 events contribute 50\%, while Q2 events account for 30\%. 
This indicates that these turbulent events, characterized by the outward and inward movement of fluid near the walls, play a dominant role in the overall turbulence dynamics of a channel flow. 
In contrast, Q1 and Q3 events contribute less to Reynolds stress in the channel flow. 
Q1 events account for 11.5\% of the Reynolds stress, and Q3 events contribute 4.9\%.

As particles become charged, they accumulate on the walls, resulting in a decrease in the number of data points.
Regarding Reynolds stress contributions, Q2 and Q4 events exhibit the most significant changes.
Specifically, the contribution from Q2 events decreases from 30\% to 25\%, while the contribution from Q4 events increases from 50\% to 58\%. 
This increase in Q4 events, which involve the sweeping of particles towards the wall, is attributed to the enhanced electrostatic attraction that charged particles experience from the wall. 
Conversely, the decrease in Q2 events, which involve the ejection of particles away from the wall, results from the dominance of electrostatic forces over the turbulent structures responsible for these ejections. 
In contrast, Q1 and Q3 events remain relatively unaffected by charging, unlike in duct flow.

\captionsetup[subfigure]{labelformat=empty}
\captionsetup{justification=justified, width=1\linewidth}

\begin{figure}[t]
    \centering
    \begin{subfigure}{0.45\textwidth}
        \centering
        \raisebox{-0.5\height}{\stackinset{l}{0.1ex}{t}{0.1ex}{\text{(a)}}{\includegraphics[width=\textwidth , height= 4.7cm]{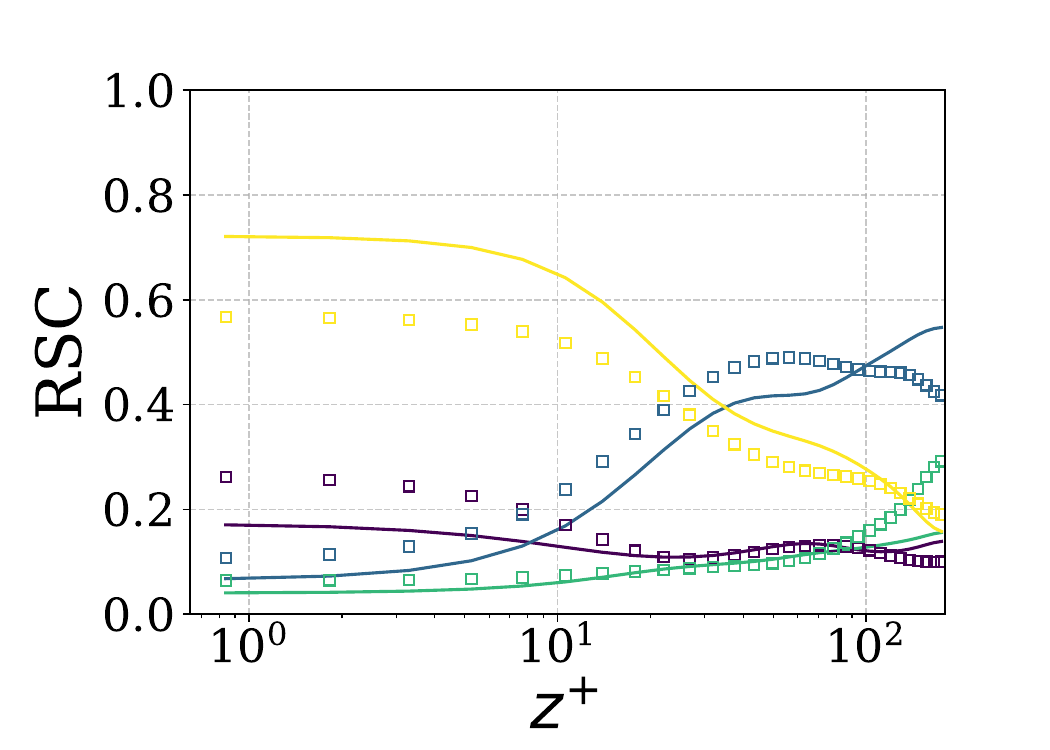}}}
        \caption{}
        \label{fig:QA_wallnormal_channel}
    \end{subfigure}
    \hfill
    \begin{subfigure}{0.45\textwidth}
        \centering
        \raisebox{-0.5\height}{\stackinset{l}{0.1ex}{t}{0.1ex}{\text{(b)}}{\includegraphics[width=\textwidth , height= 4.7cm]{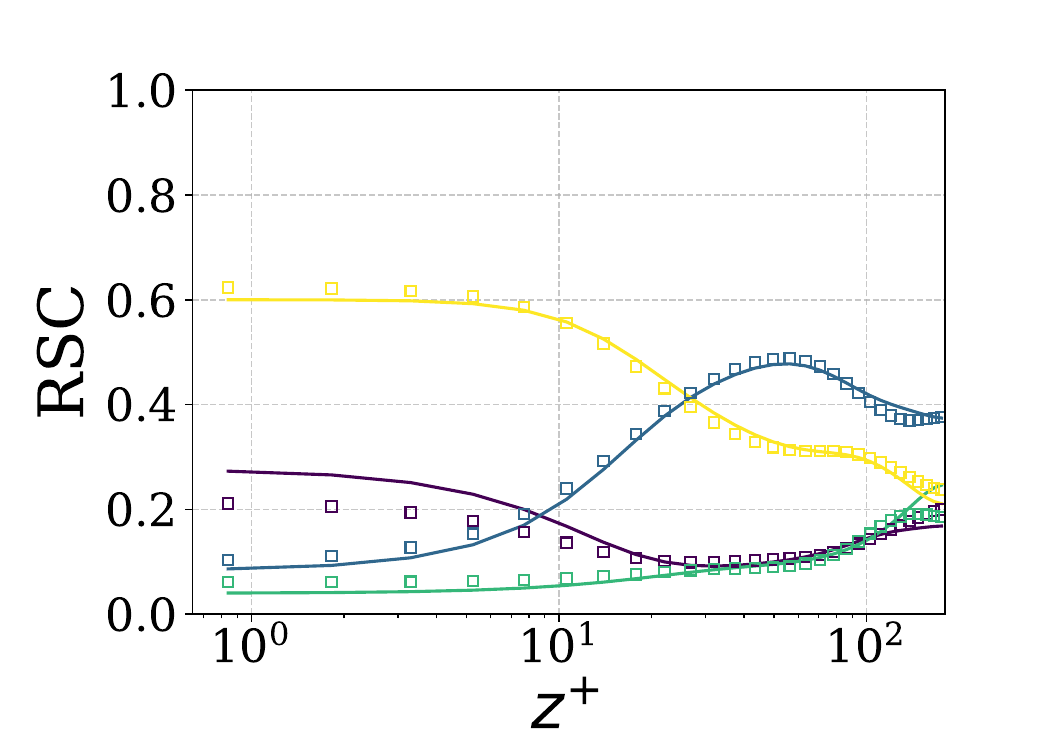}}}
        \caption{}
        \label{fig:QA_wallnormal_duct}
    \end{subfigure} 
    \caption{Wall normal profiles of contribution to particle Reynolds stress for (a) channel flow and (b) duct flow. The wall-normal profiles along the bisector,  $y^{+\textup{}}= 180$, are depicted for duct flow. Lines represent the charged, and the data points represent the uncharged case. For the charged case, the average powder charge in the domain is half of the equilibrium charge, $q^*_{\textup{avg}}=0.5$. Colors indicate different quadrants: 
        (\,\textcolor{QI}{\rule[0.2ex]{0.5cm}{1pt}}\,) QI\,, 
        (\,\textcolor{QII}{\rule[0.2ex]{0.5cm}{1pt}}\,) QII\,, 
        (\,\textcolor{QIII}{\rule[0.2ex]{0.5cm}{1pt}}\,) QIII\,
        (\,\textcolor{QIV}{\rule[0.2ex]{0.5cm}{1pt}}\,) QIV\,.}
    \label{fig:QA_wallnormal} 
\end{figure}

Figure \ref{fig:QA_wallnormal} shows wall-normal profiles of the contribution to particle Reynolds stress in channel and duct flows, comparing charged and uncharged cases. 
When uncharged, contribution profiles from different quadrants demonstrate consistent patterns between channel and duct flows, with slight variations in magnitude. However, the influence of particle charge on the profiles varies significantly across the two flow types. Generally, the changes in the distribution of particle Reynolds stress contributions upon charging are less pronounced in duct flow compared to channel flow. This difference may be due to the presence of secondary flow structures in ducts, which can dominate the effects of electrostatic forces. In contrast, channel flow lacks these secondary structures, leading to more significant changes in the particle behavior when charged.
\captionsetup[subfigure]{labelformat=empty}
\captionsetup{justification=justified, width=1\linewidth}

\captionsetup[figure]{labelfont=default}
In both channel and duct flows, the near-wall region ($z^+<30$) is dominated by QIV events, with approximately 0.6 of the total contributions. Following this, QI events account for about 0.2, while contributions from QII and QIII events are around 0.1. In the turbulent region ($z^+>30$), contributions from QII events significantly exceed QIV events, with QIV contributions decreasing to approximately 0.2 and QII contributions rising to about 0.4. At the center ($z^+=180$), QIII events reach 0.3, surpassing the contributions from QIV in the channel flow while it stays around 0.2 for duct flow.

Upon charging, in the near-wall region, the contribution to particle Reynolds stress from QIV events increases in channel flow but slightly decreases in duct flow. Conversely, QI events show a decrease in channel flow and a slight increase in duct flow. QII events also decrease in channel flow, with no significant change in duct flow. These observations in channel flow indicate that electrostatic forces substantially affect particle dynamics. The increased contribution from QIV suggests that particles are attracted toward the wall due to mirror charges, reducing the contributions from QI and QII events. In duct flow, such behavior is absent, suggesting that secondary flow structures have a greater influence than electrostatic forces.

Charging significantly affects contributions at the center of channel flow. Upon charging, contributions from QI and QII increase, while those from QIII and QIV decrease. The decrease in contributions from QIII and QIV events results from particles depositing on the channel walls upon charging. With fewer particles present at the center, the number of particles moving toward the wall decreases, which increases the relative contribution from other quadrants toward the center. In duct flow, the effect of charging on quadrant contributions is less pronounced: QII remains unchanged between charged and uncharged states, QI and QIV slightly reduce, and QIII increases.

\begin{figure}[t]
    \centering
    \begin{subfigure}{0.49\textwidth}
        \centering
        \raisebox{-0.5\height}{\stackinset{l}{0.1ex}{t}{0.1ex}{\text{(a)}}{\includegraphics[width=\textwidth , height= 4.7cm]{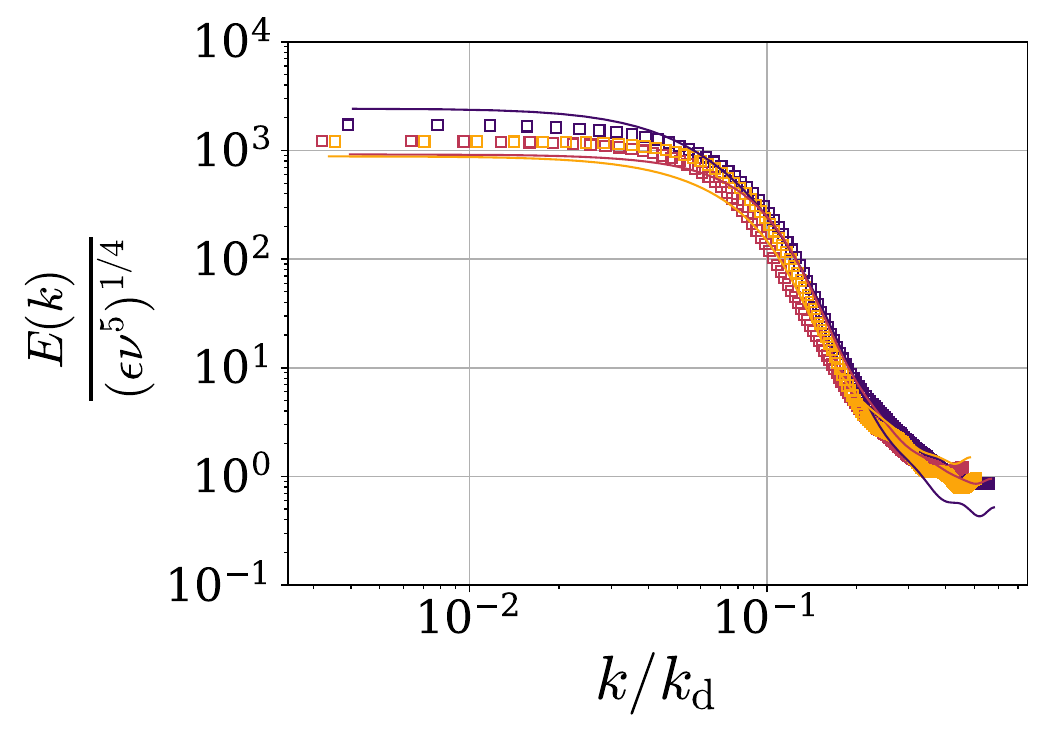}}}
        \caption{}
        \label{fig:EnergySpectra_channel}
    \end{subfigure}
    \hfill
    \begin{subfigure}{0.49\textwidth}
        \centering
        \raisebox{-0.5\height}{\stackinset{l}{0.1ex}{t}{0.1ex}{\text{(b)}}{\includegraphics[width=\textwidth , height= 4.7cm]{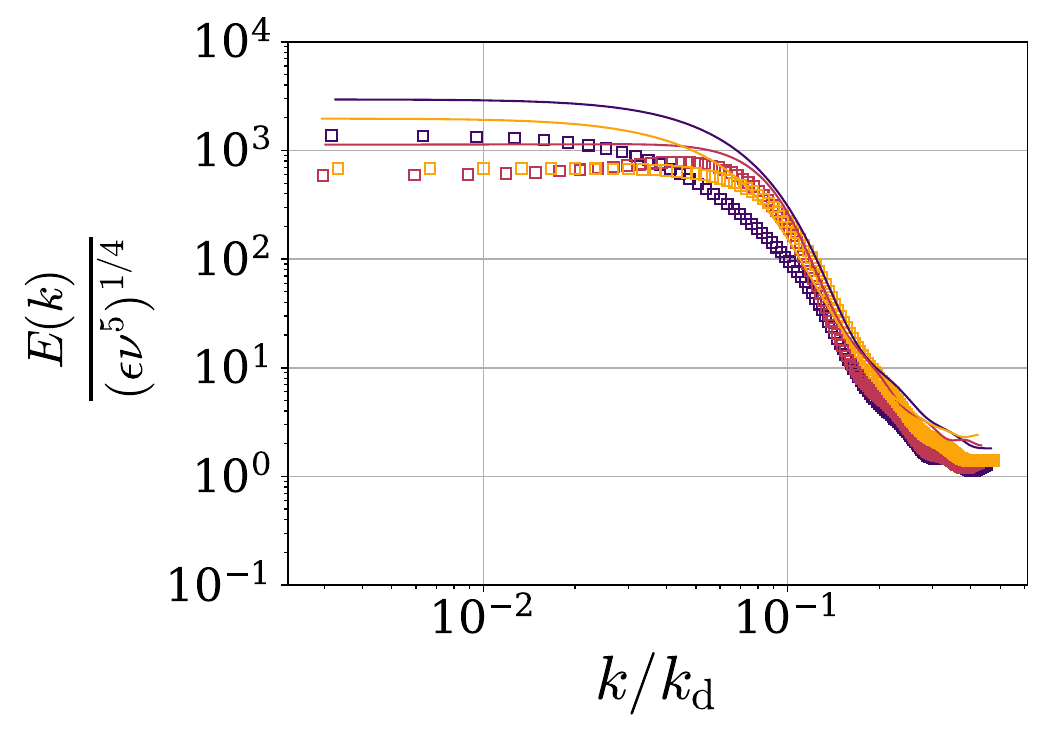}}}
        \caption{}
        \label{fig:EnergySpectra_duct}
    \end{subfigure}   
    \vspace{-0.5cm}
    \caption{One-dimensional energy spectra for (a) channel flow and (b) duct flow, normalized by energy dissipation $\epsilon$ and kinematic viscosity $\nu$, plotted against wavenumber $k$ normalized by the Kolmogorov length scale $k_\mathrm{d}$. The spectra are plotted for $y^+=180$, and $z^+=180$. Lines represent the charged, and the data points represent the uncharged case. For the charged case, the average powder charge is half of the equilibrium charge, $q^*_{\textup{avg}}=0.5$. Colors show different velocity components of fluid flow: 
        (\,\textcolor{uColor}{\rule[0.2ex]{0.5cm}{1pt}}\,) $u$\,, 
        (\,\textcolor{vColor}{\rule[0.2ex]{0.5cm}{1pt}}\,) $v$\,, 
        (\,\textcolor{wColor}{\rule[0.2ex]{0.5cm}{1pt}}\,) $w$\,.}
    \label{fig:EnergySpectra}
    
\end{figure}

Figure~\ref{fig:EnergySpectra} shows the one-dimensional energy spectra for channel and duct flow with charged and uncharged particles. In both flow types, turbulent kinetic energy in the stream-wise direction increases when particles are charged. In duct flow, this increase is more pronounced compared to channel flow. In the channel flow, turbulent kinetic energy is higher for the charged case at low $(k/k_\mathrm{d} = 10^{-2})$ and intermediate $(k/k_\mathrm{d} = 10^{-1})$ wave numbers, while in the duct flow, it spans low, intermediate, and high wave numbers$(k/k_\mathrm{d} > 10^{-1})$. In the channel flow, the kinetic energy for the charged case falls below that of the uncharged case for high wave numbers $k/k_\mathrm{d} =0.2$. Additionally, in duct flow, turbulent kinetic energy in the spanwise and wall-normal directions also increases with particle charging, whereas it decreases in channel flow.

The increase in turbulent kinetic energy, when particles are charged, is caused by particle migration from the center to the walls, where they deposit due to the effect of electrostatic forces, as explained by Figures~\ref{fig:cu_duct} and~\ref{fig:cu_chan}. This migration reduces particle interference with large eddies in the center, leading to an increase in turbulent kinetic energy. Similar findings have been reported in DNS studies of turbulent pipe flows~\citep{Gupta18}, which compare the one-dimensional energy spectra of laden and unladen flows.

\captionsetup[subfigure]{labelformat=empty}

\raggedbottom 
\captionsetup{justification=justified, width=1\linewidth}
\captionsetup[figure]{labelfont=default}

\vspace{0.5cm}
\section {Conclusions}

In this study, we investigated how flow patterns impact the charging of powder flows. 
To separate this effect from the complex nature of particle charging, we employed a generic charging model based solely on the precharge of the particles. 
As a result, powder charging is affected by particle trajectories, which are shaped by the surrounding flow. 
Specifically, powder charging depends on two main factors: the frequency of particle charging at the walls and the rate at which charge spreads from the walls to the bulk via particle-bound transport.

Our study highlights the effect of secondary flow structures on powder charging rates and charge distribution.
We revealed that powder charges faster in duct flow than in channel flow carrying the same particles. 
Furthermore, the charge is more uniformly distributed across the duct cross-section than the channel cross-section. 
This is due to the secondary flow structures in duct flows, which accelerate the particles to the walls.
As a result, in the duct flow, collision rates are higher, which means that charging events are more frequent.
But more importantly, secondary flows also bring particles to the corners, then to bisectors, and finally to the center, promoting particle-bound charge transfer.
Therefore, the charge transferred during each collision event is higher in duct flow.
Conversely, in channel flow, particles tend to trap near the wall and make repeating collisions, limiting the charge transferred during each collision event.

Stokes number of particles significantly impacts their charging behavior in both channel and duct flow scenarios. 
Typically, particles with higher Stokes number, tend to move uniformly along the wall-normal direction, resulting in faster and more uniform charging. 
However, in duct flow, high inertia particles tend to accumulate at the corners, resulting in some particles remaining uncharged.

Upon charging, electrostatic forces draw more particles toward the wall, leading to higher wall accumulation at the walls and fewer particles in the center for both channel and duct flows. 
This shift in particle distribution has impacts on the fluid phase.
Due to the accumulation of particles, stream-wise fluid velocity fluctuations were reduced at the walls while they increased at the center.

The results of this study demonstrate the fundamental influence of flow scenarios on the charging pattern of two-phase flows. 
Understanding these influences gives us new insights into how we can control the charging process in handling powder flows.
The capability to control or limit the accumulation of electric charging of powder flows would enable a better understanding of environmental processes, optimizing industrial devices, and mitigating hazards to their operational safety.



\backsection[Funding]{This research received financial support from the European Research Council (ERC) under the European Union’s Horizon 2020 research and innovation program (grant agreement No. 947606 Pow-FEct).}

\backsection[Declaration of interests]{The authors report no conflict of interest.}



\bibliographystyle{jfm}


\end{document}